\documentclass[runningheads]{llncs}
\usepackage[english]{babel}
\usepackage{amsmath}
\usepackage{amssymb}
\usepackage{colortbl}
\usepackage{txfonts}
\usepackage{xypic}
\usepackage{multirow}
\usepackage{textcomp}
\usepackage{calc}
\usepackage{stmaryrd}
\usepackage{marvosym}
\usepackage{pgf}
\usepackage{ifthen}
\usepackage{mdwlist}
\usepackage{enumerate}
\usepackage{paralist}
\usepackage{xcolor}
\usepackage{bm}
\usepackage{cancel}
\usepackage{pdfpages}
\usepackage{hyperref}
\usepackage{todonotes}

\usepackage{tikz}
\usetikzlibrary{shapes.geometric}
\usetikzlibrary{arrows,automata}
\usetikzlibrary{matrix}
\usetikzlibrary{decorations.pathmorphing}
\usetikzlibrary{decorations.markings}
\usetikzlibrary{decorations.shapes}

\newcommand{\derives}[1]{ \stackrel {{#1}} \Rightarrow}
\DeclareRobustCommand{\newvdash}{\shortmid\joinrel\relbar\joinrel\relbar}
\newcommand{\transition}[1]{\stackrel {{#1}} \newvdash}

\DeclareMathOperator{\sux}{succ}
\DeclareMathOperator{\Sux}{Succ}
\DeclareMathOperator{\Sing}{Sing}

\newcommand{\suchthat}{\;\ifnum\currentgrouptype=16 \middle\fi|\;}

\newcommand{\mrk}[1]{{#1}'}
\newcommand{\oldstack}[3]{%
{\ifthenelse{\equal{#1}{1}}{%
\mrk{#2}
}%
{#2}}_{#3}%
}
\newcommand{\stack}[3]{%
[%
{\ifthenelse{\equal{#1}{1}}{%
\mrk{#2}
}%
{#2}}\ {#3}%
]%
}
\newcommand{\tstack}[2]{%
[#1,\ #2]%
}

\newcommand{\va}[1]{\stackrel{#1}{\longrightarrow}}

\newcommand{\flush}[1]{\stackrel{#1}{\Longrightarrow}}

\makeatletter
\newcommand{\shift}[2][\hspace*{4mm}]{\ext@arrow 0359\rightarrowfill@@{#1}{#2}}
\def\rightarrowfill@@{\arrowfill@@\relax\relbar\rightarrow}
\def\arrowfill@@#1#2#3#4{%
  $\m@th\thickmuskip0mu\medmuskip\thickmuskip\thinmuskip\thickmuskip
   \relax#4#1
   \xleaders\hbox{$#4#2$}\hfill
   #3$%
}
\makeatother

\newcommand{\nont}[4]{\langle^{#1} #2, #3 {}^{#4} \rangle}
\newcommand{\chain}[3]{{}^{#1}\!\left[ #2 \right]\!{}^{#3}}

\newcommand{\tconfig}[3]{\langle #1, \ #2, \ #3 \rangle}

\newcommand{\symb}[1]{\mathop{symbol}(#1)}

\newcommand{\state}[1]{\mathop{state}(#1)}

\newcommand{\parrow}{}
\newcommand{\iarrow}{dashed}

\newcommand{\pair}[2]{\langle {#1},{#2}\rangle} 

\newcommand{\comp}[1]{\transition{#1}}   

\DeclareMathOperator{\tree}{Tree}

\DeclareMathOperator{\xz}{Succ}
\DeclareMathOperator{\avvi}{Next}

\newcommand{\suc}[3]{\xz_{#1}(#2,#3)}
\newcommand{\nex}[3]{\avvi_{#1}(#2,#3)}
\newcommand{\flk}[3]{\nex{#1}{#2}{#3}}

\newcommand{\lp}{\llparenthesis}
\newcommand{\rp}{\rrparenthesis}

\newcommand{\g}[1]{\textbf{\textit{#1}}}

\newcommand{\opa}{OPA}

\newcommand{\markvar}{\g{B}}
\newcommand{\pushvar}{\g{A}}
\newcommand{\flushvar}{\g{C}}

\newtheorem{statement}{Statement}

\newcommand{\tagxml}[2]{%
\langle %
{\ifthenelse{\equal{#1}{1}}{%
/#2  
}%
{#2}}%
\rangle%
}

\makeatother

\tikzset{
	path/.style={dotted},
	every edge/.style={draw,solid},
	normal/.style={solid},
	siblings/.style={dashed},
	support/.style={decorate, decoration={snake,amplitude=.4mm,segment length=2mm,post length=1mm}}
}

\begin{document}

\title{Generalizing input-driven languages:\\ theoretical and practical benefits}

\author{Dino Mandrioli\inst{1}, Matteo
Pradella\inst{1,2}}

\institute{
DEIB -- Politecnico di Milano,
  via Ponzio 34/5, Milano, Italy\\
\and
  IEIIT -- Consiglio Nazionale delle Ricerche,
  via Golgi 42, Milano, Italy\\
\email{\{dino.mandrioli, matteo.pradella\}@polimi.it}
}

\maketitle

\begin{abstract}
Regular languages (RL) are the simplest family in Chomsky's hierarchy. Thanks to their simplicity they enjoy various nice  algebraic and logic properties that have been successfully exploited in many application fields. Practically all of their related problems are decidable, so that they support automatic verification algorithms. Also, they can be recognized in real-time.

Context-free languages (CFL) are another major family well-suited to formalize programming, natural, and many other classes of languages; their increased generative power w.r.t. RL, however, causes the loss of several closure properties and of the decidability of important problems; furthermore they need complex parsing algorithms. Thus, various subclasses thereof have been defined with different goals, spanning from efficient, deterministic parsing to closure properties, logic characterization and automatic verification techniques.

Among CFL subclasses, so-called \textit{structured} ones, i.e., those where the typical tree-structure is visible in the sentences, exhibit many of the algebraic and logic properties of RL, whereas deterministic CFL have been thoroughly exploited in compiler construction and other application fields.

After surveying and comparing the main properties of those various language families, we go back to \textit{operator precedence languages} (OPL), an old family through which R. Floyd pioneered deterministic parsing, and we show that they offer unexpected properties in two fields so far investigated in totally independent ways: they enable parsing parallelization in a more effective way than traditional sequential parsers, and exhibit the same algebraic and logic properties so far obtained only for less expressive language families.

\end{abstract}

\section{Introduction}\label{sec:intro}

Regular (RL) and context-free languages (CFL) are by far the most
widely studied families of formal languages in the richest literature
of the field. In Chomsky's hierarchy, they are, respectively, in
positions 2 and 3, 0 and 1 being recursively enumerable and
context-sensitive languages.

Thanks to their simplicity, RL enjoy practically all positive
properties that have been defined and studied for formal language
families: they are closed under most algebraic operations, and most of
their properties of interest (emptiness, finiteness, containment) are
decidable. Thus, they found fundamental applications in many fields of
computer and system science: HW circuit design and minimization,
specification and design languages (equipped with powerful supporting
tools), automatic verification of SW properties, etc. One of their
most relevant applications is now model-checking which exploits the
decidability of the containment problem and important
characterizations in terms of mathematical logics
\cite{Clarke1986a,Emerson90}.

On the other hand, the typical linear structure of RL sentences makes
them unsuitable or only partially suitable for application in fields
where the data structure is more complex, e.g., is tree-like. For
instance, in the field of compilation they are well-suited to drive
lexical analysis but not to manage the typical nesting of programming
and natural language features.  The classical language family adopted
for this type of modeling and analysis is the context-free one. The
increased expressive power of CFL allows to formalize many syntactic
aspects of programming, natural, and various other categories of
languages. Suitable algorithms have been developed on their basis to
parse their sentences, i.e., to build the structure of sentences as
syntax-trees.

General CFL, however, lose various of the nice mathematical properties
of RL: they are closed only under some of the algebraic operations, and
several decision problems, typically the inclusion problem, are
undecidable; thus, the automatic analysis and synthesis techniques
enabled for RL are hardly generalized to CFL. Furthermore, parsing CFL
may become considerably less efficient than recognizing RL: the
present most efficient parsing algorithms of practical use for general
CFL have an $O(n^3)$ time complexity.

The fundamental subclass of deterministic CFL (DCFL) has been
introduced, and applied to the formalization of programming language
syntax, to exploit the fact that in this case parsing is in $O(n)$. DCFL,
however, do not enjoy enough algebraic and logic properties to extend
to this class the successful applications developed for RL: e.g.,
although their equivalence is decidable, containment is not; they are
closed under complement but not under union, intersection,
concatenation and Kleene~$^*$.

From this point of view, \textit{structured CFL} are somewhat in
between RL and CFL. Intuitively, by structured CFL we mean languages
where the structure of the syntax-tree associated with a given
sentence is immediately apparent in the
sentence.\textit{ Parenthesis languages (PL)} introduced in a
pioneering paper by McNaughton \cite{McNaughton67} are the first
historical example of such languages. McNaughton showed that they
enjoy closure under Boolean operations (which, together with the
decidability of the emptiness problem, implies decidability of the
containment problem) and their generating grammars can be minimized in
a similar way as \textit{finite state automata} (FSA) are minimized
(in fact an equivalent formalism for parenthesis languages are
\textit{tree automata} \cite{Tha67,TATA}).

Starting from PL various extensions of this family have been proposed
in the literature, with the main goal of preserving most of the nice
properties of RL and PL, yet increasing their generative power; among
them \textit{input-driven languages (IDL)}
\cite{DBLP:conf/icalp/Mehlhorn80,Input-driven}, later renamed
\textit{visibly pushdown languages (VPL)} \cite{jacm/AlurM09} have
been quite successful: they are closed under all traditional
language operations (and therefore enjoy the consequent decidability
properties). Also, they are characterized in terms of a
\textit{monadic second order} (MSO) logic by means of a natural
extension of the classic characterization for RL originally and
independently developed by to B\"uchi, Elgot, and Trakhtenbrot
\cite{bib:Buchi1960a,Elg61,Tra61}.  For these reasons they are a natural
candidate for extending model checking techniques from RL. To achieve
such a goal in practice, however, MSO logic is not yet tractable due
to the complexity of its decidability problems; thus, some research is
going on to ``pair'' IDL with specification languages inspired by
temporal logic as it has been done for RL \cite{lmcs/AlurABEIL08}.
 
On the other hand being IDL structured their parsing problem trivially
scales down to the recognition problem but clearly they are not
suitable to formalize languages whose structure is hidden from the
surface sentences as, e.g., typical arithmetic expressions.

Rather recently, we resumed the study of an old class of languages which was interrupted a long time ago, namely \textit{operator precedence
  languages (OPL)}. OPL and their generating grammars (OPG) have been
introduced by Floyd \cite{Floyd1963} to build efficient deterministic
parsers; indeed they generate a large and meaningful subclass of
DCFL. In the past their algebraic properties, typically closure under
Boolean operations \cite{Crespi-ReghizziMM1978}, have been
investigated with the main goal of designing inference algorithms for
their languages \cite{Crespi-ReghizziCACM73}. After that, their
theoretical investigation has been abandoned because of the advent of
more powerful grammars, mainly LR ones \cite{KnuthLR65,Harrison78},
that generate all DCFL (although some deterministic parsers based on
OPL's simple syntax have been continuously implemented at least for
suitable subsets of programming languages \cite{GruneJacobs:08}).

The renewed interest in OPG and OPL has been ignited by two seemingly
unrelated remarks: on the one hand we realized that they are a proper
superclass of IDL and that all results that have been obtained for
them (closures, decidability, logical characterization) extend
naturally, but not trivially, to OPL; on the other hand new motivation
for their investigation comes from their distinguishing property of
\textit{local parsability}: with this term we mean that their
deterministic parsing can be started and led to completion from any
position of the input string unlike what happens with general
deterministic pushdown automata, which must necessarily operate
strictly left-to-right from the beginning of the input. This property
has a strong practical impact since it allows for exploiting modern
parallel architectures to obtain a natural speed up in the processing
of large tree-structured data. An automatic tool that generates
parallel parsers for these grammars has already been produced and is
freely available. The same local parsability property can also be
exploited to incrementally analyze large structures without being
compelled to reparse them from scratch after any modification thereof.

This renewed interest in OPL has also led to extend their study to
\emph{$\omega$-languages}, i.e., those consisting of infinite strings:
in this case too the investigation produced results that perfectly
parallel the extension of other families, noticeably RL and IDL, from
the finite string versions to the infinite ones.

In this paper we follow the above ``story'' since its beginning to these
days and, for the first time, we join within the study of one single language family two different application
domains, namely parallel parsing on the one side, and algebraic and logic
characterization finalized to automatic verification on the other
side.  To the
best of our knowledge, OPL is the largest family that enjoys all of such
properties.

The paper is structured as follows: initially we resume some
background on the two main families of formal languages. 
Section \ref{sec:reg} introduces RL and various formalisms widely used
in the literature to specify them; special attention is given to their
logic characterization which is probably less known to the wider
computer science audience and is crucial for some fundamental results
of this paper. Section \ref{sec:cf} introduces CFL in a parallel way
as for regular ones. Then, two sections are devoted to different
subclasses of CFL that allowed to obtain important properties
otherwise lacking in the larger original family: precisely, Section
\ref{sec:strCF} deals with \textit{structured CFL}, i.e., those
languages whose tree-shaped structure is in some way immediately
apparent from their sentences with no need to parse them; it shows
that for such subclasses of CFL important properties of RL, otherwise
lacking in the larger original family, still hold. Section
\ref{sec:DCFPars}, instead, considers the problem of efficient
deterministic parsing for CFL.  Finally, Section \ref{sec: OPG} ``puts
everything together'' by showing that OPL on the one hand are significantly
 more expressive than traditional structured languages, but enjoy the
same important properties as regular and structured CFL and, on the
other hand, enable exploiting parallelism in parsing much more
naturally and efficiently than for general deterministic CFL.

Since all results presented in this paper have already appeared in the
literature, we based our presentation more on intuitive explanation
and simple examples than on detailed technical constructions and
proofs, to which appropriate references are supplied for the
interested reader. However, to guarantee self-containedness, some
fundamental definitions have been necessarily given in full formality.
For convenience we do not add a final 's' to acronyms when used as
plurals so that, e.g., CFL denotes indifferently a single language,
the language family and all languages in the family.


\section{Regular Languages}\label{sec:reg}

The  family of regular languages (RL) is one of the most important families in
computer science. In the traditional Chomsky's hierarchy it is the least powerful language family. Its importance stems from both its
simplicity and its rich set of properties. 

RL is a very robust family, as it enjoys many properties and
practically all of them are decidable. It is defined through several
different devices, both operational and descriptive. Among them we
mention {\em Finite State Automata} (FSA), which are used for various
applications, not only in computer science, {\em Regular Grammars},
{\em Regular Expressions}, often used in computing for describing
the lexical elements of programming languages and in many programming
environments for managing program sources, and various logic
classifications that support automatic verification of their
properties.

\subsection{Regular Grammars and Finite State Automata}\label{sec:reg:grammars}

Chomsky introduced his seminal hierarchy in 1956. The interested
reader may refer to a manual of formal languages, such as
\cite{Harrison78,CBM13} for a more complete description.

Grammars are generative rewriting systems, and consist of a set of rewriting
rules, also called productions. Intuitively, a production $x \to y$ means that a
string $x$ can be replaced with a string $y$ in the rewriting.

\begin{definition}\label{def:grammar}
    A {\em grammar} $G$ is a tuple $(V_N, \Sigma, P, S)$, 
    where 
    \begin{itemize}
        \item $V_N$, called {\em nonterminal alphabet}, 
        and $\Sigma$, called {\em terminal alphabet},
    are finite sets of characters such that
    $V_N \cap \Sigma = \emptyset$; 
     $V_N \cup \Sigma$ is named $V$; 
\item P is a finite set of {\em productions} or {\em rules} having the form
    $V^* V_N V^* \to V^*$;  in a production $x \to y$, $x$ is called the {\em
      left-hand side} (abbreviated as lhs), and $y$ the {\em
            right-hand side} (abbreviated as rhs);
\item $S \in V_N$ is the {\em starting character} (or {\em axiom}).
    \end{itemize}
\end{definition}

The following \emph{naming conventions} are adopted for letters and strings, unless otherwise specified: 
lowercase Latin letters at the beginning of the alphabet $a,b,\ldots$ denote terminal characters; uppercase Latin
letters $A,B, \ldots$ denote nonterminal characters; lowercase Latin letters at the end of the alphabet $x, y,z \ldots$ denote terminal strings; lowercase Greek  letters $\alpha, \ldots, \omega$ denote strings over $V$.

To define the language of a grammar, we need the concept of {\em derivation}.
Grammars are a generative device, because with them one always starts with the
axiom (usually the $S$ character), then {\em derives} new strings by applying an iterative rewriting as specified by grammar's productions, until a string belonging to $\Sigma^*$ is obtained.
This intuition is formalized by the following definition.
 
\begin{definition}\label{def:derivation}
    Consider a grammar $G = (V_N, \Sigma, P, S)$. 
    A string $\beta$ is {\em derivable in one step} from a string $\alpha$, denoted by $\alpha \derives{}_G \beta$ 
    iff $\alpha = \alpha_1 \alpha_2 \alpha_3$, $\beta = \alpha_1 \alpha'_2 \alpha_3$, and $\alpha_2 \to \alpha'_2 \in P$.\\
    The reflective and transitive closure of the relation $\derives{}_G$ is denoted by
    $\derives{*}_G$.
    The language L(G) generated by G is the set $\{ x \in \Sigma^* \mid S \derives{*}_G x\}$.
\end{definition}


Regular languages are defined by restricted grammars, called {\em regular grammars} (also {\em left-(or right-)linear grammars}).

\begin{definition}\label{def:reg-grammar}
A grammar $G = (V_N, \Sigma, P, S)$, such that 
$
P \subseteq \{ X \to Y a \mid X \in V_N, Y \in V_N \cup \{\varepsilon\}, a \in
\Sigma \},
$ is called {\em left-linear}.
Symmetrically, if its rules have the form $X \to a Y$,
$G$ is called {\em right-linear}.\\
A grammar is called {\em regular} if it is either left-linear or right-linear. A language is regular if it is generated by some regular grammar.
\end{definition}

\begin{example}\label{ex:reg:gram}
Consider a right-linear grammar 
$G_1 = (V_N,$ $\Sigma,$ $P, S)$, where $V_N = \{S, B\}$, $\Sigma = \{a,b\}$, 
$P = \{S \to a S, S \to a B, S \to a, B \to b B, B \to b \}$.

For simplicity and shortness, it is customary to group together productions with
the same lhs, writing all the corresponding right parts separated by $|$. In
this case, 
$P = \{ S \to a S \mid a B \mid a, B \to b B \mid b \}$. It is also customary to drop
$G$ from the relation $\derives{}_G$, if the grammar is clear from the context.

Here is an example derivation: 
$
S 
\derives{} a S 
\derives{} a a S 
\derives{} a a a B
\derives{} a a a b B
\derives{} a a a b b.
$
Hence we see that $aaabb \in L(G_1)$. By simple reasoning we obtain $L(G_1) = \{a^+ b^*\}$.  
\end{example}


Probably the most known and used notation to define RL is that of {\em
Finite State Automata}. An automaton is a simple operational device, which
assumes a {\em state} from a finite set, and its behavior is based on the concept
of {\em state transition}.

\begin{definition}\label{reg:aut}
A {\em Finite State Automaton} (FSA) $\mathcal A$ is a tuple $(Q, \Sigma,
\delta, q_0, F)$, where 
\begin{itemize}
\item $Q$ is a finite set of {\em states};
\item $\Sigma$ is the {\em input alphabet};
\item $\delta \subseteq Q  \times \Sigma \times Q$ is the {\em transition relation};
\item $I \subseteq Q$ is the set of {\em initial states};
\item $F \subseteq Q$ is the set of {\em final states}.
\end{itemize}
If $I$ is a singleton and $\delta$ is a function, i.e., $\delta : Q \times \Sigma \to Q$, $\mathcal{A}$ is called {\em
  deterministic} (or DFSA).
\end{definition}  
Whereas grammars \emph{generate} strings through derivations, automata
\emph{recognize} or \emph{accept} them through sequences of \emph{transitions}
or \emph{runs}. In the case of FSA their recognition is formalized as follows.
\begin{definition}\label{reg:aut:trans}
Let the FSA $\mathcal A$ be $(Q, \Sigma, \delta, I, F)$. 
The {\em transition sequence} of $\mathcal A$ starting from a state $q \in Q$,
 reading a string $x \in \Sigma^*$, and ending in a state $q'$ is written $(q, x, q') \in \delta^*$ where the relation $\delta^* \subseteq Q \times \Sigma^* \times Q$ is inductively defined as follows:
\begin{itemize}
\item $(q, \varepsilon, q)\in \delta^*$;
\item $(q, y, q')\in \delta^* \land  (q', a, q'') \in \delta \implies (q, y a,
  q'') \in \delta^*$, for $q, q', q'' \in Q$, $a \in \Sigma, y \in \Sigma^*$.
\end{itemize}

\noindent The {\em language} accepted by $\mathcal{A}$ is defined as:
$L(\mathcal{A}) = \{ x \mid \exists q_0 \in I, \exists q_F \in F: (q_0, x, q_F) \in \delta^*\}.$
\end{definition}

\begin{example}\label{ex:reg:aut}
Figure~\ref{ex:reg:a1} shows a classical representation of an example automaton, where  $Q$ is
$\{q_0, q_1\}$, $q_0$ is marked as initial by the leftmost arrow; $F = \{q_1\}$
and the ball of the final state has a double line. Transitions are depicted
through labeled arrows, e.g. the arrows connecting $q_0$ to $q_0$ itself and to $q_1$ mean that
$(q_0, a, q_0) \in \delta$ and  $(q_0, a, q_1)\in \delta$. 

\begin{figure}
\begin{center}
\begin{tikzpicture}[every edge/.style={draw,solid}, node distance=3cm, auto, 
                    every state/.style={draw=black!100,scale=0.5}, >=stealth]

\node[initial by arrow, initial text=,state] (q0) {{\huge $q_0$}};
\node[state] (q1) [right of=q0, accepting] {{\huge $q_1$}};

\path[->]
(q0) edge [above] node {$a$} (q1)
(q0) edge [loop above]  node {$a$} (q0)
(q1) edge [loop above]  node {$b$} (q1);
\end{tikzpicture}
\caption{The FSA $\mathcal A_1$ recognizing the language $L(G_1)$ of Example~\ref{ex:reg:gram}.}\label{ex:reg:a1}
\end{center}
\end{figure}
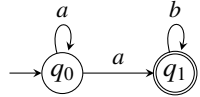
\end{example}

It is easy to see that $L(\mathcal A_1) = \{a^+ b^*\}$. 
E.g.,
$ (q_0, a, q_0)\in \delta$ implies $(q_0, a, q_0)\in \delta^*$ and 
 $(q_0, a, q_1) \in \delta$ implies $(q_0, a, q_1)\in \delta^*$; 
from the same transitions we have 
$\{(q_0, aa, q_0)$,
$(q_0, aa, q_1),$
$(q_0, aaa, q_0),$ 
$(q_0, aaa, q_1)\} \subseteq \delta^*$; 
then, $(q_1, b, q_1)\in$ $\delta$ and $(q_0, aaa,$ $q_1)$ $\in$ $\delta^*$ 
imply $(q_0, aaab, q_1)\in \delta^*$; 
$(q_1, b, q_1) \in  \delta$ and $(q_0, aaab,$ $ q_1)\in \delta^*$ 
imply $(q_0, aaabb, q_1)\in \delta^*$.
Hence, $aaabb \in L(\mathcal{A}_1)$, since $q_1 \in F$. 


By comparing the grammar of Example \ref{ex:reg:gram} with the automaton of Example \ref{ex:reg:aut} it is immediate to generalize their equivalence to the whole class RL: for any regular grammar an equivalent FSA can be built and conversely.

An important result on FSA is that they are {\em determinizable}, i.e., given a
generic FSA, it is always possible to define an equivalent DFSA.
The main idea is to note that the power set of a finite set is still
finite, so if we have a FSA with a set of states $Q$, we can define another
automaton with a set of states $\mathcal{P}(Q)$ so that the original transition
relation can be made a function. 
More formally, given an FSA $(Q, \Sigma, \delta, I, F)$,
a deterministic equivalent automaton is 
$\mathcal A_D = (\mathcal{P}(Q), \Sigma, \delta_D, \{I\}, F_D)$, where 
$\delta_D\left(q, a, \bigcup_{q' \in q, \ (q',a,q'')\in \delta} q''\right)$,
$q \in \mathcal{P}(Q)$, $a \in \Sigma$,
 and 
$F_D = \{Q' \subseteq Q \mid Q' \cap F \ne \emptyset \}.$

 We illustrate this construction with an example.

\begin{example}\label{ex:reg:aut3}
The automaton $\mathcal A_1 = (Q, \Sigma, \delta, \{q_0\}, F)$ of Example \ref{ex:reg:aut} is not deterministic. 
A deterministic equivalent automaton is shown in Figure~\ref{ex:reg:a2}.

\begin{figure}[h]
\begin{center}
\begin{tikzpicture}[every edge/.style={draw,solid}, node distance=3.5cm, auto, 
                    every state/.style={draw=black!100,scale=0.5}, >=stealth]

\node[initial by arrow, initial text=, state] (q0) {{\huge $\{q_0\}$}};
\node[state] (q1) [right of=q0, accepting] {{\Large $\{q_0,q_1\}$}};
\node[state] (q2) [right of=q1, accepting] {{\huge $\{q_1\}$}};

\path[->]
(q0) edge [above] node {$a$} (q1)
(q1) edge [loop above]  node {$a$} (q1)
(q1) edge [above]  node {$b$} (q2)
(q2) edge [loop above]  node {$b$} (q2);
\end{tikzpicture}
\caption{The DFSA $\mathcal A_2$ recognizing the language $L(G_1)$ of Example~\ref{ex:reg:gram}.}\label{ex:reg:a2}
\end{center}
\end{figure}
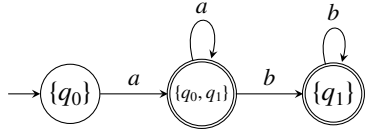

\end{example}

RL enjoy many interesting closure properties, for instance they are closed
w.r.t. all Boolean operations (union, intersection, and complement).

As far as complement is concerned, to build up the related automaton, one must
start with a deterministic automaton (if not, we have just shown how to define
an equivalent deterministic automaton).
First, we modify the original automaton so
that it reads any possible string -- in this way, if the input string is not in
the automaton's language, it changes its state to an additional ``error'' state,
if needed, so that the function $\delta$ is total. 
Then, we exchange the role between accepting and non-accepting states, so that a
rejected string becomes part of the language, and vice versa.
We illustrate this construction with our running example.

\begin{example}\label{ex:reg:complement}
We start with the automaton recognizing $L = a^+ b^*$ of Figure~\ref{ex:reg:a2}, since
it is deterministic; we also rename its states, for brevity.

Then, we introduce an error state, called $q_e$, and swap accepting with
non-accepting roles: the resulting automaton is shown in Figure~\ref{ex:reg:a3}.
It is easy to see that $L(\mathcal A_3) = \neg L(\mathcal A_2) = \neg L(G_1)$.

\begin{figure}[h]
\begin{center}
\begin{tikzpicture}[every edge/.style={draw,solid}, node distance=3cm, auto, 
                    every state/.style={draw=black!100,scale=0.5}, >=stealth]

\node[initial by arrow, initial text=, state, accepting] (q0) {{\huge $q_0$}};
\node[state] (q1) [right of=q0] {{\huge $q_1$}};
\node[state] (q2) [right of=q1] {{\huge $q_2$}};
\node[state] (qE) [below of=q1, accepting] {{\huge $q_e$}};

\path[->]
(q0) edge [above] node {$a$} (q1)
(q1) edge [loop above]  node {$a$} (q1)
(q1) edge [above]  node {$b$} (q2)
(q2) edge [loop above]  node {$b$} (q2)
(q0) edge [below]  node {$b$} (qE)
(q1) edge [left]  node {$b$} (qE)
(q2) edge [left]  node {$a$} (qE)
(qE) edge [loop right]  node {$a, b$} (qE)
;
\end{tikzpicture}
\caption{The DFSA $\mathcal A_3$ recognizing the complement of language $L(G_1)$ of Example~\ref{ex:reg:gram}.}\label{ex:reg:a3}
\end{center}
\end{figure}
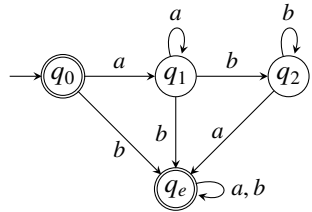
\end{example}

To prove that RL are closed w.r.t. intersection, we start from two 
FSA 
$\mathcal A' = (Q', \Sigma,$ $\delta', I', F')$, 
$\mathcal A'' = (Q'', \Sigma,$ $\delta'', I'', F'')$,
and build a
``product'' automaton,
\[
\mathcal A_p = (Q' \times Q'', \Sigma, \delta_p, I' \times I'',
F' \times F''),
\] 
where
$(\langle q_1, q_1' \rangle, a, \langle q_2, q_2' \rangle)\in \delta_p$, iff 
$(q_1, a, q_2)\in \delta'$, and
$(q_1', a, q_2')$ $\in$ $\delta''$.

Closure w.r.t. union follows from De Morgan's laws.

\begin{example}\label{ex:reg:intersection}
Let us consider the automaton of Figure~\ref{ex:reg:a1} and apply the
construction for intersection with the automaton of Figure~\ref{ex:reg:a4}.
The resulting automaton is in Figure~\ref{ex:reg:aprod}.

\begin{figure}[h]
\begin{center}
\begin{tikzpicture}[every edge/.style={draw,solid}, node distance=3cm, auto, 
                    every state/.style={draw=black!100,scale=0.5}, >=stealth]

\node[initial by arrow, initial text=,state] (q0) {{\huge $q_0$}};
\node[state] (q1) [right of=q0] {{\huge $q_1$}};
\node[state] (q2) [right of=q1, accepting] {{\huge $q_2$}};

\path[->]
(q0) edge [loop above]  node {$b$} (q0)
(q0) edge [above] node {$a$} (q1)
(q1) edge [above] node {$a$} (q2)
(q2) edge [loop above]  node {$a,b$} (q2);
\end{tikzpicture}
\caption{The FSA $\mathcal A_4$, recognizing language $b^* a a \, \{a, b\}^*$.}\label{ex:reg:a4}
\end{center}
\end{figure}
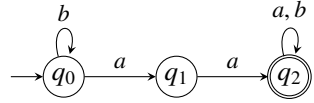

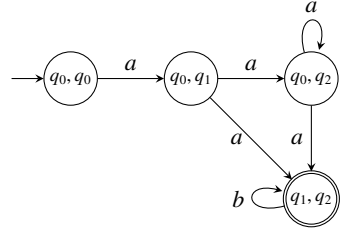
\begin{figure}[h]
\begin{center}
\begin{tikzpicture}[every edge/.style={draw,solid}, node distance=3.2cm, auto, 
                    every state/.style={draw=black!100,scale=0.5}, >=stealth]

\node[initial by arrow, initial text=, state] (q00) {{\Large $q_0, q_0$}};
\node[state] (q01) [right of=q00] {{\Large $q_0, q_1$}};
\node[state] (q02) [right of=q01] {{\Large $q_0, q_2$}};
\node[state] (q12) [below of=q02, accepting] {{\Large $q_1,q_2$}};

\path[->]
(q00) edge [above] node {$a$} (q01)
(q01) edge [above]  node {$a$} (q02)
(q02) edge [loop above]  node {$a$} (q02)
(q01) edge [left]  node {$a$} (q12)
(q02) edge [left]  node {$a$} (q12)
(q12) edge [loop left]  node {$b$} (q12)
;
\end{tikzpicture}
\caption{FSA recognizing $(a^+ b^*) \cap (b^* a a \, \{a, b\}^*)$, obtained from
  $\mathcal A_1$ and $\mathcal A_4$.}\label{ex:reg:aprod}
\end{center}
\end{figure}
\end{example}
Another fundamental property of RL is that their emptiness problem, i.e. whether or not $L = \emptyset$, is decidable. This in turn is an immediate consequence of the so-called \textit{pumping lemma} which, informally, states that, due to the finiteness of the state space, the behavior of any FSA must eventually be periodic. More formally, this property is stated as follows:

\begin{lemma}\label{lemPumpReg}
 Let $\mathcal A$ be a FSA; for every string $x\in L(\mathcal A)$, with $|x| >  |Q|$, there exists a factorization $x = ywz$ such that $x = yw^nz \in L(G)$ for every $n \geq 0$.
\end{lemma}

As a consequence of this lemma, if in a regular language there exists a sentence of any length, then there exists also a string of length $\leq |Q|$, hence the decidability of the emptiness problem and, in turn, of the containment problem (is $L(\mathcal A_1) \subseteq L(\mathcal A_2)$?) and of the equivalence problem, thanks to the closure w.r.t. Boolean operations.
The lemma can also be exploited to show that no FSA can recognize $L = \{a^nb^n \mid n \geq 0\}$.

Other notable closures, not of major interest in this paper,  are w.r.t. {\em concatenation}, $Kleene~^*$, \textit{string
homomorphism}, and {\em inverse string homomorphism};
FSA are {\em minimizable}, i.e. given a FSA, there is an algorithm to
build an equivalent automaton with the minimum possible number of states.

\subsection{Logic characterization}\label{sec:reg:logic}

From the very beginning of formal language and automata theory the investigation
of the relations between defining a language through some kind of abstract
machine and through a logic formalism has produced challenging theoretical
problems and important applications in system design and verification. A
well-known example of such an application is the classical Hoare's method to
prove the correctness of a Pascal-like program w.r.t. a specification stated as
a pair of pre- and post-conditions expressed through a first-order theory \cite{DBLP:journals/cacm/Hoare69}.

Such a verification problem is undecidable if the involved formalisms have the
computational power of Turing machines but may become decidable for less
powerful formalisms as in the important case of RL. In particular, B\"uchi, in
his seminal work \cite{bib:Buchi1960a}, provided a complete characterization of
regular languages in terms of a suitable logic, so that the decidability
properties of this class of languages could be exploited to achieve automatic
verification; later on, in fact a major breakthrough in this field has been
obtained thanks to advent of model checking.  The logic proposed by B\"uchi to
characterize regular languages is a Monadic Second Order (MSO) one that is
interpreted over the natural numbers representing the position of a character
in the string: the string is accepted by a FSA iff it satisfies a sentence in
the corresponding logic.
The basic elements of the logic defined by B\"uchi are summarized here:
\begin{itemize} 
    
    \item First-order variables, denoted as lowercase letters at
        the end of the alphabet, $\bm{x}$, $\bm{y}$, \ldots are interpreted over the natural
        numbers $\mathbb N$ (these variables are written in boldface to avoid
        confusion with strings);

    \item Second-order variables, denoted as uppercase letters at the end of
        the alphabet, written in boldface,  $\bm X$, $\bm Y$, \ldots are interpreted over the power set of
        natural numbers $\mathcal P(\mathbb N)$;

    \item For a given input alphabet $\Sigma$, the monadic predicate $a(\cdot)$
        is defined for each $a \in \Sigma$: $a(\bm x)$ evaluates to true in
        a string iff the character at position $\bm x$ is $a$;

    \item The {\em successor} predicate is denoted by $\sux$, i.e. $\sux(\bm x,
      \bm y)$ means that $\bm y = \bm x +1$.  

    \item The well-formed formulas are defined by the following syntax:
        \[
\varphi := a(\bm x) \mid \bm x \in \bm X \mid \sux(\bm x, \bm y) 
\mid \neg \varphi \mid \varphi \lor \varphi 
\mid \exists \bm x (\varphi)
\mid \exists \bm X (\varphi).
\]
        
    \item They are interpreted in the natural way. Furthermore the usual
        predefined abbreviations are introduced to denote the remaining
        propositional connectives, universal quantifiers, arithmetic relations ($=, \ne, <, >$), sums and subtractions between first order variables and numeric constants. E.g. $\bm x = \bm y$ is an abbreviation for $\forall \bm X (\bm x \in \bm X \iff \bm y \in \bm X)$; $ \bm  x = \bm z - 2$ stands for $\exists \bm y (\sux(\bm z,\bm y) \land \sux(\bm y,\bm x))$

    \item A sentence is a closed formula of the MSO logic; also, the notation
        $w \models \varphi$  denotes that string $w$ satisfies sentence $\varphi$. 
        For a given sentence $\varphi$, the language $L(\varphi)$ is defined as
        \footnote{When specifying languages by means of logic formulas, the empty string must be excluded because formulas refer to string positions.}
        \[
        L(\varphi) = \{ w \in \Sigma^+ \mid w \models \varphi\}.
        \]
\end{itemize}

For instance formula 
$\forall \bm x, \bm y ( a(\bm x) \land \sux(\bm x, \bm y) \Rightarrow b(\bm y))$ 
defines the language of strings where every occurrence of character $a$
is immediately followed by an occurrence of $b$.

B\"uchi's seminal result is synthesized by the following theorem.

\begin{theorem}
A language $L$ is regular iff there exists a sentence
$\varphi$ in the above MSO logic such that $L = L(\varphi)$.
\end{theorem}

The proof of the theorem is constructive, i.e., it
provides an algorithmic procedure that, for a given FSA builds an equivalent
sentence in the logic, and conversely; next we offer an intuitive explanation
of the construction, referring the reader to, e.g., \cite{bib:Thomas1990a} for a complete and
detailed proof.  

\subsection*{From FSA to MSO logic}

The key idea of the construction consists in representing each state $q$ of the
automaton as a second order variable $\bm{X}_q$, which is the set of all
string's positions where the machine is in state $q$. Without lack of generality
we assume the automaton to be deterministic, and that $Q = \{0, 1, \ldots, m\}$, with
0 initial, for some $m$. Then we
encode the definition of the FSA recognizing $L$ as the conjunction of several clauses each one
representing a part of the FSA definition:

\begin{itemize}
    \item The transition $\delta(q_i,a) = q_j$ is formalized by 
        $\forall \bm x, \bm y(\bm x \in \bm X_i \land a(\bm x) \land
        \sux(\bm x, \bm y) \Rightarrow \bm y \in \bm X_{j})$.
    \item The fact that the machine starts in state $0$ is represented as
$\exists \bm z (\not\exists \bm x (\sux(\bm x, \bm z) ) \land \bm z \in \bm X_{0})$.
    \item Being the automaton deterministic, for each pair of distinct second order
        variables $\bm X_i$ and $\bm X_j$ we need the subformula $\not\exists \bm y (\bm
        y \in \bm X_i \land \bm y \in \bm X_j)$. 
    \item Acceptance by the automanton, i.e. $\delta(q_i,a) \in F$, is formalized by:
$\exists \bm y( \not\exists \bm x (\sux(\bm y, \bm x) )  \land
 \bm y \in \bm X_i \land a(\bm y))$.
      \item Finally the whole language $L$ is the set of strings that satisfy
        the global sentence $\exists \bm X_0, X_1, \ldots \bm X_m (\varphi)$,
        where $\varphi$ is the conjunction of all the above clauses.
\end{itemize}

At this point it is not difficult to show that the set of strings satisfying the
above global formula is exactly $L$.

\subsection*{From MSO logic to FSA}

The construction in the opposite sense has been
proposed in various versions in the literature. Here we summarize its main
steps along the lines of \cite{bib:Thomas1990a}. 
First, the MSO sentence is translated into a
standard form using only second-order variables, the $\subseteq$ predicate, and
variables $\bm W_a$, for each $a \in \Sigma$, denoting the set of all the positions of the word
containing the character $a$.
Moreover, we use $\Sux$, which has the same meaning of $\sux$, but, syntactically, 
has second order variable arguments that are singletons. 
This simpler, equivalent logic, is defined by the following syntax:
        \[
\varphi := \bm X \subseteq \bm W_a \mid \bm X \subseteq \bm Y \mid \Sux(\bm X, \bm Y) 
\mid \neg \varphi \mid \varphi \lor \varphi 
\mid \exists \bm X (\varphi).
\]
As before, we also use the standard abbreviations, e.g. $\land$, $\forall$, $=$.
To translate first order variables to second order variables we need to
  state that a (second order) variable is a singleton. Hence we introduce the
  abbreviation: $\Sing(\bm X)$ for 
$\exists \bm Y ( \bm Y \subseteq \bm X 
\land \bm Y \ne \bm X 
\land \not\exists \bm Z (\bm Z \subseteq \bm X \land \bm Z \ne \bm Y \land \bm
Z \ne \bm X)
)$. Then, $\Sux(\bm X, \bm Y)$ is defined only for $\bm X$ and $\bm Y$ singletons.

The following step entails the inductive construction of the equivalent
automaton. This 
is built by
associating a single automaton to each elementary subformula and by composing
them according to the structure of the global formula.
This inductive approach requires to use open formulas. Hence, we are going to consider
words on the alphabet $\Sigma \times \{0, 1\}^k$, so that $\bm X_1$, $\bm X_2$,
\ldots $\bm X_k$ are the free variables used in the formula; 1 in the,
say, $j$-th component means that the considered position belongs to $\bm X_j$, 0 vice
versa. For instance, if $w = (a,0,1)(a,0,0)(b,1,0)$, then $w \models 
\bm X_2 \subseteq \bm W_a$, $w \models \bm X_1 \subseteq \bm W_b$, with $\bm X_1$ and $\bm X_2$ singletons.

\subsubsection*{Formula transformation}

\begin{enumerate}
\item First order variables are translated in the following way:
$\exists \bm x (\varphi(\bm x))$ becomes\\ $\exists \bm X ( \Sing(\bm X) \land \varphi'(\bm X))$,
where $\varphi'$ is the translation of $\varphi$, and $\bm X$ is a fresh variable.

\item Subformulas having the form $a(\bm x)$, $\sux(\bm x, \bm y)$ are
  translated into $\bm X \subseteq \bm W_a$, $\Sux(\bm X, \bm Y)$, respectively. 
  \item The other parts remain the same.
\end{enumerate}

\subsubsection*{Inductive construction of the automaton}

We assume for simplicity that $\Sigma = \{a, b\}$, and that $k = 2$, i.e. two
variables are used in the formula. Moreover we use the shortcut symbol $\circ$ to mean
all possible values. 

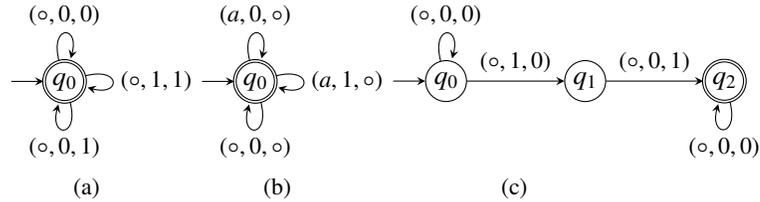
\begin{figure*}
  \centering
\begin{tabular}{m{0.2\textwidth}m{0.2\textwidth}m{0.3\textwidth}}
  \begin{tikzpicture}[every edge/.style={draw,solid}, node distance=3cm, auto, 
                    every state/.style={draw=black!100,scale=0.5}, >=stealth]

\node[initial by arrow, initial text=,state,accepting] (q0) {{\huge $q_0$}};

\path[->]
(q0) edge [loop above] node {$(\circ,0,0)$} (q1)
(q0) edge [loop below] node {$(\circ,0,1)$} (q1)
(q0) edge [loop right] node {$(\circ,1,1)$} (q1);
\end{tikzpicture}
&
\begin{tikzpicture}[every edge/.style={draw,solid}, node distance=3cm, auto, 
                    every state/.style={draw=black!100,scale=0.5}, >=stealth]

\node[initial by arrow, initial text=,state,accepting] (q0) {{\huge $q_0$}};

\path[->]
(q0) edge [loop above] node {$(a,0,\circ)$} (q1)
(q0) edge [loop below] node {$(\circ,0,\circ)$} (q1)
(q0) edge [loop right] node {$(a,1,\circ)$} (q1);
\end{tikzpicture}
&
\begin{tikzpicture}[every edge/.style={draw,solid}, node distance=3.7cm, auto, 
                    every state/.style={draw=black!100,scale=0.5}, >=stealth]

\node[initial by arrow, initial text=,state] (q0) {{\huge $q_0$}};
\node[state] (q1) [right of=q0] {{\huge $q_1$}};
\node[state] (q2) [right of=q1, accepting] {{\huge $q_2$}};

\path[->]
(q0) edge [loop above]  node {$(\circ,0,0)$} (q0)
(q0) edge [above] node {$(\circ,1,0)$} (q1)
(q1) edge [above] node {$(\circ,0,1)$} (q2)
(q2) edge [loop below]  node {$(\circ,0,0)$} (q2);
\end{tikzpicture}
  \\
  \centering{(a)} 
  &
  \centering{(b)} 
  &
  \centering{(c)} 
\end{tabular}
  \caption{Automata for the construction from MSO logic to FSA.}
  \label{fig:inductive-automata}
\end{figure*}

\begin{itemize}
\item The formula $\bm X_1 \subseteq \bm X_2$ is translated into an automaton
  that checks that there are 1's for the $\bm X_1$ component only in positions where there
  are also 1's for the $\bm X_2$ component (Figure~\ref{fig:inductive-automata}~(a)).

\item The formula $\bm X_1 \subseteq \bm W_a$ is analogous: the automaton checks that
  positions marked by 1 in the $\bm X_1$ component must have symbol $a$ (Figure~\ref{fig:inductive-automata}~(b)).

\item The formula $\Sux(\bm X_1,\bm X_2)$ considers two singletons, and checks
  that the 1 for component $\bm X_1$ is immediately followed by a 1 for
  component $\bm X_2$ (Figure~\ref{fig:inductive-automata}~(c)).

\item Formulas inductively built with $\neg$ and $\lor$ are covered by the closure of regular languages w.r.t.
  complement and union, respectively.
  
\item For $\exists$, we use the closure under
  alphabet projection: we start with an automaton with input alphabet, say, $\Sigma
  \times \{0, 1\}^2$, for the formula $\varphi(\bm X_1,\bm X_2)$; we need to
  define an automaton for the formula $\exists \bm X_1 (\varphi(\bm X_1, \bm
  X_2))$. But in this case the alphabet is $\Sigma \times \{0, 1\}$, where the
  last component represents the only free remaining variable, i.e. $\bm X_2$. \\
The
  automaton $\mathcal A_\exists$ is built by starting from the one for $\varphi(\bm X_1,\bm X_2)$,
  and changing the transition labels from 
$(a,0,0)$ and $(a,1,0)$ to $(a,0)$;   
$(a,0,1)$ and $(a,1,1)$ to $(a,1)$, and those with $b$ analogously.   
The main idea is that this last automaton nondeterministically ``guesses'' the quantified component
(i.e. $\bm X_1$) when reading its input, and the  resulting word $w \in (\Sigma
\times \{0,1\}^2)^*$ is such that $w \models \varphi(\bm X_1,\bm X_2)$. Thus,
$\mathcal A_\exists$ recognizes $\exists \bm X_1 (\varphi(\bm X_1, \bm
  X_2))$.
\end{itemize}

We refer the reader to the available literature for a full proof of equivalence
between the logic formula and the constructed automaton. Here we illustrate the
rationale of the above construction through the following example.

\begin{example}\label{ex:reg:log2aut}
 Consider the language $L = \{a,b\}^* a a \{a,b\}^*$: it consists of the strings satisfying
 the formula:\\
$
\varphi_L = \exists \bm x \exists \bm y (\sux(\bm x, \bm y) \land a(\bm x) \land
a(\bm y))$.

As seen before, first we translate this formula into a version using only
second order variables:
$ \varphi'_L = \exists \bm X, \bm Y (
\Sing(\bm X) \land \Sing(\bm Y) \land \Sux(\bm X, \bm Y) \land 
\bm X \subseteq \bm W_a \land
\bm Y \subseteq \bm W_a)$.

Here are the automata for $\Sing(\bm X)$ and $\Sing(\bm Y)$:
\[
\begin{array}{cc}
\begin{tikzpicture}[every edge/.style={draw,solid}, node distance=3.7cm, auto, 
                    every state/.style={draw=black!100,scale=0.5}, >=stealth]

\node[initial by arrow, initial text=,state] (q0) {{\huge $q_0$}};
\node[state] (q1) [right of=q0, accepting] {{\huge $q_1$}};

\path[->]
(q0) edge [loop above]  node {$(\circ,0,\circ)$} (q0)
(q0) edge [above] node {$(\circ,1,\circ)$} (q1)
(q1) edge [loop above]  node {$(\circ,0,\circ)$} (q1);
\end{tikzpicture}
&
\begin{tikzpicture}[every edge/.style={draw,solid}, node distance=3.7cm, auto, 
                    every state/.style={draw=black!100,scale=0.5}, >=stealth]

\node[initial by arrow, initial text=,state] (q0) {{\huge $q'_0$}};
\node[state] (q1) [right of=q0, accepting] {{\huge $q'_1$}};

\path[->]
(q0) edge [loop above]  node {$(\circ,\circ,0)$} (q0)
(q0) edge [above] node {$(\circ,\circ,1)$} (q1)
(q1) edge [loop above]  node {$(\circ,\circ,0)$} (q1);
\end{tikzpicture}
\end{array}
\]
The above automata could also be obtained by expanding the definition of $\Sing$ and
then projecting the quantified variables.
By intersecting the automata for $\Sing(\bm X)$, $\Sing(\bm Y)$, $\Sux(\bm
X, \bm Y)$ we obtain an automaton which is identical to the one we defined for
translating formula $\Sux(\bm X_1, \bm X_2)$, where here $\bm X$ takes the role
of $\bm X_1$ and $\bm Y$ of $\bm X_2$.
Combining it with those for $\bm X \subseteq W_a$ and $\bm Y \subseteq W_a$
produces:
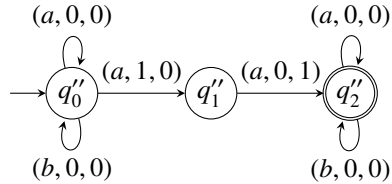
\begin{center}
\begin{tikzpicture}[every edge/.style={draw,solid}, node distance=3.7cm, auto, 
                    every state/.style={draw=black!100,scale=0.5}, >=stealth]

\node[initial by arrow, initial text=,state] (q0) {{\huge $q''_0$}};
\node[state] (q1) [right of=q0] {{\huge $q''_1$}};
\node[state] (q2) [right of=q1, accepting] {{\huge $q''_2$}};

\path[->]
(q0) edge [loop above]  node {$(a,0,0)$} (q0)
(q0) edge [loop below]  node {$(b,0,0)$} (q0)
(q0) edge [above] node {$(a,1,0)$} (q1)
(q1) edge [above] node {$(a,0,1)$} (q2)
(q2) edge [loop above]  node {$(a,0,0)$} (q2)
(q2) edge [loop below]  node {$(b,0,0)$} (q2);
\end{tikzpicture}
\end{center}

Finally, by projecting on the quantified variables $\bm X$ and $\bm Y$ we obtain
the automaton for $L$.
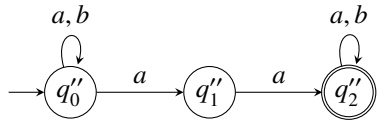
\begin{center}
\begin{tikzpicture}[every edge/.style={draw,solid}, node distance=3.7cm, auto, 
                    every state/.style={draw=black!100,scale=0.5}, >=stealth]

\node[initial by arrow, initial text=,state] (q0) {{\huge $q''_0$}};
\node[state] (q1) [right of=q0] {{\huge $q''_1$}};
\node[state] (q2) [right of=q1, accepting] {{\huge $q''_2$}};

\path[->]
(q0) edge [loop above]  node {$a,b$} (q0)
(q0) edge [above] node {$a$} (q1)
(q1) edge [above] node {$a$} (q2)
(q2) edge [loop above]  node {$a,b$} (q2);
\end{tikzpicture}
\end{center}

\end{example}

The logical characterization of a class of languages, together with the decidability of the containment problem, is the main door towards automatic verification techniques: suppose in fact to have a logic formalism $\mathfrak{L}$ recursively equivalent to an automaton family $\mathfrak{A}$; then, one can use a language $L_\mathfrak{L}$ in $\mathfrak{L}$ to specify the requirements of a given  system and an abstract machine $\mathcal{A}$ in $\mathfrak{A}$ to implement the desired system: the correctness of the design defined by $\mathcal{A}$ w.r.t. to the requirements stated by $L_\mathfrak{L}$ is therefore formalized by $L(\mathcal{A}) \subseteq L_\mathfrak{L}$, i.e., all behaviors realized by the machine are also satisfying the requirements. This is just the case with FSA and MSO logic for RL; in practice, however, the decision algorithms based on the above MSO logic have been proved of intractable complexity; thus, the great success of model-checking as the fundamental technique for automatic verification has been obtained by resorting to less powerful, typically temporal, logics which compensated the limited loss of generality by a complexity considered, and empirically verified in many practical experiments, as more tractable, despite a theoretical PSPACE-completeness \cite{Emerson90}.



\section{Context-free Languages}\label{sec:cf}

Context-free languages (CFL), with their generating context-free
grammars (CFG) and recognizing pushdown automata (PD\-A), are,
together with RL, the most important chapter in the literature of
formal languages. CFG have been introduced by Noam Chomsky in the
1950s as a formalism to capture the syntax of natural languages.
Independently, essentially the same formalism has been developed to
formalize the syntax of the first high level programming language,
FORTRAN; in fact it is also referred to as Backus-Naur form (BNF)
honoring the chief scientists of the team that developed FORTRAN and
its first compiler. It is certainly no surprise that the same
formalism has been exploited to describe the syntactic aspects of both
natural and high level programming languages, since the latter ones
have exactly the purpose to make algorithm specification not only
machine executable but also similar to human description.

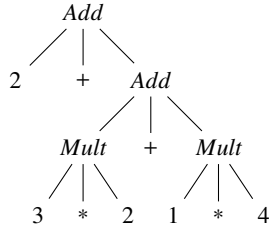
\begin{figure}
\begin{center}
\begin{tikzpicture}[
 scale=0.6,
 level 2/.style={sibling distance=1.5cm},
 level 3/.style={sibling distance=1cm}]
\node{$Add$}
child{ 
 node{2}
}
child{
 node{$+$}
}
child{
 node{$Add$}
 child{
   node{$Mult$}
   child{node{3}}
   child{node{$*$}}
   child{node{2}}
 }
 child{
   node{$+$}
 }
 child{
   node{$Mult$}
   child{node{1}}
   child{node{$*$}}
   child{node{4}}
 }
} 
; 
\end{tikzpicture}
\caption{A tree structure that shows the precedence of multiplication over addition in arithmetic expressions.}\label{fig:tree2}
\end{center}
\end{figure}

The distinguishing feature of both natural and programming languages
is that complex sentences can be built by combining simpler ones in an
a priori unlimited hierarchy: for instance a conditional sentence is
the composition of a clause specifying a condition with one or two
sub-sentences specifying what to do if the condition holds, and
possibly what else to do if it does not hold.
%
%
Such a typical nesting of sentences suggests a natural representation
of their $structure$ in the form of a tree shape.
%
The possibility of giving a sentence a tree structure which hints at
its semantics is a sharp departure from the rigid linear structure of
regular languages. As an example, consider a simple arithmetic
expression consisting of a sequence of operands with either a $+$ or a
$*$ within any pair of them, as in $2+3*2+1*4$. Sentences of this type
can be easily generated by, e.g., the following regular grammar:
\[
\begin{array}{lcl}
S  \to 1 \mid 2 \mid \ldots 0 \mid 1A \mid 2A \mid \ldots 0A \\
A \to + S \mid *S 
\end{array}
\] 
However, if we compute the value of the above expression by following
the linear structure given to it by the grammar either by associating
the sum and the multiply operations to the left or to the right we
would obtain, respectively, $44$ or $20$ which is not the way we learned to compute the value of the expression at school. On the
contrary, we first compute the multiply
operations and then the sum of the three partial results, thus obtaining
$12$; this, again, suggests to associate the semantics of the sentence
--in this case the value of the expression-- with a syntactic
structure that is more appropriately represented by a tree, as
suggested in Figure~\ref{fig:tree2}, than by a flat sequence of
symbols.

CFG are more powerful than regular ones in that not only they assign in a natural way a tree structure to their sentences but also the class of languages they generate strictly includes RL (see Section ~\ref{CompCFLRL}). CFG are defined as a special case of Chomsky's general ones in the following way.

 \begin{definition}
	\label{def:CFG and CFL}
  A grammar $G=(V_N,\Sigma, P, S)$ is \emph{context-free} iff the lhs of all its productions is a single nonterminal.

 \end{definition}
 Among all derivations associated with CFG it is often useful to distinguish the leftmost and rightmost ones:
 \begin{definition}
 A CFG derivation $S \derives{*}_G x$ is a \emph{leftmost}, denoted as $S \derives{l}_G x$ (resp. \emph{rightmost}, denoted as $S \derives{r}_G x$) one iff all of its immediate derivation steps are of the form
 $xA\beta \derives{}_G x \alpha \beta$ (resp. $\beta Ax\derives{}_G \beta \alpha x$), with $x \in \Sigma^*, A \in V_N, \beta, \alpha \in V^*$. 
  \end{definition}
  In other words in a leftmost derivation, at every step the leftmost nonterminal character is the lhs of the applied rule. It is immediate to verify that for every derivation of a CFG there are an equivalent leftmost and an equivalent rightmost one, i.e., derivations that produce the same terminal string.\footnote{This property does not hold for more general classes of grammars.}
 \begin{example}\label{GAE}
 The following CF grammar $GAE_1$ generates the same numerical arithmetic expressions as above but assigns them the appropriate structure exemplified in Figure~\ref{fig:tree2}. In fact there is an obvious correspondence between grammar derivations and the trees whose root is the axiom and for every internal node, labeled by a nonterminal, its children are, in order, the rhs of a production whose lhs is the label of the father. In particular, a leftmost (resp. rightmost) derivation corresponds to a left (resp. right), top-down traversal of the tree; symmetrically a left bottom-up tree traversal produces the mirror image of a rightmost derivation. The tree that represents the derivation of a sentence $x$ by a CFG is named \emph{syntax-tree} or \emph{syntactic tree} of $x$. Notice that in the following, for the sake of simplicity, in arithmetic expressions we will make use of a unique terminal symbol $e$ to denote any numerical value.
 \[
 	\begin{array}{ll}
 	GAE_1: &	S \to E \mid T \mid F\\
 	&	E \to  E + T \mid T * F \mid e  \\  	
 	&	T \to  T * F \mid e      \\
 	&	F \to  e  \\
 	\end{array}
 \]
It is an easy exercise augmenting the above grammar to let it generate more general arithmetic expressions including more arithmetic operators, parenthesized sub-expressions, symbolic operands besides numerical ones, etc. 

 Consider now the following slight modification $GAE_2$ of $GAE_1$:
  \[
  	\begin{array}{ll}
  	GAE_2: &
  		S \to E \mid T \mid F\\
  	& 	E \to  E * T \mid T + F \mid e  \\  	
  	& 	T \to  T + F \mid e      \\
  	& 	F \to  e  \\
  	\end{array}
  \]
 
$GAE_1$ and $GAE_2$ are equivalent in that they generate the same language; however,  $GAE_2$ assigns to the string $2+3*2+1*4$ the tree represented in Figure~\ref{fig:tree3}: if we executed the operations of the string in the order suggested by the tree --first the lower ones so that their result is used as an operand for the higher ones-- then we would obtain the value $60$ which is not the ``right one'' $12$.
 
\begin{figure}
\begin{center}
\begin{tikzpicture}[
 scale=0.6,
 level 4/.style={sibling distance=0.7cm},
 level distance=1.1cm
  ]
\node{$S$}
child{
 node{$E$}
 child{
   node{$E$}
   child{ 
     node{$E$} 
     child{ node{$T$} child{ node{2} }}
     child{ node{$+$} }
     child{ node{$F$} child{ node{3} }}
   }
   child{ node{$*$} }
   child{ node{$T$} 
     child{ node{$T$} child{ node{2} }}
     child{ node{$+$} }
     child{ node{$F$} child{ node{1} }}
   }
 }
 child{
   node{$*$}
 }
 child{
   node{$T$}
   child{ node{4} }
 }
} 
; 
\end{tikzpicture}
\caption{A tree that reverts the traditional precedence between arithmetic operators.}\label{fig:tree3}
\end{center}
\end{figure}

 Last, the grammar:
  \[
   	\begin{array}{ll}
   	GAE_{amb}: & 
   		S \to  S * S \mid S + S \mid e  \\  	
   	\end{array}
   \]
  is equivalent to $GAE_2$ and $GAE_1$ as well, but it can generate the same string by means of different left derivations which correspond to the traversal of the syntax-trees of Figure~\ref{fig:tree2}, of Figure~\ref{fig:tree3}, and even more. In such a case we say that the grammar is \emph{ambiguous} in that it associates different syntax-trees to the same sentence.
  \end{example}

 The above examples, inspired by the intuition of arithmetic expressions, further emphasize the strong connection between the structure given to a sentence by the syntax-tree and the semantics of the sentence: as it is the case with natural languages too, an ambiguous sentence may exhibit different meanings.

%
%
The examples also show that, whereas a sentence of a regular language has a fixed --either right or left-linear  structure, CFL sentences have a tree structure which in general, is not immediately ``visible'' by looking at the sentence, which is the frontier of its associated tree.
As a consequence, in the case of CFL analyzing a sentence is often not only a matter of deciding whether it belongs to a given language, but in many cases it is also necessary to build the syntax-tree(s) associated with it: its structure often drives the evaluation of its semantics as intuitively shown in the examples of arithmetic expressions and systematically applied in all major applications of CFL, including designing their compilers. The activity of jointly accepting or rejecting a sentence and, in case of acceptance, building its associated syntax-tree(s) is usually called \emph{parsing}.

\subsection{Pushdown automata as parsers}
It is well known that a typical data structure to build and visit trees is the \textit{stack} or \textit{pushdown store}. Thus, it is no surprise that the abstract machines adopted in formal language theory to analyze CFL are \textit{pushdown automata}, i.e., devices supplied with a finite state control augmented with a LIFO auxiliary memory. Among the many versions of such automata proposed in the literature to parse CFL or subfamilies thereof we report here a classical one.
\begin{definition}\label{def:PDA}
A pushdown automaton (PDA) $\mathcal A$ is a tuple $(\Sigma, Q,$ $\Gamma, 
\delta,$ $q_0, Z_0, F)$ where
	\begin{itemize}
	\item $\Sigma$ is the finite terminal alphabet;
	\item $Q$ is the finite set of states;
	\item $\Gamma$ is the finite stack alphabet;
	\item $\delta \subseteq_F  Q \times \Gamma\times (\Sigma \cup \{\varepsilon\}) \times Q\times \Gamma^*$ is the transition relation, where the notation $A \subseteq_F B$ means that $A$ is a \emph{finite subset} of $B$;
	\item $I \subseteq Q$ is the set of initial states;
	\item $Z_0 \in \Gamma$ is the initial stack symbol;
	\item $F \subseteq Q$ is the set of final or accepting states.
	\end{itemize}
	
A PDA $\mathcal A$ is deterministic (DPDA) iff
\begin{itemize}
  \item $I$ is a singleton;
\item
$\delta$ is a function $\delta: Q \times \Gamma\times (\Sigma \cup \{\varepsilon\}) \to Q \times \Gamma^*$;
\item
$\forall (q,A)$ if $\delta (q, A, a)$ is defined for some $a$, then $\delta (q, A, \varepsilon)$ is not defined; in other words, for any pair $(q, A)$ the automaton may perform either an $\varepsilon$-move or a reading move but not both of them.

\end{itemize}	

A \textit{configuration} of a PDA is a triple 	 $(x , q, \gamma)$ where $x$ is the string to be analyzed, $q$ is the current state of the automaton, and $\gamma$ is the content of the stack.

The \textit{transition} relation between two configurations $c_1 \transition{} c_2$ holds iff  $c_1 =  (x_1, q_1, \gamma_1)$, $c_2 =  (x_2, q_2, \gamma_2)$ and, either:

\begin{itemize}
\item $x_1 = ay, x_2 = y, \gamma_1 = A\beta , \gamma_2 = \alpha\beta $, and $(q_1, A, a,q_2, \alpha) \in \delta$, or
\item $x_1 = x_2, \gamma_1 = A \beta , \gamma_2 = \alpha \beta $, and $(q_1, A, \varepsilon,q_2, \alpha) \in \delta$.

\end{itemize}
A string $x$ is accepted by a PDA $\mathcal A$ iff
$c_0 = (x, q_0, Z_0) \transition{*} (\varepsilon, q_F, \gamma)$ for some $q_0 \in I$, $q_F \in F$, and $\gamma \in \Gamma^*$. The language accepted or recognized by $\mathcal A$ is the set $L(\mathcal A)$ of all strings accepted by $\mathcal A$.
\end{definition}
A transition of the second type above, i.e. where no input character is consumed because the applied transition relation is of type $(q_1, A, \varepsilon,q_2, \alpha)$ is called an \emph{$\varepsilon$-move}.
%

PDA as defined above are recursively equivalent to CFG, i.e, for any given CFG $G$ a PDA $\mathcal A$ can effectively be built such that  $L(\mathcal A)$ = $L(G)$ and conversely. 
DPDA define deterministic CFL (DCFL).
Example~\ref{ex:PDAAE} illustrates informally the rationale of this correspondence by presenting a PDA equivalent to grammar $GAE_1$. It also provides a hint on how a PDA \textit{recognizing} a CFL can be augmented as a pushdown \textit{transducer} which also builds the syntax-tree of the input sentence (if accepted).

\begin{example}\label{ex:PDAAE}
Consider the following PDA $\mathcal A$:\\
 $Q = \{q_0, q_1, q_F \}$,
 $\Gamma = \{E, T, F, +, *, e, Z_0\}$,\\
$\delta =  
 \{ (q_0, Z_0, \varepsilon, q_1, EZ_0), (q_0, Z_0, \varepsilon, q_1, TZ_0), (q_0, Z_0, \varepsilon, q_1, FZ_0) \} \cup$\\
 $ \{(q_1, E, \varepsilon,q_1, E+T), (q_1, E, \varepsilon,q_1, T*F), (q_1, E, \varepsilon,q_1, e)\} \cup$\\
  $ \{ (q_1, T, \varepsilon, q_1, T*F), (q_1, T, \varepsilon, q_1, e)\} \cup 
  \{ (q_1, F, \varepsilon, q_1, e)\}\cup $\\
  $ \{ (q_1,+, +, q_1, \varepsilon), 
	(q_1,*,*, q_1, \varepsilon), 
	 (q_1,e, e, q_1, \varepsilon) \}\cup$ \\
	 $ \{(q_1, Z_0, \varepsilon, q_F, \varepsilon)\}$.
         
$\mathcal A$ accepts, e.g., the input string $e+e*e$ through the following sequence of configuration transitions

$(e+e*e,q_0,  Z_0) \transition{}(e+e*e,q_1, EZ_0) \transition{}(e+e*e,q_1,  E+TZ_0)\transition{} (e+e*e, q_1, e+TZ_0 )\transition{}(+e*e,q_1, +TZ_0 )\transition{}(e*e,q_1, TZ_0
)\transition{}(e*e,q_1, T*FZ_0 )\transition{}(e*e,q_1, e*FZ_0 )\transition{}(*e,q_1, *FZ_0 )\transition{}(e ,q_1, FZ_0)\transition{}(e,q_1,  eZ_0)\transition{}(\varepsilon,q_1, Z_0 )\transition{}(\varepsilon,q_F,\varepsilon)$
 
In fact $\mathcal A$ has been naturally derived from the original grammar, $GAE_1$ in this case, in such a way that its transitions perfectly parallel a grammar's leftmost derivation: at any step if the top of the stack stores a nonterminal it is nondeterministically replaced by the corresponding rhs of a production of which it is the lhs, in such a way that the leftmost character of the rhs is put on top of the stack; if it is a terminal it is compared with the current input symbol and, if they match the stack is popped and the reading head advances.
Notice however, that, being $\mathcal A$ nondeterministic, there are also several sequences of transitions beginning with the same $c_0$ that do not accept the input. 	

Strictly speaking the above automaton is not a real $parser$ for the corresponding grammar since it simply accepts all and only the strings generated thereby. However, it is quite easy to make it a complete parser by adding a few translation components: precisely, a $pushdown$ $transducer$ is a pushdown automaton provided with an output alphabet $O$ which augments the range of $\delta$ in such a way that at every transition the transducer appends to the output, initially empty, a string in $O^*$; the output string (string\textit{s} in case of nondeterminism) produced at the end of the computation applied to the input $x$ is the translation $\tau(x)$ produced by the transducer.

For instance, it is not difficult to build a full parser for grammar $GAE_1$ by adding to the automaton of Example~\ref{ex:PDAAE} an output alphabet consisting of labels that uniquely identify G's productions: for every transition of the automaton that has been built in one-to-one correspondence with G's production set it suffices to augment $\delta$ by adding as output component of each triple the label of the corresponding production. Assume that productions 
          $E \to  E + T \mid T * F$
 are labeled $1, 2$, respectively; then $\delta$ is enriched accordingly with 
 $\{(q_0,$ $E, $ $\varepsilon,q_0, T+E, 1), (q_0, E, \varepsilon,q_0, F*T, 2), 
 ...\} \subseteq \delta $. Conversely, transitions that simply verify that the 
 element on top of the stack coincides with the current input character do not 
 output anything. Clearly, the transducer built in this way from the original 
 grammar for any string of the language delivers an output that corresponds to 
 a top-down, leftmost visit of the syntax-tree(s) of the input string where any 
 internal node is the label of the production used by the grammar to expand it. 
 Thus, we can assume (nondeterministic) transducers as general parsers for CFG; 
 if the original grammar is ambiguous the corresponding transducer produces 
 several translations for the same input.
 \end{example}

%
%


\subsection{A comparative analysis of CFL and RL properties}\label{CompCFLRL}
 CFG and PDA are more powerful formalisms than the corresponding regular ones. Such a greater power allows for their application in larger fields such as formalizing the syntax and semantics of programming languages and many other tree-struc\-tured data. 
  The greater power of the CFL family than the RL one abides not only in the structure of their sentences but even in sets of sentences that constitute their languages. For instance, the language $L_1 = \{a^{n}b^{n}\mid n \geq 0\}$ is easily recognized by a PDA but no FSA can accept it due its finite memory formalized by the state space $Q$, whereas, intuitively, an unbounded memory is necessary to count the number of $a$'s to be able to compare their number with that of the following $b$'s; this claim can be easily proved formally by exploiting the fundamental pumping Lemma \ref{lemPumpReg}.
  
  Not surprisingly, however, such an increased power of CFL w.r.t. RL comes at a price in terms of loss of various properties and of increased complexity of related analysis algorithms. Herewith we briefly examine which properties of RL still hold for CFL and which ones are lost.
  
 The original pumping lemma for regular languages can be naturally extended to a more general version holding for CFL:
  \begin{lemma}
  Let $G$ be a CFG; there is a constant $k$, recursively computable as a function of $G$, such that, for every string $x\in L(G)$, with $|x| >  k$, there exists a factorization $x = yw_1zw_2u$ such that $x = yw_1^nzw_2^nu \in L(G)$ for every $n \geq 0$.
  
  \end{lemma}
  A first important and natural consequence of the CF version of the pumping lemma is that the emptiness problem for CFL is still decidable; this can be easily shown in a parallel way as for regular languages. Another application of the lemma analogous to that used for RL can show that various languages, such as, e.g.,   $L_2 = \{a^{n}b^{n}c^{n} \mid n \geq 0\} $ are not CF.
  
  On the opposite side the fundamental effective equivalence between deterministic and nondeterministic FSA does not hold for PDA. A formal proof of this statement can be found in most texts on formal language theory, e.g. in \cite{Harrison78}; here we only offer an intuitive explanation based on a simple counterexample.
  Consider the language $L = \{ww^R \mid w \in \{a, b\}^*\} $, where $w^R$ denotes the mirror image of $w$. It is quite easy to build a nondeterministic PDA that accepts $L$: initially it pushes the read symbols onto the stack up to a point when it nondeterministaclly changes state and, from that point on it compares the symbol on top of the stack with the read one: if it reaches the end of the input and empties the stack at the same time, through a series of successful comparisons, then that particular computation accepts the input string; of course there will also be many other computations that fail due to a wrong guess about the middle of the input string. It is also $intuitively$ apparent that there is no way to decide deterministically where is the middle of the input sentence unless a device other than a PDA is adopted, e.g., a Turing machine.
  
  This substantial difference in recognition power between general PDA and DPDAs has several consequences on the properties of the class of languages they accept, resp. CFL and DCFL. Here we list the most important ones:
  \begin{itemize}
  \item
  CFL are closed w.r.t. union but not w.r.t. complement and intersection. In fact closure w.r.t. union is easily shown by building, for two languages $L_1$, $L_2$ two corresponding grammars $G_1$ and $G_2$ with disjoint nonterminal alphabets; then simply adding a new axiom $S$ and two productions 
   		$S \to  S_1 \mid S_2$,
  we obtain a grammar that generates all and only the strings in $L_1 \cup L_2$.
  
  On the other hand consider again the above language $L_2$, which is not a CFL: it is the intersection of $L_3 = \{a^nb^nc^* \mid n \geq 0\}$ with  $L_4 = \{a^*b^nc^n \mid n \geq 0\}$ which are clearly both DCFL. By De Morgan's laws, CFL are not closed w.r.t. complement as well.
  \item
  DCFL are closed under complement but neither under union nor under intersection. Closure under complement can be obtained by exploiting determinism and switching $F$ and $Q\setminus F$ in a similar way as done for FSA, provided that a suitable normal form of the original DPDA is built before applying the switch of the accepting states. We refer the reader to the more specialized literature, e.g., \cite{Harrison78}, for the technicalities of the necessary normal form. The previous counterexample also shows the non-closure of DCFL w.r.t. intersection and therefore w.r.t. union, again thanks to De Morgan's laws.
  \item
  CFL are closed under concatenation,  Kleene~$^*$, reversal or mirror image, homomorphism, whereas DCFL are not closed under anyone of these operations. We justify only the claim about homomorphism since the other operations are of minor interest for the purpose of this paper.
 Given an alphabet homomorphism $h: \Sigma_1 \to \Sigma_2^*$ with its natural extension $h: \Sigma_1^* \to \Sigma_2^*$, it is immediate, for a grammar $G$, to build a $G'$ such that $L(G') = h(L(G))$: it suffices to replace every occurrence of an element $a \in \Sigma_1$ in $G$'s definition by $h(a)$, including the particular case $h(a) = \varepsilon$. On the other side consider the language  $ L = \{wcw^R \mid w \in \{a, b\}^*\} $: it is immediate to build a DPDA recognizing it thanks to the new $c$ character which marks the center of the input string, but it has already been shown that $h(L)$, with $h$ defined as $h(a) = a, h(b) = b, h(c) = \varepsilon$,   cannot be recognized by any DPDA.
 \item
 The lack of the above closure properties for both CFL and DCFL, prevents their natural exploitation to obtain decidability of the inclusion property; in fact this problem has been proved to be undecidable in both cases \cite{Harrison78}.\footnote{As side remark, the equivalence problem is undecidable for general CFL and has been proved to be decidable for DCFL after remaining open for a fairly long time \cite{eqdcf}.} 
 
   \end{itemize}

 \subsection{Logic characterization}\label{subsec: MSOCFL}
 The lack of the above closure properties also hampers a natural extension of the logic characterization of regular languages to CFL: in particular the classic inductive construction outlined in Section~\ref{sec:reg:logic} strongly hinges on the correspondence between logical connectives and Boolean operations on sub-languages. Furthermore, the linear structure of RL allows any move of a FSA to depend only on the current state which is associated with a position $\bm{x}$ of the input string and on the input symbol located at position $\bm{x}+1$; on the contrary the typical nested structure of CFL sentences imposes that the move of the PDA may depend on information stored in the stack, which in turn may depend on information read from the input much earlier than the current move.
 
 Despite these difficulties some interesting results concerning a logic characterization of CFL have been obtained. In particular it is worth mentioning the characterization proposed in \cite{Lautemann94}. The key idea is to enrich the syntax of the second order logic with a \emph{matching relation symbol} $M$ which takes as arguments two string position symbols $\bm{x}$ and $\bm{y}$: a matching relation interpreting $M$ must satisfy the following axioms:
 
 \begin{itemize}
 \item
 $M(\bm{x}, \bm{y}) \Rightarrow \bm{x} < \bm{y} $: $\bm {y}$ always follows $\bm {x}$; 
 \item
 $M(\bm{x}, \bm{y})  \Rightarrow \not\exists \bm{z}  (\bm{z}\not\eq\bm{x}
  \land  \bm{z}\not\eq\bm{y}
 \land 
 (M(\bm{x}, \bm{z}) \lor M(\bm{z},\bm{y})\lor M(\bm{z},\bm{x})\lor M(\bm{y},\bm{z})))$: $M$ is one-to-one;
 
 \item
  $\forall \bm{x},\bm{y},\bm{z},\bm{w}((M(\bm{x}, \bm{y})\land M(\bm{z}, \bm{w}) \land \bm{x} < \bm{z} < \bm{y}) \Rightarrow \bm{x} < \bm{w} < \bm{y})$: $M$ is \textit{nested}, i.e., if we represent graphically  $M(\bm{x}, \bm{y})$ as an edge from $\bm {x}$ to $\bm {y}$ such edges cannot cross.
 
  \end{itemize} 
  
 The matching relation is then used to represent the tree structure(s) associated with a CF language sentence: for instance consider the (ambiguous) grammar $G_{amb}$
  \[ 	\begin{array}{ll}
   		G_{amb}: &
   		S \to  A_1 \mid A_2  \\
   &		A_1 \to  a a A_1 b b \mid aabb\\
   &		A_2 \to a A_3 b\\
   &		A_3 \to a A_2 b \mid ab\\	   
   	\end{array}
   \]

\begin{figure}
	\centering
	\begin{tikzpicture}
	\matrix (m) [matrix of nodes]
	{
		$a$ & $a$ & $a$ & $a\ $ & $\ b$ & $b$ & $b$ & $b$ \\
	};
	
	\draw[->] (m-1-1)   [out=75, in=105] to (m-1-8);
	\draw[->] (m-1-2)   [out=75, in=105] to (m-1-7);
	\draw[->] (m-1-3)   [out=75, in=105] to (m-1-6);
	\draw[->] (m-1-4)   [out=75, in=105] to (m-1-5);
  
	\draw[->] (m-1-1)   [out=-75, in=-105] to (m-1-8);
	\draw[->] (m-1-3)   [out=-75, in=-105] to (m-1-6);
	\end{tikzpicture}
	\caption{Two matching relations for $aaaabbbb$, one is depicted above and the
    other below the string.}\label{fig:doublematch}
\end{figure}
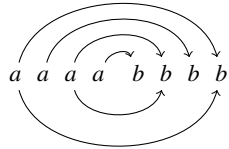

   $G_{amb}$ induces a natural matching relation between the positions of characters in its strings. For instance Figure~\ref{fig:doublematch} shows the two relations associated with the string $aaaabbbb$.
   
   More generally we could state that for a grammar $G$ and a sentence $x =
   a_1a_2 ... a_n \in L(G)$ with $\forall k, a_k \in \Sigma$, $(i,j) \in M$ iff  $ S \stackrel * {\Rightarrow}_G   a_1a_2 ... a_{i-1}Aa_{j+1} ... a_n  \stackrel * {\Rightarrow}_G  a_1a_2 ... a_n$. It is immediate to verify, however, that such a definition in general does not guarantee the above axioms for the matching relation: think e.g., to the previous grammar $GAE_1$ which generates arithmetic expressions. For this reason \cite{Lautemann94} adopts the so-called \emph{double Greibach normal form} (DGNF) which is an effective but non-trivial transformation of a generic CFG into an equivalent one where the first and last characters of any production are terminal.\footnote{To be more precise, the normal form introduced in \cite{Lautemann94} is further specialized, but for our introductory purposes it is sufficient to consider any DGNF.}
         It is now immediate to verify that a grammar in DGNF does produce a matching relation on its strings that satisfies all of its axioms.
   
   Thanks to the new relation and the corresponding symbol \cite{Lautemann94} defines a second order logic CF$\mathcal{L}$ that characterizes CFL: the sentences of CF$\mathcal{L}$ are first-order formulas prefixed by a single existential quantifier applied to the second order variable $M$ representing the matching relation. Thus, intuitively, a CF$\mathcal{L}$ sentence such as $\exists M (\phi)$ where $M$ occurs free in $\phi$ and $\phi$ has no free occurrences of first-order variables is satisfied iff there is a structure defined by a suitable matching relation such that the positions that satisfy the occurrences of $M$ in $\phi$ also satisfy the whole~$\phi$. For instance the sentence
   \begin{small}
\[ 
\exists  M, \bm z 
\left(
\begin{array}{l}
  \not\exists \bm x (\sux(\bm z, \bm x)) \land M(0, z) \land \\
    \forall \bm x, \bm y (M(\bm x,\bm y) \Rightarrow a(\bm x) \land b(\bm
  y)) \land \\
  \exists \bm y \forall \bm x \left(
  \begin{array}{l}
(0 \le \bm x < \bm y \Rightarrow  a(\bm x)) \land \\
  (\bm x \ge \bm y \ge \bm z \Rightarrow  b(\bm x))
\end{array}
\right)  
\land \\
  \left(
\begin{array}{c}
   \forall \bm x, \bm y
\left(
   M(\bm x,\bm y) \Rightarrow 
\begin{array}{c}
   (\bm x > 0 \Rightarrow  M(\bm x-1, \bm y+1)) \land \\  
   (\bm x < \bm y-2 \Rightarrow  M(\bm x+1, \bm y-1)))\\
\end{array}
\right)
\\
\lor
\\
\forall \bm x, \bm y
\left(
   M(\bm x,\bm y) \Rightarrow 
\begin{array}{c}
   (\bm x > 1 \Rightarrow  M(\bm x-2, \bm y+2) \land \\
   \neg M(\bm x-1, \bm y+1)) \land \\
   (\bm x < \bm y - 4 \Rightarrow  M(\bm x+2, \bm y-2) \land \\ 
   \neg M(\bm x+1, \bm y-1))
\end{array}
\right)
\end{array}
\right)
\end{array}
\right)
\]
      \end{small}
   is satisfied by all and only the strings of $L(G_{amb})$ with both the $M$ relations depicted in Figure~\ref{fig:doublematch}.
   
   After having defined the above logic, \cite{Lautemann94} proved its equivalence with CFL in a fairly natural way but with a few non-trivial technicalities: with an oversimplification, from a given CFG in DGNF a corresponding logic formula is built inductively in such a way that $M(\bm{x},\bm{y})$ holds between the positions of leftmost and rightmost leaves of any subtree of a grammar's syntax-tree; conversely, from a given logic formula a tree-language, i.e., a set of trees, is built such that the frontiers of its trees are the sentences of a CFL. However, as the authors themselves admit, this result has a minor potential for practical applications due to the lack of closure under complementation. The need to resort to the DGNF puts severe constraints on the structures that can be associated with the strings, a priori excluding, e.g., linear structures typical of RL; nonetheless the introduction of the $M$ relation opened the way for further important developments as we will show in the next sections.  
\subsection {Parsing} \label{parsing}
PDA are the natural abstract machines to recognize CFL as well as FSA are for RL; we have also seen that augmenting PDA with a suitable output device makes them pushdown transducers (PDT) which can be used as parsers, i.e. to build the syntax-tree(s) associated with a given string of the language. Unlike the case of RL, however, DPDA are not able to recognize all CFL. Thus, if we want to recognize or parse a generic CFL, in principle we must simulate all possible computations of a nondeterministic PDA (or PDT); this approach clearly raises a critical complexity issue. For instance, consider the analysis of any string in $\Sigma^*$ by the PDA accepting  $L = \{ww^R \mid w \in \{a, b\}^*\}$: in the worst case at each move the automaton ``splits'' its computation in two branches by guessing whether it reached the middle of the input sentence or not; in the first case the computation proceeds deterministically to verify that from that point on the input is the mirror image of what has been read so far, whereas the other branch of the computation proceeds still waiting for the middle of the string and splitting again and again at each step. Thus, the total number of different computations equals the length of the input string, say $n$, and each of them has in turn an $O(n)$ length; therefore, simulating the behavior of such a nondeterministic machine by means of a deterministic algorithm to check whether at least one of its possible computations accepts the input has an $O(n^2)$ total complexity.

The above example can be generalized in the following way: on the one hand we have

\begin{statement} \label{RTPDA}
Every CFL can be recognized in real-time, i.e. in a number of moves equal to the length of the input sentence, by a, possibly nondeterministic, PDA.
\end{statement}

\noindent One way to prove the statement is articulated in two steps:

\noindent 1) First, an original CFG generating $L$ is transformed into the \emph{Greibach normal form (GNF)}\footnote{The previous DGNF is clearly a ``symmetric variant'' of the original GNF but is not due to the same author and serves different purposes.}:
\begin{definition}\label{def: GNF}
A CFG is in Greibach normal form \cite{Harrison78} iff the rhs of all its
productions belongs to $\Sigma V_N^*$.
\end{definition}
The procedure given in \cite{Greibach65} to transform any CF grammar into the normal form essentially is based on transforming any \emph{left recursion}, i.e. a derivation such as $A \derives{*} A\alpha$ into an equivalent right one $B \derives{*} \alpha B$.

\noindent 2) Starting from a grammar in GNF the procedure to build an equivalent PDA therefrom can be ``optimized'' by:
\begin{itemize}
\item
restricting $\Gamma$ to $V_N$ only; 
\item
when a symbol $A$ is on top of the stack, a single move reads the next input symbol, say $a$, and replaces $A$ in the stack with the string $\alpha$ of the rhs of a production $A \to a\alpha$, \textit{if any} (otherwise the string is rejected). 
\end{itemize}
Notice that such an automaton is real-time since there are no more $\varepsilon$-moves but, of course, it may still be nondeterministic. If the grammar in GNF is such that there are no two distinct productions of the type $A \to a\alpha$, $A \to a\beta$, then the automaton built in this way is a real-time DPDA that is able to build leftmost derivations of the grammar. In Section~\ref{TDDetPars} we will go back to the issue of building deterministic parsers for CFL.
   
  Statement \ref{RTPDA}, on the other hand, leaves us with the natural
  question: ``provided that purely nondeterministic machines are not physically
  available and at most can be approximated by parallel machines which however
  cannot exhibit an unbounded parallelism, how can we come up with some
  deterministic parsing algorithm for general CFL and which complexity can
  exhibit such an algorithm?''. In general it is well-known that simulating
  a nondetermistic device with complexity\footnote{As usual we assume as the
  complexity of a nondeterministic machine the length of the shortest
  computation that accepts the input or of the longest one if the string is
  rejected.} $f(n)$ by means of a deterministic algorithm may expose to the
  risk of even an exponential complexity $O(2^{f(n)})$. For this reason on the
  one hand many applications, e.g., compiler construction, have restricted
  their interest to the subfamily of DCFL; on the other hand intense research
  has been devoted to design efficient deterministic parsing algorithms for
  general CFL by departing from the approach of simulating PDA. The case of
  parsing DCFL will be examined in  Section \ref{sec:DCFPars}; parsing general
  CFL, instead is of minor interest in this paper, thus we simply mention  the
  two most famous and efficient of such algorithms, namely the one due to
  Cocke, Kasami, and Younger, usually referred to as CKY and described, e.g.,
  in \cite{Harrison78}, and the one by Early reported in \cite{CBM13}; they
  both have an $O(n^3)$ time complexity.\footnote{We also mention (from
  \cite{Harrison78}) that in the literature there exists a variation of the CKY
  algorithm due to Valiant that is completely based on matrix multiplication
  and therefore has the same asymptotic complexity of this basic problem, at
  the present state of the art $O(n^{2.37})$.}
    
To summarize, in this section we have seen that CFL are considerably more general than RL
but, on the one side they require parsing algorithms to assign a given sentence an appropriate (tree-shaped
and usually not immediately visible) structure and, on the other side, they lose several closure
and decidability properties typical of the simpler RL. Not surprisingly, therefore, much, largely
unrelated, research has been devoted to face both such challenges; in both cases major successes
have been obtained by introducing suitable subclasses of the general language family: on the one
hand parsing can be accomplished for DCFL much more efficiently than for nondeterministic ones
(furthermore in many cases, such as e.g, for programming languages, nondeterminism and even
ambiguity are features that are better to avoid than to pursue); on the other hand various subclasses
of CFL have been defined that retain some or most of the properties of RL yet increasing their
generative power. In the next sections we resume, in the reverse order, the major results obtained
on both sides, so far within different research areas and by means of different subfamilies of CFL.
As anticipated in the introduction, however, we will see in the following sections that one of such
families allows for major improvements on both sides.

%


\section{Structured context-free languages}\label{sec:strCF}

RL sentences have a fixed, right or left, linear structure; CFL sentences have a more general tree-structure, of which the linear one is a particular case, which normally is not immediately visible in the sentence and, in case of ambiguity, it may even happen that the same sentence has several structures.
R. McNaughton, in his seminal paper \cite{McNaughton67}, was probably the first one to have the intuition that, if we ``embed the sentence structure in the sentence itself'' in some sense making it visible from the frontier of the syntax-tree (as it happens implicitly with RL since their structure is fixed a priori), then many important properties of RL still hold for such special class of CFL.

Informally, we name such CFL {\em structured} or {\em visible structure} CFL. The first formal definition of this class given by Mc\-Naugh\-ton is that of {\em parenthesis languages}, where each subtree of the syntax-tree has a frontier embraced by a pair of parentheses; perhaps the most widely known case of such a language, at least in the field of programming languages is the case of LISP. Subsequently several other equivalent or similar definitions of structured languages, have been proposed in the literature; not surprisingly, an important role in this field is played by \textit{tree-languages} and their related recognizing machines, \textit{tree-automata}.
Next we browse through a selection of such language families and their properties,
starting from the seminal one by McNaughton.

\subsection{Parenthesis grammars and languages}
\begin{definition}
For a given terminal alphabet $\Sigma$ let $[$, $]$ be two symbols  $\not\in \Sigma$. A parenthesis grammar (PG) with alphabet $\Sigma \cup \{[, ]\}$ is a CFG whose productions are of the type $A \to [\alpha]$, with $\alpha \in V^*$.
\end{definition}
It is immediate to build a parenthesis grammar naturally associated with any CFG: for instance the following PG is derived from the $GAE_1$ of Example \ref{GAE}:
\[
 	\begin{array}{ll}
 	GAE_{[]}: &
 		S \to [E] \mid [T] \mid [F]\\
 &		E \to  [E + T] \mid [T * F] \mid [e]  \\  	
 &		T \to  [T * F] \mid [e]      \\
 &		F \to  [e]  \\
 	\end{array}
 \]
 It is also immediate to realize that, whereas $GAE_1$ does not make immediately visible in the sentence $e + e * e$ that $e*e$ is the frontier of a subtree of the whole syntax-tree, $GAE_{[]}$ generates $[[[e] + [[e] * [e]]]]$ (but not $[[[[e] + [e]] * [e]]]$), thus making the structure of the syntax-tree immediately visible in its parenthesized frontier.
 
 Sometimes it is convenient, when building a PG from a normal one, to omit parentheses in the rhs of $renaming$ $rules$, i.e., rules whose rhs reduces to a single nonterminal, since such rules clearly do not significantly affect the shape of the syntax-tree. In the above example such a convention would avoid the useless outermost pair of parentheses.
  
  Given that the trees associated with sentences generated by PG are isomorphic to the sentences, the parsing problem for such languages disappears and scales down to a simpler recognition problem as it happens for RL. Thus, rather than using the full power of general PDA for such a job, \emph{tree-automata} have been proposed in the literature as a recognition device equivalent to PG as well as FSA are equivalent to regular grammars.
  Intuitively, a tree-automaton (TA) traverses either top-down or bottom-up a labeled tree to decide whether to accept it or not, thus it is a tree-recognizer or tree-acceptor.

\subsection{Tree Automata}\label{TrAut}

Here we give an introductory view of the tree automata formalism: our formal definition is inspired by the
traditional ones, e.g., in \cite{Tha67,TATA} but with minor adaptations for convenience and homogeneity with the notation adopted in this paper; to avoid possible terminological confusion with other literature on tree-automata we give a different name to such automata.

\begin{definition}
  A \emph{stencil} of a terminal alphabet $\Sigma$ is a string in $(\Sigma \cup \{N\})^*$. 
  A \emph{stencil or tree alphabet} is a finite set of stencils. The number of occurrences of the symbol $N$ in a stencil $st$ is called the \emph{stencil's arity} and is denoted $n(st)$.
  \end{definition}
   \begin{definition}\label{def:TDSTA}
    A \emph{top-down stencil automaton} is a tuple $(Q,$ $ST_\Sigma,$ $\Delta, I)$, where
    \begin{itemize}
    \item
    $Q$ is a finite set of states
    \item $ST_\Sigma$ is a finite set of stencils of $\Sigma$; $ST_{\Sigma,n}$ denotes the subset of $ST_\Sigma$ of the stencils with arity $n$ 
    \item $\Delta$ is a collection of relations $\{\delta_n \subseteq Q\times ST_{(\Sigma,n)} \times Q^n\}$
    \item $I \subseteq Q$ is the set of initial states.
     \end{itemize}
   A top-down stencil automaton is deterministic iff
   \begin{itemize}
   \item
   For each $n, \delta_n$ is a function $\delta_n: Q \times ST_{(\Sigma,n)} \to Q^n$
   \item
   $I$ is a singleton
    \end{itemize}
    \end{definition}
     \begin{definition}
     A tree on a given pair  ($ST_\Sigma$, $Q$) is a tree whose internal nodes are labeled by elements of $Q$, leaves are labeled by elements of $\Sigma$, and the homomorphism $h(q) = N$, with $q\in Q$ projects all strings of children of any node of the tree into an element of $ST_\Sigma$
   
     A tree on a pair ($ST_\Sigma$, $Q$) is accepted by a top-down stencil automaton $\mathcal{A} = (Q, ST_\Sigma, \Delta, I)$ iff
     \begin{itemize}
     \item 
     the root is labeled by an element $\in I$
     \item
     each pair ($q_f,\alpha)$ where $q_f$ is the label of a father node, $\alpha$ is the string of its children in $(\Sigma \cup Q)^*$ is such that $(q_f, h(\alpha), q_1, ..., q_n)\in \delta_n$, where $n$ is the arity of $ h(\alpha)$ and $q_1, ..., q_n$ are the states ocurring in $\alpha$, in the same order.
     \end{itemize}
    \end{definition} 
    It is immediate to state a one-to-one, effective correspondence between parenthesis grammars and top-down stencil automata; in doing so it may be convenient to preliminarily eliminate from the grammar renaming rules \cite{Harrison78} which do not have a real meaning in terms of sentence structure; such a normal form also requires that, instead of having just one axiom with possible renaming rules with it as lhs, $S$ is a subset of $V_N$ and there is a one-to-one correspondence between $S$ and $I$.
    \begin{example}\label{ex:STAAE}
    The following stencil automaton $\mathcal{A} = (Q,$ $ST_\Sigma,$ $\Delta, I)$ recognizes the syntax-trees of the grammar $GAE_1$ whose structure, up to the labeling of the internal nodes, is also described by the sentences of the parenthesis grammar $GAE_{[]}$.
    \begin{itemize}
    \item
    $Q = \{E,T, F\}$
    \item
    $ST_\Sigma = \{e, N+N, N*N \}$
    \item
    $\Delta = \{(E, e), (T, e), (F, e), (T, N*N, T, F), (E, N*N, T, F),$ $(E, N+N, E, T)\}$
    \item
    $I = \{E, T, F\}$
    \end{itemize}
    \end{example}
    If we compare the automata of Examples~\ref{ex:PDAAE} and \ref{ex:STAAE} we notice that both of them, resp., build and recognize the syntax-trees generated by grammar $GAE_1$ in top-down order; precisely, the computations of the automaton of Example~\ref{ex:PDAAE} follow a leftmost derivation of the grammar, whereas the automaton of Example~\ref{ex:STAAE} simply states a partial order in the application of the $\delta$ relation going from father nodes to children. In both cases, however, it may be convenient to use automata that work in a symmetric way, i.e., resp. build or traverse the syntax-trees bottom up. Two important cases of bottom up parsers are described in Sections~\ref{LRPars} and \ref{sec: OPParPars}; here instead we introduce \emph{bottom-up stencil automata} and compare them with the top-down ones.
     \begin{definition}\label{def:BUSTA}
     A \emph{bottom-up stencil automaton} is a tuple $(Q,$ $ST_\Sigma,$ $\Delta, F)$, where $Q$ and $ ST_\Sigma$ are as in Definition \ref{def:TDSTA},
     \begin{itemize}
     \item
      $\Delta$ is a collection of relations $\{\delta_n \subseteq Q^n \times ST_{(\Sigma,n)} \times Q\}$ 
       \item $F \subseteq Q$ is the subset of $Q$'s final states.
     \end{itemize}
     The automaton is deterministic iff for each $n$, $\delta_n$ is a function $\delta_n: Q^n \times ST_{(\Sigma,n)} \to Q$.
     
     A tree on a pair ($ST_\Sigma$, $Q$) is accepted by a bottom-up stencil automaton $\mathcal{A} = (Q, ST_\Sigma, \Delta, F)$ iff
     \begin{itemize}
    
       \item
          each pair ($\alpha,q_f)$ where $q_f$ is the label of a father node, $\alpha$ is the string of its children in $(\Sigma \cup Q)^*$ is such that $(q_1, ..., q_n, h(\alpha),q_f)\in \delta_n$, where $n$ is the arity of $ h(\alpha)$ and $q_1, ..., q_n$ are the states ocurring in $\alpha$, in the same order;
        \item 
         the root is labeled by an element $\in F$.
      \end{itemize}    
     \end{definition}
     Structured CFL, of which parenthesis languages are a first major example, enjoy some properties that hold for RL but are lost by general CFL. In particular, we mention their closure under Boolean operations, which has several major benefits, including the decidability of the containment problem. They key milestones that allowed McNaughton to prove this closure are:
     \begin{itemize}
     \item
     The possibility to apply all operations within a\textit{ structure universe}, i.e., to a universe of syntax-trees rather than to the ``flat'' $\Sigma^*$; in the case of parenthesis languages the natural universe associated with a given set of stencils is generated by the corresponding \emph{stencil grammar}, i.e. the grammar with the only nonterminal $N$ and one production for every stencil.
     \item
     The possibility of building a normal form of any parenthesis grammar \emph{with no repeated rhs}, i.e., such that for every rhs there is only one production rewriting it. Such a construction has been defined by McNaughton by referring directly to the grammars but can be explained even more intuitively by noticing that, if we transform a bottom-up stencil automaton into a parenthesis grammar and conversely, the grammar has no repeated rhs iff the automaton is deterministic. In both cases the procedure to obtain such a normal form is a natural extension of the basic one applied to FSA and is based on defining a new set of states (resp. nonterminals) that is the power set of the original one; for instance, suppose that a grammar contains two rules $A \to [ab]$ and $B \to [ab]$; then they are ``collapsed'' into the unique one $AB \to [ab]$; then the procedure is iterated by replacing both $A$ and $B$ by the new nonterminal $AB$ in every rhs where they occur and so on until no new nonterminal is generated. It is then an easy exercise to prove the equivalence between the original grammar (resp. automaton) and the new one by means of a natural induction.
     
     It is important to emphasize, however, that such a way to \emph{determinize} a nondeterministic bottom-up stencil automaton does not work in the case of the top-down version of these automata: deterministic top-down tree (or stencil) automata in fact have been proven to be less powerful than their bottom-up counterpart. We refer the reader to a more specific literature, e.g. \cite{TATA}, for a proof of this claim; here we simply anticipate that something similar occurs for the parsing of DCFL.
     \end{itemize}
     On the basis of these two fundamental properties deriving effective closure w.r.t. Boolean operations within a given universe of stencils is a natural extension of the operations already outlined for RL (notice that RL are a special case of structured languages whose stencils are only linear, i.e., of the type $aN, a$ (or $Na, a$) for $a \in \Sigma$).
     \begin{itemize}
     \item
     The complement language w.r.t. the universe of syntax-trees associated with a given set of stencils is obtained by 
     \begin{itemize}
      \item
      Adding a new conventional state $q_{err}$ to a deterministic bottom up stencil automaton and completing each $\delta_n$ with it on the whole domain $Q^n \times ST_{(\Sigma,n)}$ in the same way as for RL.
      \item
      Complementing the set of accepting states.
      \end{itemize}
      \item
     The intersection between two languages sharing the same set of stencils is obtained by building a new set of states that is the cartesian product of the original ones and extending the $\delta$ function in the usual way.
      \end{itemize}
     As usual, an immediate corollary of these closure properties is the decidability of the containment problem for two languages belonging to the same universe.
     
     We just mention another important result that is obtained as an extension of its version that holds for RL, namely, the possibility of minimizing the grammar generating a parenthesis language (resp. the automaton recognizing a tree language), w.r.t. the number of necessary nonterminals (resp. states); this type of results, however, is not in the scope of this paper; thus we refer the interested reader to the original paper by McNaughton or other subsequent literature.

  

     On the basis of the important results obtained by Mc\-Naugh\-ton in his seminal paper, many other families of CFL
       have been defined in several decades of literature with the general goal of extending (at least some of the) closure and decidability properties, and logic characterizations that make RL such a ``nice'' class despite its limits in terms of generative power. Most of such families maintain the key property of being ``structured'' in some generalized sense w.r.t. parenthesis languages. In the following section we introduce so called \emph{input-driven languages}, also known as \emph{visibly pushdown languages} which received much attention in recent literature and exhibit a fairly complete set of properties imported from RL.
       
     \subsection{Input-driven or visibly pushdown languages}\label{sec: IDL}
     The concept of \emph{input-driven CF language} has been introduced in \cite{DBLP:conf/icalp/Mehlhorn80} in the context of building efficient recognition algorithms for DCFL: according to \cite{DBLP:conf/icalp/Mehlhorn80} a DPDA is input-driven if the decision of the automaton whether to execute a \emph{push move}, i.e. a move where a new symbol is stored on top of the stack, or a \emph{pop move}, i.e. a move where the symbol on top of the stack is removed therefrom, or a move where the symbol on top of the stack is only updated, depends exclusively on the current input symbol rather than on the current state and the symbol on top of the stack as in the general case. Later, several equivalent definitions of the same type of pushdown automata, whether deterministic or not, have been proposed in the literature; among them, here we choose an early one proposed in \cite{AluMad04}, which better fits with the notation adopted in this paper than the later one in \cite{jacm/AlurM09}.
     \begin{definition}\label{def: VPL}
     Let the input alphabet $\Sigma$ be partitioned into three disjoint alphabets, $\Sigma = \Sigma_c \cup \Sigma_r \cup \Sigma_i$, named, respectively, the \emph{call} alphabet, \emph{return} alphabet, \emph{internal} alphabet. A \emph{visibly pushdown automaton (VPA)} over $(\Sigma_c , \Sigma_r , \Sigma_i)$ is a tuple $(Q, I, \Gamma,$ $Z_0,$ $\delta, F)$, where
     \begin{itemize}
     \item
     $Q$ is a finite set of states;
     \item
     $I \subseteq Q$ is the set of initial states;
     \item
      $F \subseteq Q$ is the set of final or accepting states;
      \item
      $\Gamma$ is the finite stack alphabet;
      \item 
      $Z_0 \in \Gamma$ is the special bottom stack symbol;
      \item
      $\delta$ is the transition relation, partitioned into three disjoint subrelations:
      \begin{itemize}
      \item
      Call move: $\delta_c \subseteq Q \times \Sigma_c \times Q \times (\Gamma \setminus  \{Z_0\})$,
      \item
      Return move: $\delta_r \subseteq Q \times \Sigma_r \times \Gamma \times Q$,
       \item
       Internal move: $\delta_i \subseteq Q \times \Sigma_i \times Q$.
      \end{itemize}
      \end{itemize}
A VPA is deterministic iff
      $I$ is a singleton, and
      $\delta_c$, $\delta_r$, $\delta_i$ are functions: \\
      $\delta_c : Q \times \Sigma_c \to Q \times (\Gamma \setminus  \{Z_0\})$, 
      $\ \delta_r :Q \times \Sigma_r \times \Gamma \to Q$,
      $\ \delta_i : Q \times \Sigma_i \to Q$.

     \end{definition}

     The automaton configuration, the transition relation between two configurations, the acceptance of an input string, and the language recognized by the automaton are then defined in the usual way: for instance if the automaton reads a symbol $a$ in $\Sigma_c$ while is in the state $q$ and has $C$ on top of the stack, it pushes onto the stack a new symbol $D$ and moves to state $q'$ provided that $(q, a, q', D)$ belongs to $\delta_c$; notice that in this way the special symbol $Z_0$ can occur only at the bottom of the stack, during the computation. A language over an alphabet $\Sigma = \Sigma_c \cup \Sigma_r \cup \Sigma_i$ recognized by some VPA is called a \emph{visibly pushdown language (VPL)}. 
     
     The following remarks help put IDL alias VPL in perspective with other families of CFL.

     \begin{itemize}
\item
Once PDA are defined in a standard form, with respect to the general one given in Definition \ref{def:PDA} where their moves either push a new symbol onto the stack or remove it therefrom or leave the stack unaffected, the two definitions of IDL and VPL are equivalent. Both names for this class of languages are adequate: on the one side, the attribute input-driven emphasizes that the automaton move is determined exclusively on the basis of the current input symbol\footnote{We will see, however, that the same term can be interpreted in a more general way leading to larger classes of languages.}; on the other side we can consider VPL as \emph{structured languages} since the structure of their sentences is immediately \emph{visible} thanks to the partitioning of $\Sigma$.
      \item
VPL generalize McNaughton's parenthesis languages:\\ open parentheses are a special case of  $\Sigma_c$ and closed ones of  $\Sigma_r$; further generality is obtained by the possibility of performing an unbounded number of internal moves, actually ``purely finite state'' moves between two matching call and return moves and by the fact that VPA can accept unmatched return symbols at the beginning of a sentence as well as unmatched calls at its end; a motivation for the introduction of such a generality is offered by the need of modeling systems where a computation containing a sequence of procedure calls is suddenly interrupted by a special event such as an exception or an interrupt.
     \item
     Being VPL essentially structured languages, their corresponding automata are just recognizers rather than real parsers.
     \item
     VPL are \emph{real-time languages}; in fact VPA read one input symbol for each move. We have mentioned that this property can be obtained for nondeterministic PDA recognizing any CFL but not for deterministic ones.
      \end {itemize}
      
      VPL have obtained much attention since they represent a major step in the path aimed at extending many, if not all, of the important properties of RL to structured CFL. They are closed w.r.t. all major
      language operations, namely the Boolean ones, concatenation,  Kleene~$^*$ and others; this also implies the decidability of the containment problem, which, together with the characterization in terms of a MSO logic, again extending the result originally stated for RL, opens the door to applications in the field of automatic verification.
      
      A key step to achieve such important results is the possibility of effectively determinizing nondeterministic VPA. The basic idea is similar to the classic one that works for RL, i.e, to replace the uncertainty on the current state of a nondeterministic automaton with the subset of $Q$ containing all such possible states (see Example~\ref{ex:reg:aut3}). Unlike the case of FSA however, when the automaton executes a return move it is necessary to ``match'' the current state with the one that was entered at the time of the corresponding call; to do so the key idea is to ``pair'' the set of states nondeterministically reached at the time of a return move  with those that were entered at the time of the corresponding call; intuitively, the latter ones are memorized and propagated through the stack, whose alphabet is enriched with pairs of set states.
      As a result at the moment of the return it is possible to check whether some of the states memorized at the time of the call ``match'' with some of the states that can be currently held by the nondeterministic original automaton.
      
      We do not go into the technical details of this construction, referring the reader to \cite{AluMad04} for them; we just mention that, unlike the case of RL, the price to be paid in terms of size of the automaton to obtain a deterministic version from a nondetermistic one is $2^{O(s^2)}$, where $s$ is the size of the original state set. In \cite{jacm/AlurM09} the authors also proved that such a gap is unavoidable since there are VPL that are recognized by a nondeterminitic VPA with a state set of cardinality $s$ but are not recognized by any deterministic VPA with less than $2^{s^2}$ states.
In Section \ref{sec: OPAlg-Log} we will provide a similar proof of determinization for a more general class of automata.
      
      Once a procedure to obtain a deterministic VPA from a nondeterministic one is available, closure w.r.t. Boolean operations follows quite naturally through the usual path already adopted for RL and parenthesis languages. Closure under other major language operations such as concatenation and  Kleene~$^*$ is also obtained without major difficulties but we do not report on it since those operations are not of major interest for this paper. Rather, we wish here to go back to the issue of logical characterization.
      \subsubsection{The logic characterization of visibly pushdown languages}\label{sec:MSOVPL}
      We have seen in Section \ref{subsec: MSOCFL}
      that attempts to provide a logic characterization of general CFL produced only partial results due to the lack of closure properties and to the fact that CFL do not have an a priori fixed structure; in fact the characterization offered by \cite{Lautemann94} and reported here requires an existential quantification w.r.t. relation $M$ that represents the structure of a string. Resorting to structured languages such as VPL instead allowed for a fairly natural generalization of the classical B\"uchi's result for RL.
      
      The key steps to obtain this goal are \cite{jacm/AlurM09}:
      \begin{itemize}
      \item
      Using the same relation $M$ introduced in \cite{Lautemann94},\footnote{Renamed \emph{nesting relation} and denoted as $\leadsto$ or $\nu$ in later literature.} adding its symbol as a basic predicate to the MSO logic's syntax given in Section \ref{sec:reg:logic} for RL, and extending its interpretation in the natural way. This turns out to be simpler and more effective in the case of structured languages since, being such languages a priori unambiguous (the structure of the sentence is the sentence itself), there is only one such relation among the string positions and therefore there is no need to quantify it. Furthermore the relation is obviously one-to-one with a harmless exception due to the existence of unmatched closed parentheses at the beginning of the sentence and unmatched open ones at the end: in such cases conventional relations $M(- \infty, \bm{x})$, $M(\bm{x}, + \infty)$ are stated.
      \item
       Repeating exactly the same path used for RL both in the construction of an automaton from a logic formula and in the converse one. This only requires the managing of the new $M$ relation in both constructions; precisely:
      
     \par In the construction from the MSO formula to VPA, the elementary  automaton associated with the atomic formula $M(\bm{X},
      \bm{Y})$, where $\bm X$ and $\bm Y$ are the usual singleton second order
      variables for any pair of first order variables $\bm x$ and $\bm y$, is
      represented by the diagram of Figure~\ref{MSOVPA} where, like in the same construction for RL, $\circ$ stands for any value of $\Sigma$ for which the transition can be defined according to the alphabet partitioning, so that
      the automaton is deterministic, the second component of the triple
      corresponds to $\bm X$, and the third to $\bm Y$.\footnote{We use here the
      following notation for depicting VPA: a label $a / B$ stands for a push move,
      $a, B$ for a pop move, and $a$ alone for an internal move.}

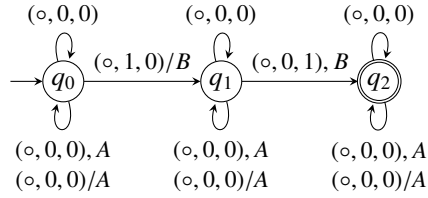
\begin{figure} [h]

\begin{center}
\begin{tikzpicture}[every edge/.style={draw,solid}, node distance=4.2cm, auto, 
                    every state/.style={draw=black!100,scale=0.5}, >=stealth]

\node[initial by arrow, initial text=,state] (q0) {{\huge $q_0$}};
\node[state] (q1) [right of=q0] {{\huge $q_1$}};
\node[state] (q2) [right of=q1, accepting] {{\huge $q_2$}};

\path[->]
(q0) edge [loop above]  node {$(\circ,0,0)$} (q0)
(q0) edge [loop below]  node 
{$\begin{array}{c}
(\circ,0,0), A\\
(\circ,0,0)/ A
 \end{array}$
} (q0)
(q0) edge [above] node {$(\circ,1,0) / B$} (q1)
(q1) edge [loop above]  node {$(\circ,0,0)$} (q1)
(q1) edge [above] node {$(\circ,0,1), B$} (q2)
(q1) edge [loop below]  node 
{$\begin{array}{c}
(\circ,0,0), A\\
(\circ,0,0)/ A
 \end{array}$
} (q1)
(q2) edge [loop above]  node {$(\circ,0,0)$} (q2)
(q2) edge [loop below]  node 
{$\begin{array}{c}
(\circ,0,0), A\\
(\circ,0,0)/ A
 \end{array}$
} (q2)
;
\end{tikzpicture}
\end{center}
\caption{ VPA associated with $M(\bm{X},
      \bm{Y})$ atomic formula.} \label{MSOVPA}
\end{figure}

      \par In the construction from the VPA to the MSO formula, besides variables $\bm X_i$ for encoding states, we also need
        variables to encode the stack. We introduce variables $\bm C_A$ and $\bm
        R_A$, for $A \in \Gamma$, to encode, respectively, the set of positions in
        which a {\em call} pushes 
        $A$ onto the stack, and in which a {\em return} pops 
        $A$ from the stack. 

The following formula
states that every pair $(\bm x, \bm y)$ in $M$ must belong, respectively, to exactly one $\bm C_A$ and exactly one $\bm R_A$:
\[
\begin{array}{l}
\forall \bm x, \bm y \left(
M(\bm x, \bm y) \Rightarrow \bigvee_{A \in \Gamma} \bm x
\in \bm C_A \land \bm y \in \bm R_A  
\right)
\land\\
\forall \bm x 
\bigwedge_{A \in \Gamma}
\left(
\begin{array}{c} 
\bm x \in \bm C_A \Rightarrow
\neg \bigvee_{B \ne A} \bm x \in \bm C_B \\
  \land\\
\bm x \in \bm R_A \Rightarrow
\neg \bigvee_{B \ne A} \bm x \in \bm R_B
\end{array}
\right).
\end{array}
\]

      The following subformulas express the conditions for call and return
      transitions, respectively.
They are added to the ones already described in
Section~\ref{sec:reg:logic} (without loss of
generality, we assume that the original VPA is deterministic).
Subformulas for internal transitions are almost identical to those for FSA.
\[
\forall \bm{x}, \bm{y}  
\bigwedge_{0 \le i \le m}
\bigwedge_{A \in \Gamma}
\bigwedge_{a \in \Sigma} 
\left(
\begin{array}{c}
\bm x \in \bm{X}_i \land 
\sux(\bm x, \bm{y}) \land \\
a(\bm{y}) \land
\delta_c(i, a) = (j, A) 
\\ \Rightarrow \\
   \bm y \in \bm C_A \land
\bm y \in \bm{X}_j
\end{array}
\right)
\]
      \[ \forall \bm{x}, \bm{y}, \bm{z} 
\bigwedge_{0 \le i \le m}
\bigwedge_{A \in \Gamma}
\bigwedge_{a \in \Sigma} 
\left(
\begin{array}{c}
\bm y \in \bm{X}_i \land 
\sux(\bm y, \bm{z}) \land \\
M(\bm{x}, \bm{z}) \land
\bm z \in \bm{R}_A \land 
a(\bm{z})\\ 
\Rightarrow \\
\bm z \in \bm{X}_j \land j = \delta_r(i, a, A)
\end{array}
\right).\]

\end{itemize}



      \begin{example}
        Consider the alphabet $\Sigma = (\Sigma_c = \{a\}$, $\Sigma_r = \{b\},$ $\Sigma_i = \emptyset)$ and the VPA depicted in Figure~\ref{fig:vpa-anbn}.
      
   \begin{figure} 
     \begin{center}
     \begin{tikzpicture}[every edge/.style={draw,solid}, node distance=3.6cm, auto, 
                         every state/.style={draw=black!100,scale=0.5}, >=stealth]
     
     \node[initial by arrow, initial text=,state] (q0) {{\huge $0$}};
     \node[state] (q1) [right of=q0] {{\huge $1$}};
     \node[state] (q2) [right of=q1] {{\huge $2$}};
     \node[state] (q3) [right of=q2, accepting] {{\huge $3$}};
     
     \path[->]
     (q0) edge [above] node {$a / B$} (q1)
     (q1) edge [loop above]  node {$a / A$} (q1)
     (q1) edge [above] node {$b , A$} (q2)
     (q1) edge [bend right, below] node {$b , B$} (q3)
     (q2) edge [loop above]  node {$b , A$} (q2)
     (q2) edge [above] node {$b , B$} (q3)
     ;
     \end{tikzpicture}
     \end{center}
     \caption{A VPA recognizing $\{a^n b^n \mid n > 0\}$.}\label{fig:vpa-anbn}
     \end{figure}
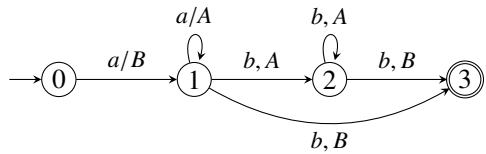
         
      The MSO formula 
 $\exists \bm X_0 
  , \bm X_1 
  , \bm X_2 
  , \bm X_3 
  , \bm C_A
  , \bm C_B
  , \bm R_A
  , \bm R_B
 (\varphi_{\mathcal A} \land \varphi_M)$ is built on the basis of such an
 automaton, where $\varphi_M$ is the conjunction of the formulas defined above, and
 $\varphi_{\mathcal A}$ is:
 \begin{small}
\[
   \begin{array}{l}
\exists \bm z (\not\exists \bm x (\sux(\bm x, \bm z) ) \land \bm z \in \bm X_{0})
\,\land \\
\exists \bm y( \not\exists \bm x (\sux(\bm y, \bm x) )  \land
 \bm y \in \bm X_3) 
\,\land \\
\forall \bm{x}, \bm{y}  
\left(
\bm x \in \bm{X}_0 \land 
\sux(\bm x, \bm{y}) \land 
a(\bm{y}) 
\Rightarrow 
   \bm y \in \bm C_B \land
\bm y \in \bm{X}_1 
\right)
\land \\
\forall \bm{x} , \bm{y}  
\left(
\bm x \in \bm{X}_1 \land 
\sux(\bm x, \bm{y}) \land 
a(\bm{y})
\Rightarrow 
   \bm y \in \bm C_A \land
\bm y \in \bm{X}_1 
\right) 
\land \\
 \forall \bm{x} , \bm{y} , \bm{z} 
\left(
\bm y \in \bm{X}_1 \land 
\sux(\bm y, \bm{z}) \land 
M(\bm{x}, \bm{z}) \land
\bm z \in \bm{R}_A \land 
b(\bm{z}) \Rightarrow 
\bm z \in \bm{X}_2
\right)
\land \\
 \forall \bm{x} , \bm{y} , \bm{z} 
\left(
\bm y \in \bm{X}_2 \land 
\sux(\bm y, \bm{z}) \land 
M(\bm{x}, \bm{z}) \land
\bm z \in \bm{R}_A \land 
b(\bm{z}) \Rightarrow 
\bm z \in \bm{X}_2
\right)
\land \\
 \forall \bm{x} , \bm{y} , \bm{z} 
\left(
\bm y \in \bm{X}_2 \land 
\sux(\bm y, \bm{z}) \land 
M(\bm{x}, \bm{z}) \land
\bm z \in \bm{R}_B \land 
b(\bm{z}) \Rightarrow 
\bm z \in \bm{X}_3
\right)
\land \\
 \forall \bm{x} , \bm{y} , \bm{z} 
\left(
\bm y \in \bm{X}_1 \land 
\sux(\bm y, \bm{z}) \land 
M(\bm{x}, \bm{z}) \land
\bm z \in \bm{R}_B \land 
b(\bm{z}) \Rightarrow 
\bm z \in \bm{X}_3
\right).
 \end{array}
\]
\end{small}
\end{example}

Other studies aimed at exploiting less complex but also less powerful logics, such as first-order and temporal ones to support a more practical automatic verification of VPL \cite{lmcs/AlurABEIL08} as it happened with great success with model-checking for RL; algorithmic model-checking, however, is not the main focus of this paper and we do not go deeper into this issue.

\subsection{Other structured context-free languages}

As we said, early work on parenthesis languages and tree-automata ignited many attempts to enlarge those classes of languages and to investigate their properties. Among them VPL have received much attention in the literature and in this paper, probably thanks to the completeness of the obtained results --closure properties and logic characterization. To give an idea of the vastness of this panorama and of the connected problems, and to help comparison among them, in this section we briefly mention a few more of such families with no attempt for exhaustiveness.

\subsubsection{Balanced grammars}\label{sec:BalGr}
\textit{Balanced grammars} (BG) have been proposed in \cite{Berstel:2001:BGT} as a first approach to model mark-up languages such as XML by exploiting suitable extensions of parenthesis grammars. Basically a (BG) has a partitioned alphabet exactly in the same way as VPL; on the other hand any production of a BG has the form $A \to a\alpha b$ where $a \in \Sigma_c$, $b \in \Sigma_r$, and $\alpha$ is a regular expression\footnote{A regular expression over a given alphabet $V$ is built on the basis of alphabet's elements by applying union, concatenation, and  Kleene~$^*$ symbols; it is well-known that the class of languages definable by means of regular expressions is RL.} over $V_N \cup \Sigma_i$.

Since it is well-known that the use of regular expressions in the rhs of CF grammars can be replaced by a suitable expansion by using exclusively ``normal'' rules, we can immediately conclude that balanced languages, i.e. those generated by BG, are a proper subclass of VPL (e.g. they do not admit unmatched elements of $\Sigma_c$ and $\Sigma_r$). Furthermore they are not closed under concatenation and  Kleene~$^*$ \cite{Berstel:2001:BGT}; we are not aware of any logic characterization of these languages.

\subsubsection{Height-deterministic languages}\label{sec:HDPDL}
Height-deterministic languages, introduced in \cite{conf/mfcs/NowotkaS07}, are a more recent and fairly general way of describing CFL in terms of their structure. In a nutshell the hidden structure of a sentence is made explicit by making  $\varepsilon$-moves visible, in that the original $\Sigma$ alphabet is enriched as $\Sigma \cup \{\varepsilon\}$; if the original input string in $\Sigma^*$ is transformed into one over $\Sigma \cup \{\varepsilon\}$ by inserting an explicit $\varepsilon$ wherever a recognizing automaton executes an $\varepsilon$-move, we obtain a linear representation of the syntax-tree of the original sentence, so that the automaton can be used as a real parser. We illustrate such an approach by means of the following example.
\begin{example}\label{ex: hda}
Consider the language $L = L_1 \cup L_2$, with $L_1 = \{a^nb^nc^* \mid n \geq 0\}$, $L_2 = \{a^*b^nc^n \mid n \geq 0\}$ which is well-know to be inherently ambiguous since a string such as  $ a^nb^nc^n$ can be parsed both according to $L_1$'s structure and according to $L_2$'s one. A possible nondeterministic PDA $ \mathcal A $ recognizing $L$ could act as follows:
\begin {itemize}
\item
$ \mathcal A $ pushes all $a$'s until it reaches the first $b$; at this point it makes a nondeterministic choice:
\begin {itemize}
\item
in one case it makes a ``pause'', i.e., an $\varepsilon$-move and enters a new state, say $q_1$;
 \item
 in the other case it directly enters a different state, say $q_2$ with no ``pause''; 
\end {itemize} 
 \item
 from now on its behavior is deterministic; precisely:
 \begin {itemize}
 \item
 in $q_1$ it checks that the number of $b$'s equals the number of $a$'s and then accepts any number of $c$'s;
 \item
 in $q_2$ it pushes the $b$'s to verify that their number equals that of the $c$'s.
 \end {itemize}
 \end {itemize}
Thus, the two different behaviors of $ \mathcal A $ when analyzing a string of
the type $ a^nb^nc^n$ would result into two different strings in the extended
alphabet $\Sigma \cup \{\varepsilon\}$\footnote{This could be done explicitly
  by means of a nondeterministic transducer that outputs a special marker in
  correspondence of an $\varepsilon$-move, but we stick to the original \cite
  {conf/mfcs/NowotkaS07} formalization where automata are used exclusively as
  acceptors without resorting to formalized transducers. }, namely  $
a^n\varepsilon b^nc^n$ and  $ a^nb^nc^n$; it is now simple, if needed, to state
a one-to-one correspondence between strings augmented with explicit
$\varepsilon$ and the different syntax-trees associated with the original input:
in this example, $ a^n\varepsilon b^nc^n$ corresponds to the structure of 
Figure~\ref{fig:treeHD}~(a) and  $ a^nb^nc^n$ to that of Figure~\ref{fig:treeHD}~(b). 
It is also easy to build other nondeterministic PDA recognizing $L$ that ``produce'' different strings associated with different structures.

\begin{figure}
\begin{center}
\begin{tabular}{ccc}
\begin{tikzpicture}[
 scale=0.6,
 level distance=1cm
  ]
\coordinate
  child{ 
     child{ 
     child{node{$a$}}
     child{ 
     child{node{$a$}}
     child{ 
        node{$\varepsilon$}
     }
     child{node{$b$}}
     }
     child{node{$b$}}
     }
   child{node{$c$}}
  }
  child{node{$c$}}
; 
\end{tikzpicture}
&
$\ \ \ \ \ \ $
&\begin{tikzpicture}[
 scale=0.6,
 level distance=1cm
  ]
\coordinate
  child{node{$a$}}
  child{ 
   child{node{$a$}}
   child{
     child{node{$b$}}
     child{ 
     child{node{$b$}}
     child{node{$c$}}
     }
     child{node{$c$}}
   }
  }
; 
\end{tikzpicture}
\\
(a) &  & (b) 
\end{tabular}
\end{center}
\caption{Different structures for 
$a^n \varepsilon b^n c^n$ (a)
and 
$a^n b^n c^n$ (b), for $n = 2$. 
}\label{fig:treeHD}
\end{figure}
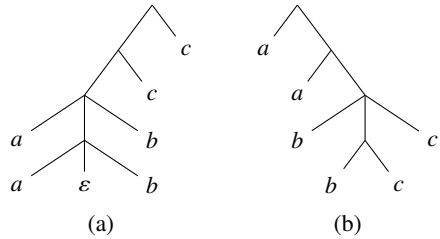

\end{example}


Once the input string structures are made visible by introducing the explicit  $\varepsilon$ character, the characteristics of PDA, of their subclasses, and of the language subfamilies they recognize, are investigated by referring to the \textit{heights of their stack}. Precisely:
\begin{itemize}
\item
Any PDA is put into a \emph{normalized form}, where
\begin{itemize}
\item
the $\delta$ relation is \emph{complete}, i.e., it is defined in such a way that for every input string, whether accepted or not, the automaton scans the whole string:\\ 
$\forall x \exists c(c_0 = (x, q_0, Z_0) \transition{*} c = (\varepsilon, q, \gamma))$;
\item
every element of $\delta$ is exclusively in one of the forms: $(q, A, o, q', AB)$, or $(q, A, o,$ $q', A)$, or $(q, A, o, q', \varepsilon)$, where $o \in (\Sigma \cup \{\varepsilon\})$ ;
\item
for every $q \in Q$ all elements of $\delta$ having $q$ as the first component either have $o\in \Sigma$ or $o = \varepsilon$, but not both of them.
\end{itemize}
\item
For any normalized $\mathcal A$ and word $w \in (\Sigma \cup \{\varepsilon\})^*$, $\mathcal N( \mathcal A,w)$ denotes the set of all stack heights reached by $\mathcal A$ after scanning $w$.
\item
 $\mathcal A$ is said \emph{height-deterministic} (HPDA) iff $\forall w \in (\Sigma \cup \{\varepsilon\})^*$, $\:|\mathcal N( \mathcal A,w)|\: = 1$.
 \item
 The family of height-deterministic PDA is named HPDA; similarly, HDPDA denotes the family of deterministic height-deterministic PDA, and HRDPDA that of deterministic, real-time (i.e., those that do not perform any $\varepsilon$-move) height-deterministic PDA. The same acronyms with a final $L$ replacing the $A$ denote the corresponding families of languages. 
\end{itemize}
It is immediate to realize (Lemma 1 of \cite {conf/mfcs/NowotkaS07}) that every PDA admits an equivalent normalized one. Example \ref{ex: hda} provides an intuitive explanation that every PDA admits an equivalent HPDA (Theorem 1 of \cite {conf/mfcs/NowotkaS07}); thus HPDL = CFL; also, any (normalized) DPDA is, by definition an HPDA and therefore a deterministic HPDA; thus HDPDL = DCFL. Finally, since every deterministic VPA is already in normalized form and is a real-time machine, VPL $\subset$ HRDPDL: the strict inclusion follows from the fact that $L = \{a^nba^n\}$ cannot be recognized by a VPA since the same character $a$ should belong both to $\Sigma_c$ and to $\Sigma_r$.

Coupling the extension of the alphabet from $\Sigma$ to $\Sigma \cup \{\varepsilon \}$ with the set  $\mathcal N( \mathcal A,w)$ allows us to consider HPDL as a generalized kind of structured languages. As an intuitive explanation, let us go back to Example \ref{ex: hda} and consider the two behaviors of $\mathcal A$ when parsing the string $aabbcc$ once it has been ``split'' into $aabbcc$ and $aa\varepsilon bbcc$; the stack heights $\mathcal N( \mathcal A,w)$ for all their prefixes are, respectively:
$(1,2,3,4,3,2)$ and $(1,2,2,1,0,0,0)$ (if we do not count the bottom of the stack $Z_0$). In general, it is not difficult to associate every sequence of stack lengths during the parsing of an input string (in $(\Sigma \cup \{\varepsilon\})^*$!) with the syntax-tree visited by the HPDA.

As a consequence, the following fundamental definition of \emph{synchronization} between HPDA can be interpreted as a form of structural equivalence.

\begin{definition}\label{def: synchpda}
Two HPDA $\mathcal A$ and $\mathcal B$ are synchronized, denoted as $\mathcal A \sim  \mathcal B$,  iff $\forall w \in (\Sigma \cup \{\varepsilon\})^*$, $\mathcal N( \mathcal A,w)$ = $\mathcal N( \mathcal B,w)$. 
\end{definition}

It is immediate to realize that  synchronization  is an equivalence relation and therefore to associate an equivalence class $[\mathcal A]_\sim$ with every HPDA; we also denote as  $\mathcal A$-HDPL the class of languages recognized by automata in $[\mathcal A]_\sim$. Then, in \cite {conf/mfcs/NowotkaS07} the authors show that:
\begin{itemize}
\item
For every \emph{deterministic} HPDA $\mathcal A$ the class $\mathcal A$-HDPL is a Boolean algebra.\footnote{If the automaton is not deterministic only closures under union and intersection hold.}
\item
Real-time HPDA can be determinized (with a complexity of the same order as for VPA), so that the class of real-time HPDL coincides with HRDPDL.
\end{itemize}

On the other hand neither HRDPDL nor HDPDL
are closed under concatenation and  Kleene~$^*$ \cite{Crespi-ReghizziM12} so that the gain obtained in terms of generative power w.r.t. VPL has a price in terms of closure properties. 
We are not aware of logic characterizations for this class of structured languages.

Let us also mention that other classes of structured languages based on some notion of synchronization have been studied in the literature; in particular \cite {conf/mfcs/NowotkaS07} compare their class with those of \cite{FisPnu01} and \cite{conf/dlt/Caucal06}. Finally, we acknowledge that our survey does not consider some logic characterization of tree or even graph languages which refer either to very specific families (such as, e.g. star-free tree languages \cite{ThomasRegularTrees}) and/or to an alphabet of basic elements, e.g., arcs connecting tree or graph nodes \cite{Caucal03}, which departs from the general framework of string (structured) languages.

 \section{Parsing context-free languages deterministically}\label{sec:DCFPars}
 We have seen in Section \ref{parsing} that general  CFL, having a hidden and sometimes ambiguous structure, need not only recognizing mechanisms but more complex parsing algorithms that, besides deciding whether a string belongs to the language or not, also produce the syntax-tree(s) formalizing its structure and possibly driving its semantics; parsing is the core of any compiler or interpreter. We noticed that not all CFL can be recognized --and therefore parsed-- by DPDA: whereas any CFL can be recognized by a nondeterministic PDA that operates in real-time, the best deterministic algorithms to parse general CFL, such as CKY and Early's ones have a $O(n^3)$ complexity, which is considered not acceptable in many fields such as programming language compilation. 
 
 For this and other reasons many application fields restrict their attention to DCFL; DPDA can be easily built, with no loss of generality, in such a way that they can operate in linear time, whether as pure recognizers or as parsers and language translators: it is sufficient, for any DPDA, to effectively transform it into an equivalent \emph{loop-free} one, i.e. an automaton that does not perform more than a constant number, say $k$, of  $\varepsilon$-moves before reading a character from the input or popping some element from the stack (see, e.g., \cite{Harrison78}). In such a way the whole input $x$ is analyzed in at most $h\cdot|x|$ moves, where $h$ is a function of $k$ and the maximum length of the string that the automaton can push on the stack in a single move. Notice, however, that in general it is not possible to obtain a DPDA that recognizes its language in real-time. Consider, for instance, the language $L = \{a^mb^nc^nd^m \mid m, n \geq 1\} \cup \{a^mb^+ed^m \mid m \geq 1\}$: intuitively, a DPDA recognizing $L$ must first push the $a$s onto the stack; then, it must also store on the stack the subsequent $b$s since it must be ready to compare their number with the following $c$s, if any; after reading the $b$s, however, if it reads the $e$ it must necessarily pop all $b$s by means of $n$ $\varepsilon$-moves before starting the comparison between the $a$s and the $d$s.
 
 Given that, in general, it is undecidable to state whether a CFL is deterministic or not \cite{DBLP:journals/jcss/HartmanisH70}, the problem of automatically building a deterministic automaton, \emph{if any}, from a given CFG is not trivial and deserved much research. In this section we will briefly recall two major approaches to the problem of deterministic parsing. We do not go deep into their technicalities, however, because the families of languages they can analyze are not of much interest from the point of view of algebraic and logical characterizations. It is Section~\ref{sec: OPG}, instead, where we introduce a class of languages that allows for highly efficient deterministic parsing \emph{and} exhibits practically all desirable algebraic and logical properties.

 \subsection{Top-down deterministic parsing}\label{TDDetPars}

 In Section~\ref{parsing} we introduced the GNF for CFG and observed that, if in such grammars there are no two distinct productions of the type $A \to a\alpha$, $A \to a\beta$, then the automaton built therefrom is deterministic.
 The above particular case has been generalized leading to the definition of \emph{LL grammars}, so called because they allow to build deterministic parsers that scan the input Left-to-right and produce the Leftmost derivation thereof. Intuitively, an $LL(k)$ grammar is such that, for any leftmost derivation 
 \[ S\#^k \derives{*} xA\alpha \#^k \] it is possible to decide deterministically which production to apply to rewrite nonterminal $A$ by ``looking ahead'' at most $k$ terminal characters of the input string that follows $x$.\footnote{ The ``tail'' of \emph{k \#} characters is a simple trick to allow for the look ahead when the reading head of the automaton is close to the end of the input.}
  Normally, the practical application of LL parsing is restricted to the case $k=1$ to avoid memorizing and searching too large tables. In general this choice allows to cover a large portion of programming language syntax even if it is well-known that not all DCFL can be generated by LL grammars. For instance no LL grammar can generate the deterministic language $\{a^nb^n \mid n \geq 1\} \cup \{a^nc^n \mid n \geq 1\}$ since the decision on whether to join the $a$'s with the $b$'s or with the $c$'s clearly requires an unbounded look-ahead.
 
 \begin{example}
 The following grammar is a simple transformation of $GAE_1$ in LL form; notice that the original left recursion $ E \derives{} E + T$ has been replaced by a right one $ E \derives{*} T + E$  and similarly for nonterminal $T$. 
    \[
     	\begin{array}{ll}
     	GAE_{LL}: &
     		S \to E\# \\
     &		E \to  TE'\\
     &		E' \to + E \mid \varepsilon \\  	
     &		T \to  FT'  \\
     &		T' \to * T \mid \varepsilon\\
     &		F \to  e.  \\
     	\end{array}
     \]
 \end{example}
 
 \subsection{Bottom-up deterministic parsing}\label{LRPars}
 So far the PDA that we used to recognize or parse CFL are working in a top-down manner by trying to build leftmost derivation(s) that produce the input string starting from grammar's axiom. We also mentioned, however, that trees can be traversed also in a bottom-up way; a typical way of doing so is visiting them in \emph{leftmost post-order}, i.e. scanning their frontier left-to right and, as soon as a string of children is identified, writing them followed by their father, then proceeding recursively until the root is reached. For instance such a visit of the syntax-tree by which $GAE_1$ generates the string $e+e*e$ is $eE+eT*eFTES$ which is the reverse of the rightmost derivation 
 $S \derives{} E \derives{} E+T \derives{} E+ T*F \derives{} E+ T*e \derives{} E+ e*e \derives{} e+ e*e $.
 
 Although Definition~\ref{def:PDA} introduces PDA in such a way that they are naturally oriented towards building leftmost derivations, it is not difficult to let them work in a bottom-up fashion: the automaton starts reading the input left-to-right and pushes the read character on top of the stack (formally, since at every transition the character on top of the stack is \textit{replaced} by a string, it rewrites the existing symbol and adds the read character over it); as soon as --by means of its finite memory control device-- it realizes that a whole rhs is on top of the stack it replaces it by the corresponding lhs (again, to be fully consistent with Definition~\ref{def:PDA}, this operation should not be formalized as a single transition but could be designed as a sort of ``macro-move'' consisting of several micro-steps that remove one character per step and whose  last one also pushes the lhs character).\footnote{A formal definition of PDA operating in the above way is given, e.g., in~\cite{conf/mfcs/NowotkaS07} and reported in Section~\ref{sec:HDPDL}}
 This type of operating by a PDA is called \emph{shift-reduce parsing} since it consists in \textit{shifting} characters from the input to the stack and \textit{reducing} them from a rhs to the corresponding lhs.
  
 Not surprisingly, the ``normal'' behavior of such a bottom-up parser is, once again, nondeterministic: in our example once the rhs $e$ is identified, the automaton must apply a reduction either to $F$ or to $T$ or to $E$. Even more critical is the choice that the automaton must take after having read the substring $e+e$ and having (if it did the correct reductions) $E+T$ on top of the stack: in this case the string could be the rhs of the rule $E \to E + T$, and in that case the automaton should reduce it to $E$ but the $T$ on the top could also be the beginning of another rhs, i.e., $T*F$, and in such a case the automaton should go further by shifting more characters before doing any reduction; this is just the opposite situation of what happens with VPL, where the current input symbol allows the machine to decide how to progress.
 
 In fact, the previous type of nondeterminism, i.e., the choice whether to reduce $e$ to $E$ or $T$ or $F$, could be resolved as we did with structured languages, by eliminating repeated rhs through a classical power set construction. In this case, instead, the automaton must build by itself the unknown structure and this may imply a lack of knowledge about having reached or not a complete rhs.
  ``Determinizing'' bottom-up parsers has been an intense research activity during the 1960's, as well as for their top-down counterpart. In Section~\ref{sec: OPG} we thoroughly examine one of the earliest practical deterministic bottom-up parsers and the class of languages they can recognize, namely Floyd's \emph{operator precedence languages}. The study of these languages, however, has been abandoned after a while due to advent of a more powerful class of grammars --the LR ones, defined and investigated by Knuth \cite{KnuthLR65}, whose parsers proceed Left-to-right as well as the LL ones but produce (the reverse of)  Rightmost derivations. LR grammars in fact, unlike LL and operator precedence ones, generate all DCFL.


\section{Operator precedence languages}\label{sec: OPG}
Operator precedence grammars (OPG) have been introduced by R. Floyd in his pioneering paper \cite{Floyd1963} with the goal of building efficient, deterministic, bottom-up parsers for programming languages. In doing so he was inspired by the hidden structure of arithmetic expressions which suggests to ``give precedence'' to, i.e, to execute first multiplicative operations w.r.t. the additive ones, as we illustrated through Example \ref{GAE}. Essentially, the goal of deterministic bottom-up parsing is to unambiguously decide when a complete rhs has been  identified so that we can proceed with replacing it with the unique corresponding lhs with no risk to apply some roll-back if another possible reduction was the right one. Floyd achieved such a goal by suitably extending the notion of precedence between arithmetic operators to all grammar terminals, in such a way that a complete rhs is enclosed within a pair \emph{(yields-precedence, takes-precedence)}. OPG obtained a considerable success thanks to their simplicity and to the efficiency of the parsers obtained therefrom; incidentally, some independent studies also uncovered interesting algebraic properties (\cite{Crespi-ReghizziMM1978}) which have been exploited in the field of grammar inference (\cite{Crespi-ReghizziCACM73}). As we anticipated in the introduction, however, the study of these grammars has been dismissed essentially because of the advent of other classes, such as the LR ones, which can generate all DCFL; OPG instead do not have such power as we will see soon, although they are able to cover most syntactic features of normal programming languages, and parsers based on OPG are still in practical use (see, e.g., \cite{GruneJacobs:08}).

Only recently we renewed our interest in this class of grammars and languages for two different reasons that are the object of this survey. On the one side in fact, OPL, despite being apparently not structured, since they require and have been motivated by parsing, have shown rather surprising relations with various families of structured languages; as a consequence it has been possible to extend to them all the language properties investigated in the previous sections. On the other side, OPG enable parallelizing their parsing in a natural and efficient way, unlike what happens with other parsers which are compelled to operate in a strict left-to-right fashion, thus obtaining considerable speed-up thanks to the wide availability of modern parallel HW architectures.

Therefore, after having resumed the basic definitions and properties of OPG and their languages, we show, in Section \ref{sec: OPAlg-Log}, that they considerably increase the generative power of structured languages but, unlike the whole class of DCFL, they still enjoy all algebraic and logic properties that we investigated for such smaller classes. In Section \ref{sec: OPParPars} we show how parallel parsing is much better supported by this class of grammars than by the presently used ones.

\begin{definition}
A grammar rule is in \textit{operator form} if its rhs has no adjacent nonterminals; an \textit{operator grammar} (OG) contains only such rules. 
\end{definition}
Notice that the grammars considered in Example \ref{GAE} are OG. Furthermore any CF grammar can be recursively transformed into an equivalent OG one~\cite{Harrison78}.

Next, we introduce the notion of precedence relations between elements of $\Sigma$: we say that $a$ is \emph{equal in precedence} to $b$ iff the two characters occur consecutively, or at most with one nonterminal in between, in some rhs of the grammar; in fact, when evaluating the relations between terminal characters for OPG, nonterminals are inessential, or ``transparent''. $a$ \emph{yields precedence} to $b$ iff $a$ can occur at the immediate left of a subtree whose leftmost \textit{terminal} character is $b$ (again whether there is a nonterminal character at the left of $b$ or not is inessential). Symmetrically, $a$ \emph{takes precedence} over $b$ iff $a$ occurs as the rightmost \textit{terminal} character of a subtree and $b$ is its following terminal character. These concepts are formally defined as follows.

\begin{definition}
For an OG $G$ and a nonterminal $A$, the \textit{left and right terminal sets} are
\begin{eqnarray*}
  \mathcal{L}_G(A) & = & \{a\in\Sigma \mid A \stackrel \ast \Rightarrow B a\alpha\} \\
  \mathcal{R}_G(A) & = & \{a\in\Sigma \mid A \stackrel \ast \Rightarrow \alpha a B\}
\text{ where } B\in V_N\cup\{\varepsilon\}.
\end{eqnarray*}

The grammar name $G$ will be omitted unless necessary to prevent confusion. 

For an OG $G$, let $\alpha, \beta$ range over  $(V_N \cup \Sigma)^{\ast}$ and $a,b\in \Sigma$. The three binary operator precedence (OP) relations are defined as follows: \label{PrecRelat}
\begin{itemize}
 \item equal in precedence:   
 $ a\doteq b  \iff $ \\
  $\exists A\to\alpha aBb\beta, B\in V_N\cup\{\varepsilon\}$,
 \item takes precedence: $a\gtrdot b  \iff$ \\ 
  $\exists A\to\alpha Db\beta, D\in V_N \text{ and } a\in \mathcal{R}_G(D)$,
 \item yields precedence: $a\lessdot b  \iff $\\
  $\exists A\to\alpha aD\beta, D\in V_N \text{ and } b\in \mathcal{L}_G(D)$.
\end{itemize}
For an OG $G$, the \textit{operator precedence matrix} (OPM) $M=OPM(G)$ is a $|\Sigma| \times |\Sigma|$ array that, for each ordered pair $(a,b)$, stores the set $M_{ab}$ of OP relations holding between $a$ and $b$. 
\end{definition}

For the grammar $GAE_1$ of Example  ~\ref{GAE} the left and right terminal sets
of nonterminals $E$, $T$ and $F$ are, respectively:

$\mathcal{L}(E)  = \{+, *,e\}$,
$\mathcal{L}(T)  = \{*,e\}$, 
$\mathcal{L}(F)  = \{e\}$,
$\mathcal{R}(E) =  \{+, *,e\}$,
$\mathcal{R}(T) =  \{*,e\}$, and 
$\mathcal{R}(F) =  \{e\}$. 

\begin{figure}
\begin{center}
$
\begin{array}{c|cccc}
    &+ & *  & e  \\
\hline
+ & \gtrdot & \lessdot & \lessdot \\
* & \gtrdot & \gtrdot  & \lessdot \\
e & \gtrdot & \gtrdot  &          \\
\end{array}
$
\end{center}
\caption{The OPM of the $GAE_1$ of Example~\ref{GAE}.}
\label{fig:OPM1}
\end{figure}

Figure~\ref{fig:OPM1} displays the OPM associated with the grammar of $GAE_1$ of Example~\ref{GAE} where, for an ordered pair $(a,b)$, $a$ is one of the symbols shown in the first column of the matrix and $b$ one of those occurring in its first row.
Notice that, unlike the usual arithmetic relations denoted by similar symbols, the above precedence relations do not enjoy any of the transitive, symmetric, reflexive properties.

\begin{definition}\label{def:OPG}
An OG $G$ is an \emph{operator precedence} or \emph{Floyd grammar} (\emph{OPG}) iff $M=OPM(G)$ is a \textit{conflict-free} matrix, i.e., $\forall a,b$, $|M_{ab}|\leq 1$. 
An \emph{operator precedence language (OPL)} is a language generated by an OPG.

An OPG  is in \emph{Fischer normal form} (FNF) iff it is \textit{invertible}, i.e., no two rules have the same rhs,  has no \textit{empty rules}, i.e., rules whose rhs is $\varepsilon$, except possibly $S\to \varepsilon$, and no \textit{renaming rules}, i.e, rules whose rhs is a single nonterminal, except possibly those whose lhs is $S$.
\end{definition}
For every OPG an equivalent one in FNF can effectively be built \cite{Fischer69,Harrison78}.
A FNF grammar (manually) derived from   $GAE_1$ of  Example~\ref{GAE} is $GAE_{FNF}$: 
\[
\begin{array}{l}
S  \to  E \mid T \mid F\\
 E \to  E + T  \ \mid \  E + F \ \mid \  T + T \mid  F + F \ \mid \ F + T  \ \mid \ T + F\\
 T \to  T * F \ \mid F * F \\
 F  \to  e  
\end{array} 
\]
We can now see how the precedence relations of an OPG can drive the deterministic parsing of its sentences: consider again the sentence $e+e*e$; add to its boundaries 
the conventional symbol $\#$ which implicitly yields precedence to any terminal character and to which every terminal character takes precedence, and evaluate the precedence relations between pairs of consecutive symbols; they are displayed below:
\[ \#\lessdot e \gtrdot +\lessdot e \gtrdot* \lessdot e \gtrdot \#. \]
The three occurrences of $e$ enclosed within the pair $(\lessdot , \gtrdot)$, are the rhs of production $F \to e$; thanks to the fact that the grammar is in FNF there is no doubt on its corresponding lhs; therefore they can be reduced to $F$. Notice that such a reduction could be applied \emph{in any order, possibly even in parallel}; this feature will be exploited later in Section \ref{sec: OPParPars} but for the time being let us consider a traditional bottom-up parser proceeding rigorously left-to-right. Thus, the first reduction produces the string $F+e*e$; if we now recompute the precedence relations on the new string, due to the ``transparency'' of nonterminals, we obtain
$ \#\lessdot F + \lessdot\ e \gtrdot * \lessdot e \gtrdot \#. $

At this point the next rhs to be reduced is the second occurrence of $e$; after a third similar reduction the original string and the corresponding precedence relations, are reduced to
$\#\lessdot F + \lessdot\ F * F \gtrdot \# $
where the (only) next rhs to be reduced in $F*F$. 

Notice that there is no doubt on whether the first nonterminal $F$ should be part of a rhs \emph{beginning} with $F$ or of another one \emph{ending} with $F$, such as
$ \#\lessdot F + F \lessdot  *\ F \gtrdot \# $.
In fact, if the rhs to be reduced were just $*F$, its corresponding lhs would be a nonterminal adjacent to the $F$ at its left, thus contradicting the hypothesis of the grammar being an OG. At this point the bottom-up shift-reduce algorithm continues deterministically until the axiom is reached and a syntax-tree of the original sentence --represented by the mirror image of the rightmost derivation-- is built.

This first introduction to OPG allows us to draw some early important properties thereof:
\begin{itemize}
\item
In some sense OPL are \emph{input-driven} even if they do not match exactly the definition of these languages: in fact, the decision of whether to apply a push operation (at the beginning of a rhs) or a shift one (while scanning a rhs) or a pop one (at the end of a rhs) depends only on terminal characters but not on a \emph{single} one, as  a look-ahead of one more terminal character is  needed.\footnote{As it happens in other deterministic parsers such as LL or LR ones (see Section~\ref{sec:DCFPars}).}
\item
The above characteristic is also a major reason why OPL, though allowing for efficient deterministic parsing of various practical programming languages \cite{GruneJacobs:08,PrologOP,Floyd1963}, do not cover the whole family DCFL; consider in fact the language $L = \{0a^nb^n \mid n \geq 0\} \cup \{1a^nb^{2n} \mid n \geq 0\}$: a DPDA can easily ``remember'' the first character in its state; then push all the $a$'s onto the stack and, when it reaches the $b$s decide whether to scan one or two $b$s for every $a$ depending on its memory of the first read character. On the contrary, it is clear that any grammar generating $L$ would necessarily exhibit some precedence conflict.\footnote{The above $L$ is instead LL (see Section~\ref{TDDetPars}); on the contrary, the language $\{a^nb^n \mid n \geq 1\} \cup \{a^nc^n \mid n \geq 1\}$ is OPL but not LL; thus, OPL and LL languages are uncomparable.}
\item
We like to consider OPL as \emph{structured languages} in a generalized sense since, once the OPM is given, the structure of their sentences is immediately defined and univocally
determinable as it happens. e.g., with VPL once the alphabet is partitioned
into call, return, and internal alphabet. However, we would be reluctant to
label OPL as \emph{visible} since there is a major difference between
paren\-thesis-like terminals which make the language sentence isomorphic to its
syn\-tax-tree, and precedence relations which help \textit{building} the tree but
are computed only during the parsing. Indeed, not all of them are immediately visible
in the original sentence: e.g., in some cases such as in the above
sentence $ \#\lessdot F + \lessdot\ F * F \gtrdot \#$ precedence relations are not even matched so that they can be assimilated to real parentheses only when they mark a complete rhs. In summary, we would  consider that OPL are structured (by the OPM) as well as PL (by explicit parentheses), VPL (by alphabet partitioning), and other families of languages; 
however, we would intuitively label them at most as ``semi-visible'' since making their structure visible requires some partial parsing, though not necessarily a complete recognition.
\end{itemize}

\subsection{Algebraic and logic properties of operator precedence languages}\label{sec: OPAlg-Log}
OPL enjoy all algebraic and logic properties that have been illustrated in the previous sections for much less powerful families of structured languages.

As a first step we introduce the notion of a \emph{chain} as a formal description of the intuitive concept of ``semi-visible structure''. To illustrate the following definitions and properties we will continue to make use of examples inspired by arithmetic expressions but we will enrich such expressions with, possibly nested, explicit parentheses as the visible part of their structure. For instance the following grammar $GAE_P$ is a natural enrichment of previous $GAE_1$ to generate arithmetic expressions that involve parenthesized subexpressions (we use the slightly modified symbols '$\lp$' and '$\rp$' to avoid overloading with other uses of parentheses).
\[
 	\begin{array}{ll}
 	GAE_P:
    & S \to E \mid T \mid F\\
 	&	E \to  E + T \mid T * F \mid e  \mid \ \lp E \rp  \\  	
 	&	T \to  T * F \mid e    \mid \ \lp E \rp   \\
 	&	F \to  e   \mid \ \lp E \rp \\
 	\end{array}
 \]

\begin{definition}[Operator precedence alphabet]
An \emph{operator precedence (OP) alphabet} is a pair $(\Sigma, M)$ where $\Sigma$ is an alphabet and 
$M$ is a conflict-free \textit{operator precedence matrix},
i.e.\ a $|\Sigma \cup \{ \# \}|^2$ array that associates at most 
one of the operator precedence relations: $\doteq$, $\lessdot$ or $\gtrdot$ with each ordered pair $(a,b)$.
As stated above the delimiter $\#$   yields precedence to other terminals and other terminals take precedence over it (with the special case $\# \doteq \#$ for the final reduction of renaming rules.) Since such relations are stated once and forever, we do not explicitly display them in OPM figures.
\end{definition}

 If $M_{ab} = \{\circ\}$, with $\circ \in \{\lessdot, \doteq, \gtrdot\}$ ,we write $a \circ b$. For $u,v \in\Sigma^*$ we write $u \circ v$ if $u = xa$ and $v = by$ with $a \circ b$.

\begin{definition}[Chains]\label{def: chains}
 Let $(\Sigma, M)$ be a precedence alphabet.
\begin{itemize}
\item A \emph{simple chain} is a word $a_0 a_1 a_2 \dots$ $a_n a_{n+1}$,
written as
$
\chain {a_0} {a_1 a_2 \dots a_n} {a_{n+1}},
$
such that:
$a_0, a_{n+1} \in \Sigma \cup \{\#\}$, 
$a_i \in \Sigma$ for every $i: 1\leq i \leq n$, 
$M_{a_0 a_{n+1}} \neq \emptyset$,
and $a_0 \lessdot a_1 \doteq a_2 \dots a_{n-1} \doteq a_n \gtrdot a_{n+1}$.

\item A \emph{composed chain} is a word 
$a_0 x_0 a_1 x_1 a_2  \dots a_n x_n a_{n+1}$, with $x_i \in \Sigma^*$, 
where $^{a_0}[a_1 a_2 \dots$ $a_n]^{a_{n+1}}$ is a simple chain, and
either $x_i = \varepsilon$ or $\chain {a_i} {x_i} {a_{i+1}}$ is a chain (simple or composed),
for every $i: 0\leq i \leq n$. 
Such a composed chain will be written as
$\chain {a_0} {x_0 a_1 x_1 a_2 \dots a_n x_n} {a_{n+1}}$.
\item The \textit{body} of a chain $\chain axb$, simple or composed, is the word $x$.
\end{itemize}
\end{definition}

\begin{figure*}
\begin{center}
\begin{tabular}{m{0.3\textwidth}m{0.01\textwidth}m{0.30\textwidth}}
$
\begin{array}{c|ccccccc}
    & +        & *        & \lp      & \rp     & e        \\
\hline
+   & \gtrdot  & \lessdot & \lessdot & \gtrdot & \lessdot \\
*   & \gtrdot  & \gtrdot  & \lessdot & \gtrdot & \lessdot \\
\lp & \lessdot & \lessdot & \lessdot & \doteq  & \lessdot \\
\rp & \gtrdot  & \gtrdot  &          & \gtrdot &          \\
e   & \gtrdot  & \gtrdot  &          & \gtrdot &          \\
\end{array}
$ 
&
&
\begin{tikzpicture}[scale=0.5,
level 1/.style={sibling distance=30mm,level distance=9mm,every child/.style={edge from parent/.style={draw=none}} },
level 2/.style={sibling distance=16mm,},
level 3/.style={level distance=16mm,every child/.style={edge from parent/.style={solid,draw=black}}}
]
\node[draw=none] (root) {}
	child {node (ml) {$\#$}}
	child {node {}
		child {node (x0) {$x_0$}
			child {node {$e$}
		  		edge from parent [path]
		  	}
		}
		child {node (plus1) {$+$}}
		child {node (x1) {$x_1$}
			child {node (y0) {$y_0$}  edge from parent [path]
				child {node {$e$}
			  		edge from parent [path]
			  	}
			}
			child {node (times) {$*$}
				edge from parent [draw=none]
			}
			child {node (y1) {$y_1$}	edge from parent [path]
				child {node (lp) {$\lp$}
					edge from parent [path]
				}
				child {node (w0) {$w_0$}	edge from parent [draw=none]
					child {node (z0) {$z_0$}	edge from parent [path]
						child {node {$e$}
					  		edge from parent [path]
					  	}
					}
					child {node (plus2) {$+$}
						edge from parent [draw=none]
					}
					child {node (z1) {$z_1$}	edge from parent [path]
						child {node {$e$}
					  		edge from parent [path]
					  	}
					}
				}
				child {node (rp) {$\rp$}
					edge from parent [path]
				}
			}
		}
	}
	child {node (mr) {$\#$}};
\draw[siblings] (ml) -- (x0);
\draw[siblings] (x0) -- (plus1);
\draw[siblings] (plus1) -- (x1);
\draw[siblings] (x1) -- (mr);
\draw[siblings] (y0) -- (times);
\draw[siblings] (times) -- (y1);
\draw[siblings] (lp) -- (w0);
\draw[siblings] (w0) -- (rp);
\draw[siblings] (z0) -- (plus2);
\draw[siblings] (plus2) -- (z1);
\end{tikzpicture}
\\
\qquad\quad(a) & &\qquad \qquad(b)
\end{tabular}
\end{center}
\caption{OPM of grammar $GAE_P$ (a) and structure of the chains in the
  expression $\# e + e * \lp e + e \rp \#$ (b).}\label{fig:chain}
\end{figure*}
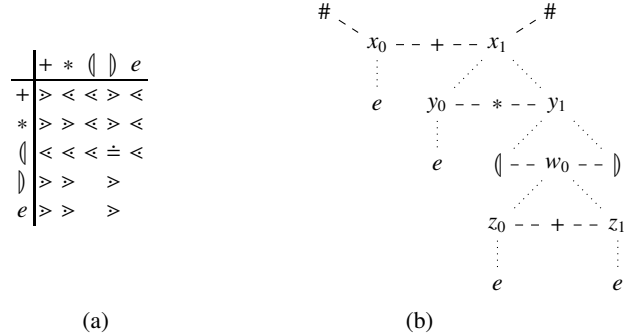

\begin{example}\label{ex:expr-chains}
Figure~\ref{fig:chain}~(a) depicts the $OPM(GAE_P$), whereas Figure~\ref{fig:chain}~(b) represents the ``semi-visible'' structure induced by the operator precedence alphabet of grammar $GAE_P$ for the expression $\# e + e * \lp e + e \rp \#$:   $\chain {\#} {x_0 + x_1 } {\#}$, $\chain {+} {y_0 * y_1 } {\#}$, $\chain {*} {\lp w_0 \rp} {\#}$, $\chain {\lp} {z_0 + z_1 } {\rp}$ are composed chains and $\chain {\#} {e} {+}$, $\chain {+} {e} {*}$, $\chain {\lp} {e} {+}$, $\chain {+} {e} {\rp}$ are simple chains.
\end{example}

\begin{definition}[Compatible word]
A word $w$ over $(\Sigma, M)$ is \emph{compatible} with $M$ iff the two following conditions hold:
\begin{itemize}
\item For each pair of letters $c, d$, consecutive in $w$, $M_{cd} \neq \emptyset$;
\item for each substring $x$ of  $\# w \#$ such that $x = a_0 x_0 a_1 x_1 a_2\dots$ $ a_n x_n a_{n+1}$,
if 
$a_0 \lessdot a_1 \doteq a_2 \dots a_{n-1} \doteq a_n \gtrdot a_{n+1}$ and,
for every $ 0\leq i \leq n$, either $x_i = \varepsilon$ or
 $\chain {a_i} {x_i} {a_{i+1}}$ is a chain (simple or composed), then
 $M_{a_0 a_{n+1}} \neq \emptyset$.
 \end{itemize}
\end{definition}

 For instance, the word  $e + e * \lp e + e \rp$ is compatible with the operator precedence alphabet of grammar $GAE_P$, whereas  $e + e * \lp e + e \rp \lp e + e \rp$ is not.

Thus, given  an OP alphabet, the set of possible chains over $\Sigma^*$ represents the universe of possible structured strings compatible with the given OPM.

\subsubsection{Operator precedence automata}\label{sec: OPAs} Despite abstract machines being the classical way to formalize recognition and parsing algorithms for any family of formal languages, and despite OPL having been invented just with the purpose of supporting deterministic parsing, their theoretical investigation  has been abandoned before a family of automata completely equivalent to their generative grammars appeared in the literature. Only recently, when the numerous still unexplored benefits obtainable from this family appeared clear to us, we filled up this hole with the herewith resumed definition (\cite{LonatiEtAl2015}).  The formal model presented in this paper is a ``traditional'' left-to-right automaton, although, as we already anticipated and will thoroughly exploit in the next Section \ref{sec: OPParPars} a distinguishing feature of OPL is that their parsing can be started in arbitrary positions with no harm nor loss of efficiency. This choice is explained by the need to extend and to compare results already reported for other language families. 
The original slightly more complicated version of this model was introduced in \cite{LonatiMandrioliPradella2011a}.

\begin{definition}[Operator precedence automaton]\label{def:OPA}
An  \emph{operator precedence automaton (OPA)} is a tuple
$\mathcal A = (\Sigma, M, Q, I, F, \delta) $ where:
\begin{itemize}
\item $(\Sigma, M)$ is an operator precedence alphabet,
\item $Q$ is a set of states (disjoint from $\Sigma$),
\item $I \subseteq Q$ is the set of initial states,
\item $F \subseteq Q$ is the set of final states,
\item $\delta \subseteq Q \times ( \Sigma \cup Q) \times Q$ is the transition relation, which is the union of three disjoint relations:
\[
\delta_{\text{shift}}\subseteq Q \times \Sigma \times Q,
\quad 
\delta_{\text{push}}\subseteq Q \times \Sigma \times Q,
\quad 
\delta_{\text{pop}}\subseteq Q \times Q \times Q.
\]
\end{itemize}
An OPA is deterministic iff
\begin{itemize}
\item
$I$ is a singleton
\item
All three components of $\delta$ are functions:
\[
\delta_{\text{shift}}: Q \times \Sigma \to Q,
\  
\delta_{\text{push}}: Q \times \Sigma \to Q,
\
\delta_{\text{pop}}: Q \times Q \to Q.
\]
\end{itemize}

\end{definition}

We represent an  OPA by a graph with $Q$ as the set of vertices and
$\Sigma \cup Q$ as the set of edge labelings. 
The edges of the graph are denoted by different shapes of arrows to distinguish the three types of transitions:
there is an edge from state $q$ to state $p$ labeled by $a \in \Sigma$ denoted by a dashed (respectively, normal) arrow iff  $(q,a,p) \in \delta_{\text{shift}}$ (respectively, $ \in \delta_{\text{push}}$) and 
there is an edge from state $q$ to state $p$ labeled by $r \in Q$ and denoted by a double arrow iff $(q,r, p) \in \delta_{\text{pop}}$.

To define the semantics of the automaton, we need some new notations.

We use letters $p, q, p_i, q_i, \dots $ to denote states in $Q$.
Let  $\Gamma$ be	$\Sigma \times Q$ and let $\Gamma' = \Gamma \cup  \{Z_0\} $ be the \textit{stack alphabet}; 
we denote symbols in $\Gamma'$ as $\tstack aq$ or $Z_0$.
We set $\symb {\tstack aq} = a$, $\symb {Z_0}=\#$, and
$\state {\tstack aq} = q$.
Given a string $\Pi =  \pi_n \dots  \pi_2  \pi_1 Z_0$, with $ \pi_i \in \Gamma$ , $n \geq 0$, 
we set $\symb \Pi = \symb{\pi_n}$, including the particular case  $\symb {Z_0} = \#$.

As usual, a \emph{configuration} of an \opa\ is a triple $c = \tconfig w q \Pi$,
where $w \in \Sigma^*\#$, $q \in Q$, and $\Pi \in \Gamma^*Z_0$.

A \emph{computation} or \emph{run} of the automaton is a finite sequence of \emph{moves} or \emph{transitions} 
$c_1 \transition{} c_2$; 
there are three kinds of moves, depending on the precedence relation between the symbol on top of the stack and the next symbol to read:

\smallskip
\noindent {\bf push move:} if $\symb \Pi \lessdot \ a$ then
$
\tconfig {ax} p  \Pi\transition{} \tconfig {x} q {\tstack   a p\Pi }$, with $(p,a, q) \in \delta_{\text{push}}$;

\smallskip
\noindent {\bf shift move:} if $a \doteq b$ then 
$
\tconfig {bx} q { \tstack a p \Pi}  \transition{} \tconfig x  r { \tstack b p \Pi}$, with $(q,b,r) \in \delta_{\text{shift}}$;

\smallskip

\noindent {\bf pop move:} if $a \gtrdot b$
then 
$
\tconfig {bx} q  { \tstack a p \Pi}\transition{} \tconfig {bx} r \Pi $, with $(q, p, r) \in \delta_{\text{pop}}$.

Observe that shift and pop moves are never performed when the stack contains only $Z_0$.

Push and shift moves update the current state of the automaton according to the transition relations $\delta_{\text{push}}$ and  $\delta_{\text{shift}}$, respectively: push moves put a new element on top of the stack consisting of the input symbol together with the current state of the automaton, whereas shift moves update the top element of the stack by \textit{changing its input symbol only}.
Pop moves remove the element on top of the stack,
and update the state of the automaton according to $\delta_{\text{pop}}$ on the basis of the pair of states consisting of the current state of the automaton and the state of the removed stack symbol; notice that in this moves the input symbol is used only to establish the $\gtrdot$ relation and it remains available for the following move.

The language accepted by the automaton is defined as:
\[
L(\mathcal A) = \left\{ x \mid  \tconfig {x\#} {q_I} {Z_0} \comp * 
\tconfig {\#} {q_F}{Z_0} , q_I \in I, q_F \in F \right\}.
\]

\begin{example}\label{ex:exprAut}
The OPA  depicted in Figure~\ref{fig:exprAut} accepts the language of arithmetic expressions generated by grammar $GAE_P$.
The same figure also shows the syntax-tree of the sentence $e + e * \lp e + e \rp$ and an accepting computation on this input.
\end{example}

\tikzset{every loop/.style={min distance=9mm,looseness=10}}

\begin{figure}
\begin{center}

\begin{tabular}{cc}
\begin{tikzpicture}[every edge/.style={draw,solid}, node distance=4cm, auto, 
                    every state/.style={draw=black!100,scale=0.5}, >=stealth]

\node[initial by arrow, initial text=,state] (q0) {{\huge $q_0$}};
\node[state] (q1) [right of=q0, xshift=0cm, accepting] {{\huge $q_1$}};
\node[state] (q2) [below of=q0, xshift=0cm] {{\huge $q_2$}};
\node[state] (q3) [right of=q2, xshift=0cm, accepting] {{\huge $q_3$}};

\path[->]
(q0) edge [below, \parrow] node {$e$} (q1)
(q0) edge [bend right, right, \parrow]  node {$\lp$} (q2)
(q1) edge [loop above, double, right]  node {$\ q_0, q_1$} (q1)
(q1) edge [bend right, above, \parrow]  node {$+, *$} (q0)

(q2) edge [below,  \parrow] node {$e$} (q3)
(q2) edge [loop left, left, \parrow] node {$\lp\ $} (q2)
(q3) edge [loop above, right, double]  node {$\ q_0, q_1, q_2, q_3$} (q3) 
(q3) edge [bend right, above, \parrow]  node {$+, *$} (q2)
(q3) edge [loop right, right, \iarrow] node {$\ \rp$} (q3);
\end{tikzpicture}

&
  
\centering\begin{tikzpicture}[scale=0.4]
\node {$E$}
child {
	node {$E$}
	child{ node{$e$} } 	
}
child { node {$+$} }
child {
	node {$T$}
	child {
		node {$T$}
		child{ node {$e$} }
	}
	child{ node {$*$} 	}
	child{
		node {$F$}
		child{ node{$\lp$} }
		child{ node{$E$}
			child{ 
				node{$E$}
				child{ node{$e$} }
			}
			child{ node{$+$} } 
			child{ 
				node{$T$}
				child{ node{$e$} }	
			}						
		}
		child{ node{$\rp$} }
	}	
};
\end{tikzpicture}

\end{tabular}
\\
\begin{small}
\[
\begin{array}{|r|c|r|}
  \hline
  \text{input}            & \text{state} & \text{stack}   											 
                                                                                                                                           \\\hline
 e + e * \lp e + e \rp \# & q_0          & Z_0                                                                                             \\ \hline 
  + e * \lp e + e \rp  \# & q_1          & \tstack {e} {q_0} Z_0                                                                           \\ \hline
  + e * \lp e + e \rp  \# & q_1          & Z_0                                                                                             \\ \hline 
  e * \lp e + e \rp  \#   & q_0          & \tstack {+} {q_1}  Z_0                                                                          \\ \hline
   * \lp e + e \rp \#     & q_1          & \tstack  e {q_0} \tstack {+} {q_1} Z_0                                                          \\ \hline
  * \lp e + e \rp  \#     & q_1          & \tstack  + {q_1}  Z_0                                                                           \\ \hline
  \lp e + e \rp  \#       & q_0          & \tstack  * {q_1}  \tstack  + {q_1}  Z_0                                                         \\ \hline
  e + e \rp  \#           & q_2          & \tstack  \lp {q_0}    \tstack  * {q_1}  \tstack  + {q_1} Z_0                                    \\ \hline
   + e \rp  \#            & q_3          & \tstack  e {q_2} \tstack  \lp {q_0}   \tstack  * {q_1} \tstack  + {q_1}  Z_0                    \\ \hline
    + e \rp  \#           & q_3          & \tstack  \lp {q_0}  \tstack  * {q_1}  \tstack  + {q_1}  Z_0                                     \\ \hline
  e \rp  \#               & q_2          & \tstack  + {q_3} \tstack  \lp {q_0}   \tstack  * {q_1}  \tstack  + {q_1}   Z_0                  \\ \hline
  \rp  \#                 & q_3          & \tstack  e {q_2}  \tstack  + {q_3}  \tstack  \lp {q_0}  \tstack  * {q_1}  \tstack  + {q_1}  Z_0 \\ \hline
   \rp  \#                & q_3          & \tstack  + {q_3} \tstack  \lp {q_0}  \tstack  * {q_1}   \tstack  + {q_1}  Z_0                   \\ \hline
  \rp  \#                 & q_3          & \tstack  \lp {q_0}   \tstack  * {q_1} \tstack  + {q_1}  Z_0                                     \\ \hline
     \#                   & q_3          & \tstack  \rp {q_0}   \tstack  * {q_1} \tstack  + {q_1}  Z_0                                     \\ \hline
    \#                    & q_3          & \tstack  * {q_1}  \tstack  + {q_1}  Z_0                                                         \\ \hline
    \#                    & q_3          & \tstack  + {q_1}  Z_0                                                                           \\ \hline
  \#                      & q_3          & Z_0                                                                                             \\ \hline
\end{array}
\]
\end{small}
\caption{Automaton and example of computation for the language of Example~\ref{ex:exprAut}. Recall that shift, push and pop transitions are denoted by dashed, normal and double arrows, respectively.}\label{fig:exprAut}
\end{center}
\end{figure} 

\medskip

Notice the similarity of the above definition of OPA with that of VPA (Definition \ref{def: VPL}) and with the normalized form for PDA given in Section~\ref{sec:HDPDL}.

Showing the equivalence between OPG and OPAs, though somewhat intuitive, requires to overtake a few non-trivial technical difficulties, mainly in the path from OPG to OPAs. Here we offer just an informal description of the rationale of the two constructions and an illustrating example; the full proof of the equivalence between OPG and OPA can be found in \cite{LonatiEtAl2015}.

For convenience and with no loss of generality, let $G$ be an OPG with no empty rules, except possibly $S\to \varepsilon$, and no renaming rules, except possibly those whose lhs is $S$,
\newcommand{\pref}{\mathbb{P}} an OPA
$\mathcal A$  equivalent to $G$ is built in such a way that a successful computation thereof corresponds to
building bottom-up the mirror image of a rightmost derivation of $G$: $\mathcal A$ performs a push transition when it reads the first terminal of a new rhs; it performs a shift transition when it reads a terminal symbol inside a rhs, i.e. a leaf with some left sibling leaf;
it performs a pop transition when it completes the recognition of a rhs, then it guesses (nondeterministically, if there are several rules with the same rhs) the nonterminal at the lhs.

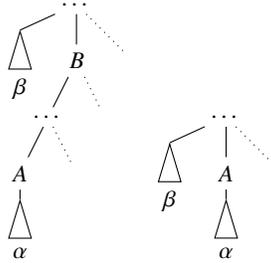
\begin{figure}[h!]
\begin{center}
  
\begin{tabular}{ccc}
\begin{tikzpicture}[scale=0.5,
  triangle/.style={isosceles triangle,draw=,shape border rotate=90,isosceles triangle stretches=true, inner sep=0,yshift={0mm}},
  small triangle/.style={triangle, minimum height = 5mm, minimum width = 3mm },
]
\node {$\dots$}
child[child anchor=north] {node[label=below:$\beta$,small triangle] {}}
child {node {$B$}
	child {node{ $\dots$}
		child {node {$A$}
			child[child anchor=north] {node[label=below:$\alpha$,small triangle]{}}
		}
		child {node {}
                	edge from parent [path]
		}
	}
	child {node {}
                edge from parent [path]
	}
}
child {node {}
	edge from parent [path]
}
;
\end{tikzpicture}
  &

&
\begin{tikzpicture}[scale=0.5,
  triangle/.style={isosceles triangle,draw=,shape border rotate=90,isosceles triangle stretches=true, inner sep=0,yshift={0mm}},
  small triangle/.style={triangle, minimum height = 5mm, minimum width = 3mm },
]
\node {$\dots$}
child[child anchor=north] {node[label=below:$\beta$,small triangle]{}}
child {node {$A$}
	child[child anchor=north] {node[label=below:$\alpha$,small triangle]{}}
}
child {node {}
	edge from parent [path]
}
;
\end{tikzpicture}
  \end{tabular}
\caption{When parsing $\alpha$, the prefix previously under construction is $\beta$.}
\label{fig:OPGOPA-trees}
\end{center}
\end{figure}

Each state of $\mathcal A$ contains two pieces of information: the first component represents the prefix of the rhs~under construction,
whereas the second component is used to recover the rhs~\emph{previously under
  construction} (see Figure~\ref{fig:OPGOPA-trees})
whenever all rhs's nested below have been completed.
Without going into the details of the construction and the formal equivalence proof between $G$ and $\mathcal A$, we further illustrate the rationale of the construction through the following Example.

\begin{example}
Consider again grammar $GAE_P$.
Figure~\ref{fig:GtoAEx} shows the first part of an accepting computation of the
automaton derived therefrom.
Consider, for instance, step 3 of the computation: at this
point the automaton has already reduced (nondeterministically) the first $e$ to
$E$ and has pushed the following $+$ onto the stack, paired with the state from
which it was coming; thus, its new state is $\pair{E +} {\varepsilon} $; at 
 step 6, instead, the state is $\pair{T *} {E +} $ because at this point the
automaton has built the $T*$ part of the current rhs and remembers that the
prefix of the suspended rhs is $E+$.
Notice that the computation partially shown in
Figure~\ref{fig:GtoAEx} is equal to that of Figure~\ref{fig:exprAut} up to a
renaming of the states; in fact the shape of syntax-trees and consequently the
sequence of push, shift and pop moves in OPL depends only on the OPM, not on
the visited states.
\begin{figure}[h!]
\begin{center}
{\small 
\[
\begin{array}{|r|r|c|r|}
\hline
\text{step} & \text{input}\      & \text{state}                     & \text{stack}                                                                   \\ \hline
0           & e+e*\lp e+e \rp \# & \pair{\varepsilon} {\varepsilon} & Z_0                                                                            \\ \hline
1           & +e*\lp e+e \rp \#  & \pair{e} {\varepsilon}           & \tstack{e} {\pair{\varepsilon} {\varepsilon}}  Z_0                             \\ \hline
2           & +e*\lp e+e \rp \#  & \pair{E} {\varepsilon}           & Z_0                                                                            \\ \hline
3           & e*\lp e+e \rp \#   & \pair{E +} {\varepsilon}         & \tstack{+} {\pair{E} {\varepsilon}}  Z_0                                       \\ \hline
4           & *\lp e+e \rp \#    & \pair{e} {\varepsilon}           & \tstack{e} {\pair{E +} {\varepsilon}}  \tstack{+} {\pair{E} {\varepsilon}} Z_0 \\ \hline
5           & *\lp e+e \rp \#    & \pair{T} {E +}                   & \tstack{+} {\pair{E} {\varepsilon}} Z_0                                        \\ \hline
6           & \lp e+e \rp \#     & \pair{T *} {E +}                 & \tstack{*} {\pair{T} {E +}}  \tstack{+} {\pair{E} {\varepsilon}} Z_0           \\ \hline
\end{array}
\]
}
\caption{Partial accepting computation of the automaton built from grammar $GAE_P$.\label{fig:GtoAEx}}
\end{center}
\end{figure}
\end{example}

The converse construction from OPAs to OPG is somewhat simpler; in essence, from a given OPA $\mathcal A = (\Sigma,M, Q, I, F, \delta)$ a grammar $G$ is built whose nonterminals are 4-tuples $(a, q, p, b) \in \Sigma \times Q \times Q \times \Sigma$, written as $\nont apqb$.
$G$'s rules are built on the basis of $\mathcal A$'s chains as follows (particular cases are omitted for simplicity):
\begin{itemize}
\item 
for every simple chain $\chain {a_0} {a_1 a_2 \dots a_n} {a_{n+1}}$, if there is a sequence of $\mathcal A$'s transitions that, while reading the body of the chain starting from $q_0$ leaves $\mathcal A$ in $q_{n+1}$, include the rule 
\[
\nont {a_0}{q_0}{q_{n+1}}{a_{n+1}}\longrightarrow a_1 a_2 \dots a_n\ 
\]


\item 
for every composed chain $\chain {a_0} {x_0 a_1 x_1 a_2 \dots a_n x_n} {a_{n+1}}$,
add the rule 
\[
\nont {a_0}{q_0}{q_{n+1}}{a_{n+1}} \longrightarrow \Lambda_0 a_1 \Lambda_1 a_2 \dots a_n \Lambda_n \ 
\]
if there is a sequence of $\mathcal A$'s transitions that, while reading the body of the chain starting from $q_0$ leaves $\mathcal A$ in $q_{n+1}$, and, for every $i = 0,1, \dots, n$, $\Lambda_i =
\varepsilon$ if $x_i = \varepsilon$, otherwise
$\Lambda_i = \nont {a_i}{q_i}{q'_i}{a_{i+1}}$ if $x_i \neq \varepsilon$ and there is a path leading from $q_i$ to $q'_i$ when traversing $x_i$.
\end{itemize}
Since the size of $G$'s nonterminal alphabet is bounded, the above procedure eventually terminates when no new rules are added to $P$.\footnote{The above claim can be easily proved if the OPM has no circularities in the $\doteq$ relation, since this implies an upper bound to the length of $P$'s rhs; in the (seldom) case where this hypothesis is not verified other ones can be exploited (see \cite{LonatiEtAl2015} for a more detailed analysis.)}

We have seen that a fundamental issue to state the properties of most abstract machines is their determinizability: in the cases examined so far we have realized that the positive basic result holding for RL extends to the various versions of structured CFL, though at the expenses of more intricate constructions and size complexity of the deterministic versions obtained from the nondeterministic ones, but not to the general CF family. Having OPL been born just with the motivation of supporting deterministic parsing, and being they structured as well, it is not surprising to find that for any nondeterministic OPA with $s$ states an equivalent deterministic one can be built with $2^{O(s^2)}$ states, as it happens for the analogous construction for VPL: in \cite{LonatiEtAl2015} besides giving a detailed construction for the above result, it is also noticed that the construction of an OPA from an OPG is such that, if the OPG is in FNF, then the obtained automaton is already deterministic since the grammar has no repeated rhs. 
As a consequence, producing a deterministic OPA from an OPG by first putting the OPG into FNF produces an automaton of an exponentially smaller size than the other way around.

%
\subsubsection{Operator precedence  vs other structured languages}
A distinguishing feature of OPL is that, on the one side they have been defined to support deterministic parsing, i.e., the construction of the syntax-tree of any sentence which is not immediately visible in the sentence itself but, on the other side, they can still be considered as structured in the sense that their syntax-trees are univocally determined once an OPM is fixed, as it happens when we enclose grammar's rhs within parentheses or we split the terminal alphabet into $\Sigma_c \cup \Sigma_r \cup \Sigma_i$. It is therefore natural to compare their generative power with that of other structured languages.

On this respect, the main result of this section is that \emph{OPL strictly include VPL}, which in turn strictly include parenthesis languages and the languages generated by balanced grammars (discussed in Section~\ref{sec:BalGr}).

This result was originally proved in \cite{Crespi-ReghizziM12}. To give an intuitive explanation of this claim consider the following remarks:
\begin{itemize}
\item
Sequences $\in \Sigma_i^*$ can be assimilated to regular ``sublanguages'' with a linear structure; if we conventionally assign to them a left-linear structure, this can be obtained through an OPM where every character, but those in $\Sigma_c$, takes precedence over all elements in $\Sigma_i$; by stating instead that all elements of $\Sigma_c$ yield precedence to the elements in $\Sigma_i$ we obtain that after every call the OPA pushes and pops all subsequent elements of $\Sigma_i$, as an FSA would do without using the stack.
\item
All elements of $\Sigma_r$ produce a pop from the stack of the corresponding element of $\Sigma_c$, if any; thus we obtain the same effect by letting them take precedence over all other terminal characters.
\item
A VPA performs a push onto its stack when (and only when) it reads an element of $\Sigma_c$, whereas an OPA pushes the elements to which some other element yields precedence; thus, it is natural to state that whenever on top of the stack there is a call symbol, possibly after having visited a subtree whose result is stored as the state component in the top of the stack together with the terminal symbol, such a symbol yields precedence to the following call (roughly, open parentheses yield precedence to other open parentheses and closed parentheses take precedence over other closed parentheses).
\item
Once the whole subsequence embraced by two matching call and return is scanned and possibly reduced, the two terminals are faced, with the possible occurrence of an irrelevant nonterminal in between, and therefore the call must be equal in precedence to the return.
\item
Finally, the usual convention that $\#$ yields precedence to everything and everything takes precedence over $\#$ enables the managing of possible unmatched returns at the beginning of the sentence and unmatched calls at its end.
\end{itemize}
In summary, for every VPA $\mathcal A$ with a given partitioned alphabet $\Sigma$, an OPM such as the one displayed in Figure \ref {fig:pmatr} and an OPA $\mathcal A'$ defined thereon can be built such that $L(\mathcal A') = L(\mathcal A)$.

\begin{figure}
    \begin{center}
        \[
        \begin{array}{ccc}
    \begin{array}{c|c|c|c|}
        & \Sigma_c & \Sigma_r & \Sigma_i \\
        \hline
\Sigma_c & \lessdot & \dot= & \lessdot \\
        \hline
\Sigma_r & \gtrdot & \gtrdot & \gtrdot \\ 
        \hline
\Sigma_i & \gtrdot & \gtrdot & \gtrdot \\ 
        \hline
\end{array}
&
\qquad
&
{\small \begin{array}{l}
\text{\textbf{Legend}}\\ 
\text{$\Sigma_c$ denotes ``calls''} \\
\text{$\Sigma_r$ denotes ``returns''}\\
\text{$\Sigma_i$ denotes internal characters}\\

\end{array}}
\end{array}
\]
\caption{A partitioned matrix, where $\Sigma_c$, $\Sigma_r$,$\Sigma_i$ are set of terminal characters.
A precedence relation in position $\Sigma_\alpha$, $\Sigma_\beta$ means that relation holds between all symbols of $\Sigma_\alpha$ and all those of $\Sigma_\beta$. 
}\label{fig:pmatr}
\end{center}
\end{figure}

In~\cite{Crespi-ReghizziM12} it is also shown the converse property, i.e., that whenever an OPM is such that the terminal alphabet can be partitioned into three disjoint sets $\Sigma_c$, $\Sigma_r$,$\Sigma_i$ such that the OPM has the shape of Figure~\ref{fig:pmatr}, any OPL defined on such an OPM is also a VPL. Strict inclusion of VPL within OPL follows form the fact that VPL are real-time whereas OPL include also non-real-time languages (see Section~\ref{sec:DCFPars}); there are also real-time OPL\footnote{When we say that an OPL $L$ is real-time we mean, as usual, that there is an abstract machine, in particular a DPDA, recognizing it that performs exactly $|x|$ moves for every input string $x$; this is not to say that an OPA accepting $L$ operates in real-time, since OPA's pop moves are defined as $\varepsilon$ moves. } such as \[ L = \{b^nc^n \mid n \geq 1\} \cup \{f^nd^n \mid n \geq 1\} \cup \{e^n(fb)^n \mid n \geq 1\} \] that are not VPL. In fact, strings of type $b^nc^n$ impose that $b$ is a call and $c$
a return; for similar reasons, $f$ must be a call and $d$ a return. Strings of type $e^n( f b)^n$ impose that at least one of $b$ and $f$
must be a return, a contradiction for a VP alphabet. In conclusion we have the following result:

\begin{theorem}
 VPL are the subfamily of OPL whose OPM is a
{\em partitioned matrix}, i.e., a matrix whose structure is depicted in Figure~\ref{fig:pmatr}.
\end{theorem}

As a corollary OPL also strictly include balanced languages and parenthesis languages. OPL are instead uncomparable with HRDPDL: we have already seen that the language $L_1 = \{a^nba^n\}$ is an HRDPDL but it is neither a VPL nor an OPL since it necessarily requires a conflict $a \lessdot a$ and $a \gtrdot a$; conversely, the previous language $L_2 = \{a^mb^nc^nd^m \mid m, n \geq 1\} \cup \{a^mb^+ed^m \mid m \geq 1\}$ can be recognized by an OPA but by no HRDPDA (see Section~\ref{sec:DCFPars}).

The increased power of OPL over other structured languages goes far beyond the mathematical containment properties and opens several application scenarios that are hardly accessible by ``traditional'' structured languages. The field of programming languages  was the original motivation and source of inspiration for the introduction of OPL; arithmetic expressions, used throughout this paper as running examples, are just a small but meaningful subset of such languages and we have emphasized from the beginning that their partially hidden structure cannot be ``forced'' to the linearity of RL, nor can always be made explicit by the insertion of parentheses.

VPL too have been presented as an extension of parenthesis languages with the motivation that not always calls, e.g. procedure calls, can be matched by corresponding returns: a sudden closure, e.g. due to an exception or an interrupt or an unexpected end may leave an open chain of suspended calls. Such a situation, however, may need a generalization that cannot be formalized by the VPL formalism, since in VPL unmatched calls can occur only at the end of a string.\footnote{Recently, such a weakness of VPL has been acknowledged in \cite{AF16} where the authors introduced \emph{colored VPL} to cope with the above problem; the extended family, however, still does not reach the power of OPL (\cite{AF16}).} Imagine, for instance, that the occurrence of an interrupt while serving a chain of calls imposes to abort the chain to serve immediately the interrupt; after serving the interrupt, however, the normal system behavior may be resumed with new calls and corresponding returns even if some previous calls have been lost due to the need to serve the interrupt with high priority.
Various, more or less sophisticated, policies can be designed to manage such systems and can be adequately formalized as OPL. The next example describes a first simple case of this type; other more sophisticated examples of the same type and further ones inspired by different application fields can be found in \cite{LonatiEtAl2015}.
\begin{example}[Managing interrupts]\label{ex:interrupt}
Consider a software system that is designed to serve requests issued by different users but subject to interrupts. Precisely, assume that the system manages ``normal operations'' according to a traditional LIFO policy, and may receive and serve some interrupts denoted as $int$. 

We model its behavior by introducing an alphabet with two pairs of calls and
returns: $call$ and $ret$ denote the call to, and return from, a normal procedure; $int$, and {\em serve} denote the occurrence of an interrupt and its serving, respectively. The occurrence of an interrupt provokes discarding possible pending $call$s not already matched by corresponding $ret$s; furthermore when an interrupt is pending, i.e., not yet served, calls to normal procedures are not accepted and consequently corresponding returns cannot occur; interrupts however, can accumulate and are served themselves along a LIFO policy. Once all pending interrupts have been served the system can accept new $call$s  and manage them normally.

Figure~\ref{fig:interrupt}~(a) shows an OPM that assigns to sequences on the above alphabet a structure compatible with the described priorities. Then, a suitable OPA can specify further constraints on such sequences; for instance the automaton of Figure~\ref{fig:interrupt}~(b) restricts the set of sequences compatible with the matrix by imposing that all $int$ are eventually served and the computation ends with no pending $call$s; furthermore unmatched \emph{serve} and $ret$ are forbidden.
E.g., the string $call.call.ret. int.serve.call.ret$ is accepted through the sequence of states $q_0  \va{call} q_1 \va{call}  q_1  \shift{ret}{q_1} \flush{q_1} q_1 \flush{q_0}  q_0 \va{int} q_2 \shift{serve}{q_2} \flush{q_0} q_0  \va{call}  {q_1} \shift{ret} q_1\flush{q_0}  q_0$; on the contrary, a sequence beginning with  $call.serve$ would not be accepted.

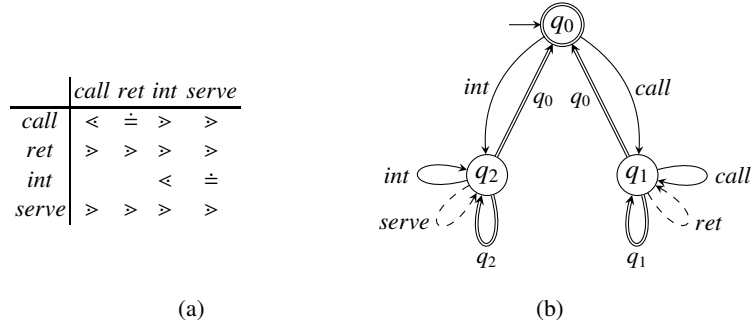
\begin{figure*}
  
\begin{center}
\begin{tabular}{m{0.3\textwidth}m{0.1\textwidth}m{0.3\textwidth}}

\begin{center}
\begin{tabular}{c|ccccc}
	& $call$     & $ret$     & $int$      & $serve$   \\
\hline
$call$  & $\lessdot$ & $\dot=$   & $\gtrdot$  & $\gtrdot$ \\
$ret$   & $\gtrdot$  & $\gtrdot$ & $\gtrdot$  & $\gtrdot$ \\
$int$   &            &           & $\lessdot$ & $\doteq$  \\
$serve$ & $\gtrdot$  & $\gtrdot$ & $\gtrdot$  & $\gtrdot$ \\
\end{tabular}
\end{center}
&

&
\begin{center}
\begin{tikzpicture}[every edge/.style={draw,solid}, node distance=4cm, auto, 
                    every state/.style={draw=black!100,scale=0.5}, >=stealth]

\node[initial by arrow, initial text=,state, accepting] (q0) {{\huge $q_0$}};
\node[state] (q1) [below of=q0, xshift=-2cm] {{\huge $q_2$}};
\node[state] (q2) [right of=q1] {{\huge $q_1$}};

\path[->]

(q0) edge [bend right, left, \parrow]  node {$int$} (q1)
(q0) edge [bend left, right, \parrow]  node {$call$} (q2)

(q1) edge [loop left, left, \parrow] node {$int$} (q1)
(q1) edge [\iarrow, in=-120, out=-150, loop, left] node {$serve$} (q1)
(q1) edge [loop below, double]  node {$q_2$} (q1)
(q1) edge [double, right]  node {$q_0$} (q0)

(q2) edge [loop right, right, \parrow] node {$call$} (q2)
(q2) edge [\iarrow, in=-30, out=-60, loop, right] node {$ret$} (q2)
(q2) edge [double, loop below] node {$q_1$} (q2)
(q2) edge [left, double]  node {$q_0$} (q0)
;
\end{tikzpicture}
\end{center}
\\
\vspace{-0.5cm}
\qquad\qquad\qquad\qquad (a)
&
&
\vspace{-0.5cm}
\qquad\qquad\qquad\quad (b)
\end{tabular}
\caption{OPM (a) and automaton (b) for the language of Example~\ref{ex:interrupt}.}\label{fig:interrupt}
\end{center}
\end{figure*}
\end{example}

\subsubsection{Closure and decidability properties}

Structured languages with \textit{compatible structures} often enjoy many closure properties typical of RL; noticeably, families of structured languages are often Boolean algebras. The intuitive notion of compatible structure is formally defined for each family of structured languages; for instance two VPL have compatible structure iff their tri-partition of $\Sigma$ is the same; two height-deterministic PDA languages (HPDL) have compatible structure if they are synchronized. In the case of OPL, the notion of structural compatibility is naturally formalized through the OPM.
\begin{definition}\label{def:OPMComp}
Given two OPM $M_1$ and $M_2$, we define set inclusion and union:
\[
   \nonumber M_1\subseteq M_2 \text{ if } \forall a,b: (M_1)_{ab}\subseteq (M_2)_{ab}
   \]
\[
\nonumber M=M_1\cup M_2   \text{ if }  \forall a,b: M_{ab}= (M_1)_{ab}\cup (M_2)_{ab}.
\]

Two matrices are \emph{compatible} if their union is conflict-free. A matrix is \emph{total} (or \emph{complete}) if it contains no empty cell.
\end{definition}

The following theorem has been proved originally in \cite{Crespi-ReghizziMM1978} by exploiting some standard forms of OPG that have been applied to grammar inference problems \cite{Crespi-ReghizziCACM73}.

\begin{theorem}\label{th:OPbool}
For any conflict-free OPM $M$ the OPL whose OPM is contained in $M$ are a Boolean algebra. In particular, if $M$ is complete, the top language of its associated algebra is $\Sigma^*$ with the structure defined by $M$.
\end{theorem}

Notice however, that the same result could be proved in a simpler and fairly standard way by exploiting OPA and their traditional composition rules (which pass through determinization to achieve closure under complement).
As usual in such cases, thanks to the decidablity of the emptiness problem for general CFL, a major consequence of Boolean closures is the following corollary.

\begin{corollary}
The inclusion problem between OPL with compatible OPM is decidable.
\end{corollary}
Closure under concatenation and  Kleene~$^*$ has been proved more recently in \cite{Crespi-ReghizziM12}; whereas such closures are normally easily proved or disproved for many  families of languages, the technicalities to achieve this result are not trivial for OPL; however we do not go into their description since closure or non-closure w.r.t. these operations is not of major interest in this paper.
\subsubsection{Logic characterization}
Achieving a logic characterization of OPL has probably been the most difficult job in the recent revisit of these languages and posed new challenges w.r.t. the analogous path
previously followed for RL and then for CFL \cite{Lautemann94}  and VPL \cite{jacm/AlurM09}. In fact we have seen that moving from linear languages such as RL to tree-shaped ones such as CFL led to the introduction of the relation $M$ between the positions of leftmost and rightmost leaves of any subtree (generated by a grammar in DGNF); the obtained characterization in terms of first-order formulas existentially quantified w.r.t. the $M$ relation (which is a representation of the sentence structure) however, was suffering from the lack of closure under complementation of CFL \cite{Lautemann94}; the same relation instead proved more effective for VPL thanks to the fact that they are structured and enjoy all necessary closures.

To the best of our knowledge, however, \textit{all previous characterizations} of formal languages in terms of logics that refer to string positions (for instance, there is significant literature on the characterization of various subclasses of RL in terms of first-order or temporal logics, see, e.g.,  \cite{Emerson90}) {\em have been given for real-time languages}. This feature is the key that allows, in the exploitation of MSO logic, to state a natural correspondence between automaton's state $q_i$ and second-order variable $\bm{X}_i$ in such a way that the value of $\bm{X}_i$ is the set of positions where the  state visited by the automaton is $q_i$.

OPL instead include also DCFL that are not real-time and, as a consequence, there are positions where the recognizing OPA traverses different configurations with different states. As a further consequence, the $M$ relation adopted for CFL and VPL is not anymore a one-to-one relation since the same position may be the position of the left/rightmost leaf of several subtrees of the whole syntax-tree; this makes formulas such as the key ones given in Section \ref{sec:MSOVPL} meaningless.

The following key ideas helped overtaking the above difficulties:
\begin{itemize}
\item
A new relation $\mu$ replaces the original $M$ adopted in \cite{Lautemann94} and \cite{jacm/AlurM09}; $\mu$ is based on the look-ahead-look-back mechanism which drives the (generalized) input-driven parsing of OPL based on precedence relations: thus, whereas in $M(\bm {x}, \bm {y})$ $\bm {x}$, $\bm {y}$ denote the positions of the extreme leaves of a subtree, in $\mu(\bm {x},\bm {y})$ they denote the position of the \textit{context} of the same subtree, i.e., respectively, of the character that yields precedence to the subtree's leftmost leaf, and of the one over which the subtree's rightmost leaf takes precedence. The new $\mu$ relation is not one-to-one as well, but, unlike the original $M$, its parameters  $\bm {x}$, $\bm {y}$ are not ``consumed'' by a pop transition of the automaton and remain available to be used in further automaton transitions of any type.
In other words,  $\mu$ holds between the positions $0$ and $n+1$ of every chain (see Definition~\ref{def: chains}). For instance, Figure~\ref{fig:log:avv} displays the $\mu$ relation, graphically denoted by arrows,  holding for the sentence $e + e * \lp e + e \rp$ generated by grammar $GAE_P$:  we have 
$\mu(0 , 2)$, $\mu(2 , 4)$, $\mu(5 , 7)$, $\mu(7 , 9)$, $\mu(5 , 9)$, $\mu(4 , 10)$, $\mu(2 , 10)$, and $\mu(0 , 10)$.
Such pairs correspond to
contexts where a reduce operation is executed  during the parsing of the string (they are listed according to their execution order).

In general $\mu(\g{x} , \g{y})$ implies $\g{y} > \g{x}+1$, and a position $\g{x}$ may be in relation $\mu$ with more than one position and vice versa.  Moreover,
if $w$ is compatible with $M$, then $\mu(0 , |w|+1)$.

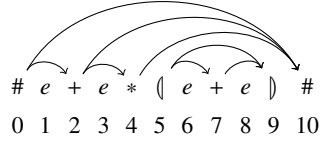
\begin{figure} 
\centering
\begin{tikzpicture}[flush/.style={double, >=stealth, thin, rounded corners}]
\matrix (m) [matrix of nodes]
 {
    \# & $e$ & $+$ & $e$ & $*$ & $\lp$ & $e$ & $+$ & $e$ & $\rp$ & \# \\
    0 & 1 & 2 & 3 & 4 & 5 & 6 & 7 & 8 & 9 & 10 \\
 };

\draw[->] (m-1-1)  to [out=60, in=120] (m-1-3);
\draw[->] (m-1-3)  to [out=60, in=120] (m-1-5);
\draw[->] (m-1-6)  to [out=60, in=120] (m-1-8);
\draw[->] (m-1-8)  to [out=60, in=120] (m-1-10);
\draw[->] (m-1-6)  to [out=60, in=120] (m-1-10);
\draw[->] (m-1-5)  to [out=60, in=120] (m-1-11);
\draw[->] (m-1-3)  to [out=60, in=120] (m-1-11);
\draw[->] (m-1-1)  to [out=60, in=120] (m-1-11);
\end{tikzpicture}
\caption{The string $e + e * \lp e + e \rp$, with positions and relation $\mu$.}\label{fig:log:avv}
\end{figure}

\begin{example}
	\label{ex:ss}
	The following sentence of the MSO logic enriched with the $\mu$ relation defines, within the universe of strings compatible with the OPM of Figure \ref{fig:chain}(a),  the language where parentheses are used only when they are needed (i.e.\ to give precedence to $+$ over $*$).
	\[
	\forall \g{x} \forall \g{y} \left(
    \begin{array}{c}
	\begin{array}{c}
	\mu(\g{x} ,\g{y}) \land
	\lp(\g{x}+1) \land \rp(\g{y}-1) 
	\end{array}
\\	\Rightarrow \\
	(*(\g{x}) \lor *(\g{y}))
	\land \\
	\exists \g{z}
	\left( 
	\begin{array}{c}
	\g{x}+1 < \g{z} < \g{y}-1 \land +(\g{z}) \ \land \\
	\neg \exists \g{u} \exists \g{v} 
	\left( 
	\begin{array}{c}
	\g{x}+1 < \g{u} < \g{z}  \land \lp(\g{u}) \land \\ 
	\g{z} < \g{v} < \g{y}-1 \ \land \ \rp(\g{v}) \land \\
	\mu(\g{u}-1 , \g{v}+1)
	\end{array}
	\right) 
	\end{array}
	\right) 
    \end{array}
	\right)
	\]
\end{example}

\item
Since in every position there may be several states held by the automaton while visiting that position, instead of associating just one second-order variable to each state of the automaton we define three different sets of second-order variables, namely, ${\pushvar}_0, {\pushvar}_1, \ldots, {\pushvar}_N$, ${\markvar}_0, {\markvar}_1, \ldots, {\markvar}_N$ and
${\flushvar}_0, {\flushvar}_1, \ldots, {\flushvar}_N$.
Set ${\pushvar}_i$ contains
those positions of word $w$ where state $q_i$ may be assumed after a
shift or push transition, i.e.\ after a transition that ``consumes'' an input symbol.
Sets ${\markvar}_i$ and ${\flushvar}_i$ encode a pop transition concluding the reading of the body of a chain $^a [w_0 a_1 w_1 \dots$ $a_l w_l]^{a_{l+1}}$ in a state $q_i$: 
set ${\markvar}_i$ contains the position of symbol $a$ that precedes the corresponding push, whereas
${\flushvar}_i$ contains the position of $a_l$, which is the symbol on top of the stack when the automaton performs the pop move relative to the whole chain.

Figure~\ref{fig:log:sets} presents such sets for the example automaton of Figure~\ref{fig:exprAut}, with the same input as in Figure~\ref{fig:log:avv}.
Notice that each position,
except the last one, belongs to exactly one ${\pushvar}_i$, whereas it may belong to several ${\markvar}_i$ and at most one ${\flushvar}_i$.

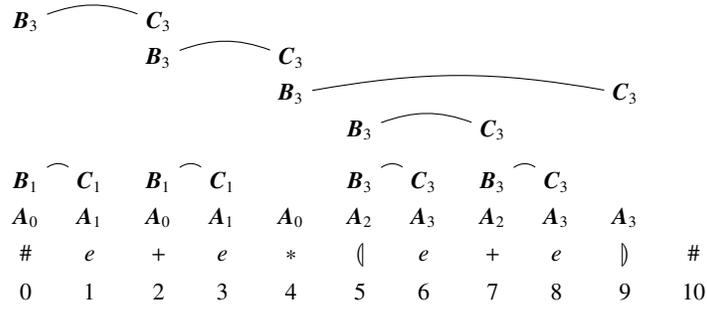
\begin{figure*} 
\centering
\begin{tikzpicture}[scale=0.5, flush/.style={double, >=stealth, thin, rounded corners}]
\matrix (m) [matrix of nodes]
 {
                     &      &               &        &               &      &               &        &               &        &              &      &               &        &               &      &               &        &               &        &       \\
 $\markvar_3$        &      &               &        & $\flushvar_3$ &      &               &        &               &        &              &      &               &        &               &      &               &        &               &        &       \\
                     &      &               &        &               &      &               &        &               &        &              &      &               &        &               &      &               &        &               &        &       \\
                     &      &               &        & $\markvar_3$  &      &               &        & $\flushvar_3$ &        &              &      &               &        &               &      &               &        &               &        &       \\
                     &      &               &        &               &      &               &        &               &        &              &      &               &        &               &      &               &        &               &        &       \\
                     &      &               &        &               &      &               &        & $\markvar_3$  &        &              &      &               &        &               &      &               &        & $\flushvar_3$ &        &       \\
                     &      &               &        &               &      &               &        &               &        &              &      &               &        &               &      &               &        &               &        &       \\
                     &      &               &        &               &      &               &        &               &        & $\markvar_3$ &      &               &        & $\flushvar_3$ &      &               &        &               &        &       \\
                     & $\;$ &               &        &               & $\;$ &               &        &               &        &              & $\;$ &               &        &               & $\;$ &               &        &               &        &  &  & \\
    $\markvar_1$     &      & $\flushvar_1$ &        & $\markvar_1$  &      & $\flushvar_1$ &        &               &        & $\markvar_3$ &      & $\flushvar_3$ &        & $\markvar_3$  &      & $\flushvar_3$ &        &               &        &  &  & \\
	$\pushvar_0$ &      & $\pushvar_1$  &        & $\pushvar_0$  &      & $\pushvar_1$  &        & $\pushvar_0$  &        & $\pushvar_2$ &      & $\pushvar_3$  &        & $\pushvar_2$  &      & $\pushvar_3$  &        & $\pushvar_3$  &        &       \\
    \#               &      & $e$           & $\ \ $ & $+$           &      & $e$           & $\ \ $ & $*$           & $\ \ $ & $\lp$        &      & $e$           & $\ \ $ & $+$           &      & $e$           & $\ \ $ & $\rp$         & $\ \ $ & \#    \\
    0                &      & 1             &        & 2             &      & 3             &        & 4             &        & 5            &      & 6             &        & 7             &      & 8             &        & 9             &        & 10    \\
 };

\draw[-] (m-10-1)  to [out=35, in=145]  (m-10-3);
\draw[-] (m-10-5)  to [out=35, in=145]  (m-10-7);
\draw[-] (m-10-11)  to [out=35, in=145] (m-10-13);
\draw[-] (m-10-15)  to [out=35, in=145] (m-10-17);

\draw[-] (m-2-1)  to [out=20, in=160]  (m-2-5); 
\draw[-] (m-8-11) to [out=20, in=160]  (m-8-15);

\draw[-] (m-4-5)  to [out=20, in=160] (m-4-9); 

\draw[-] (m-6-9)  to [out=10, in=170] (m-6-19);
\end{tikzpicture}
\caption{The string of Figure~\ref{fig:log:avv} with $\markvar_i$, $\pushvar_i$, and $\flushvar_i$ evidenced for the automaton of Figure~\ref{fig:exprAut}. Pop moves of the automaton are represented by linked pairs $\markvar_i$, $\flushvar_i$.}\label{fig:log:sets}
\end{figure*}
\end{itemize}

We can now outline how an OPA can be derived from an MSO logic formula making use of the new symbol $\mu $ and conversely.

\subsubsection*{From MSO formula to OPA}

The construction from MSO logic to OPA essentially follows the lines given originally by  B\"uchi, and reported in Section \ref{sec:reg:logic}: once the original alphabet has been enriched and the formula has been put in the canonical form in the same way as described in Section \ref{sec:reg:logic}, we only need to define a suitable automaton fragment to be associated with the new atomic formula  $\mu(\bm{X_i} , \bm{X_j})$; then, the construction of the global automaton corresponding to the global formula proceeds in the usual inductive way.

	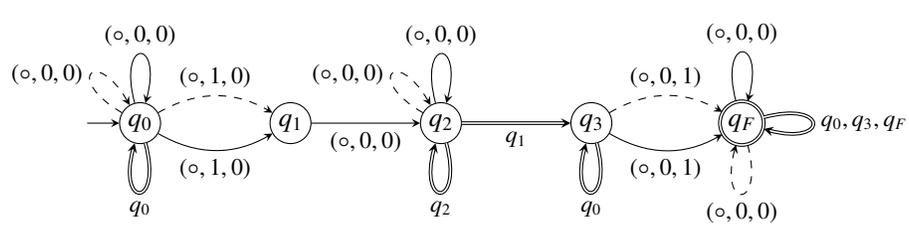
\begin{figure*}
		\begin{center}
			\begin{tikzpicture}[every edge/.style={draw,solid}, node distance=4cm, auto, 
			every state/.style={draw=black!100,scale=0.5}, >=stealth]
			
			\node[initial by arrow, initial text=,state] (Q0) {{\huge $q_0$}};
			\node[state] (Q1) [right of=Q0, xshift=0cm] {{\huge $q_1$}};
			\node[state] (Q2) [right of=Q1, xshift=0cm] {{\huge $q_2$}};
			\node[state] (Q3) [right of=Q2, xshift=0cm] {{\huge $q_3$}};
			\node[state] (QF) [right of=Q3, xshift=0cm, accepting] {{\huge $q_F$}};
			
			\path[->]
			
			(Q1) edge [\parrow, below]  node {$(\circ,0,0)$} (Q2)
			
			(Q0) edge [\iarrow, bend left, above]  node {$(\circ,1,0)$} (Q1) 
			(Q0) edge [\parrow, bend right, below]  node {$(\circ,1,0)$} (Q1) 
			
			(Q2) edge [double, below]  node {$q_1$} (Q3) 
			
			(Q0) edge [\iarrow, in=120, out=150, loop, left] node {$(\circ,0,0) $} (Q0)
			(Q0) edge [\parrow, loop above, above] node {$(\circ,0,0)$} (Q0)
			(Q0) edge [double, loop below] node {$q_0$} (Q0)
			
			(Q2) edge [double,loop below] node {$q_2$} (Q2)
			
			(Q2) edge [\iarrow, in=120, out=150, loop, left] node {$(\circ,0,0)$} (Q2)
			(Q2) edge [\parrow, loop above, above] node {$(\circ,0,0) $} (Q2)
			
			(Q3) edge [\iarrow, bend left, above]  node {$(\circ,0,1)$} (QF) 
			(Q3) edge [\parrow, bend right, below]  node {$(\circ,0,1)$} (QF) 
			(Q3) edge [double, loop below] node {$q_0$} (Q3)
			
			(QF) edge [\iarrow,loop below] node {$(\circ,0,0) $} (QF)
			(QF) edge [\parrow, loop above, above] node {$(\circ,0,0)$} (QF)
			(QF) edge [double, right, loop right]  node {$q_0, q_3, q_F$} (QF) 
			;
			
			\end{tikzpicture}
			\caption{\opa\ for atomic formula $\mu(\g{X} , \g{Y})$.}\label{fig:buchi}
		\end{center}
	\end{figure*}
	
	Figure~\ref{fig:buchi} represents the \opa\ for atomic formula $\psi = 
		\mu(\g{X}, \g{Y})$.
As before, labels are triples belonging to $\Sigma \times \{0,1\}^2$,
where the first component encodes a
character $a \in \Sigma$, the second the positions belonging to $\g X$ (with 1) or not (with
0), while the third component is for $\g Y$. The symbol $\circ$ is used as a shortcut for any value in $\Sigma$ compatible with the OPM, so that the resulting automaton is deterministic.
		
		The semantics of $\mu$ requires for	$\mu (\g{X}, \g{Y})$  that there must be a chain $\chain {a} {w_2} {b}$ in the input word, where $a$ is the symbol at the only position in $\g{X}$, and $b$ is the symbol at the only position in $\g{Y}$. By definition of chain, this means that $a$ must be read, hence in the position represented by $\g{X}$ the automaton performs either a push or a shift move (see Figure~\ref{fig:buchi}, from state $q_0$ to $q_1$), as pop moves do not consume input. After that, the automaton must read $w_2$. In order to process the chain
		$^a[w_2]^b$, reading $w_2$ must start with a push move (from state $q_1$ to state $q_2$), and it must end with one or more pop moves, 
		before reading $b$ (i.e.\ the only position in $\g{Y}$ -- going from state $q_3$ to $q_F$).
		
		This means that the automaton, after a generic sequence of moves corresponding to
		visiting an irrelevant (for $\mu(\g{X}, \g{Y})$) portion of the syntax-tree, when reading the symbol at position $\g{X}$
		performs either a push or a shift move, depending on whether $\g{X}$ is the position of a
		leftmost leaf of the tree or not. 
		Then it visits the subsequent
		subtree ending with a pop labeled $q_1$; at this point, if it reads
		the symbol at position $\g{Y}$, it accepts anything else that follows the examined fragment.

It is interesting to compare the diagram of Figure~\ref{fig:buchi} with those of Figure~\ref{fig:inductive-automata}~(c) and of Figure~\ref{MSOVPA}: the first one, referring to RL, uses two consecutive moves; the second one, referring to VPL, may perform an unbounded number of internal moves and of matching call-return pairs between the call-return pair in positions $\bm{x}$,$\bm{y}$; the OPA does the same as the VPA but needs a pair of extra moves to take into account the look-ahead-look-back implied by precedence relations.
\subsubsection*{From the OPA $\mathcal{A}$ to the MSO formula}

In this case the overall structure 	of the logic formula is the same as in the previous cases for  RL	and VPL, i.e., an existential quantification over second-order variables, which represent states through sets of positions within the string, of a global formula that formalizes a) the constraints imposed by automaton's transitions, b) the fact that in position $0$ the automaton must be in an initial state, and c) that at the end of the string it must be in an accepting state.
The complete formalization of the $\delta$ transition relation as a collection of formulas relating the various variables $\bm{A_i},\bm{B_i},\bm{C_i}$, however, is much more involved than in the two previous cases. Here we only provide a few meaningful examples of such formulas, just to give the essential ideas of how they have been built; their complete set can be found in \cite{LonatiEtAl2015} together with the equivalence proof. Without loss of generality we \textit{assume that the OPA is deterministic}.

Preliminarily, we introduce some notation to make the following formulas more understandable:
\begin{itemize}
\item
When considering a chain $\chain a w b$ we assume $w = w_0 a_1 w_1$ $\dots$ $a_\ell w_\ell$,
with $\chain a {a_1 a_2 \dots a_\ell} b$ being a simple chain (any $w_g$ may be empty).
We denote by $s_g$ the position of symbol $a_g$, for $g = 1, 2, \ldots, \ell$
and set $a_0 = a$, $s_0 = 0$,  $a_{\ell+1} = b$, and $s_{\ell+1} = |w|+1$.
\item
$\bm{x} \lessdot \bm{y}$ states that the symbol in position $\bm{x}$ yields precedence to the one in position $\bm{y}$  and similarly for the other precedence relations
\item
The fundamental abbreviation
\[
	\tree (\g{x},\g{z},\g{v},\g{y}) := 
	\mu(\g{x} , \g{y})
	\land
	\left(
	\begin{array}{c}
		( \g{x}+1 = \g{z} \ \lor\  \mu(\g{x} , \g{z}) ) \land \\ 
        \neg \exists \g{t} ( \g{z} < \g{t} < \g{y} \land \mu( \g{x} , \g{t}) ) 
		 \land \\
		( \g{v}+1 = \g{y} \ \lor\ \mu(\g{v} , \g{y} )) \land \\ 
        \neg \exists \g{t} ( \g{x} < \g{t} < \g{v}
		\land \mu (\g{t} , \g{y} )) 
	\end{array}
	\right)
\]
is satisfied, for every chain $\chain a w b$ embraced within positions $\g{x}$ and $\g{y}$ by a (unique, maximal) $\g{z}$ such that $\mu(\g{x} , \g{z})$, if  $w_0 \not = \varepsilon$, $\g{z} = \g{x}+1$ if instead $w_0  = \varepsilon$; symmetrically for $\g{y}$ and $\g{v}$.
In particular,  if $w$ is the body of  a simple chain, then $\mu(0 , \ell+1)$ and $\tree(0, 1, \ell, \ell+1)$ are satisfied;
if it is the body of a composed chain, 
then $\mu(0 , |w|+1)$ and   $\tree(0, s_1, s_\ell, s_{\ell+1})$ are satisfied. If $w_0 = \varepsilon$ then $s_1 = 1$,
and if $w_\ell = \varepsilon$ then $s_\ell = |w|$.
In the example of Figure~\ref{fig:log:avv} relations 
$\tree (2,3,3,4)$,
$\tree (2,4,4,10)$,
$\tree (4,5,9,10)$,
$\tree (5,7,7,9)$ are satisfied, among  others.

\item 
	The shortcut $Q_i(\g{x},\g{y})$ is used to represent that $\mathcal{A}$ is in state $q_i$ when at position $\g{x}$ and the next position to read, possibly after scanning a chain, is $\g{y}$. Since the automaton is not real time, we must distinguish between push and shift moves (case $\suc i {\g{x}} {\g{y}}$), and pop moves (case $ \flk i {\g{x}} {\g{y}}$).
    \[
    \begin{array}{rl}
		\suc k {\g{x}} {\g{y}}  := &  
		\g{x}+1 = \g{y} \land \g{x} \in {\pushvar}_k
		\\
		\nex k {\g{x}} {\g{y}}  := &
		\mu(\g{x} , \g{y}) \land \g{x} \in {\markvar}_k \land 
        \\
        & \exists \g{z}, \g{v}\  (\tree(\g{x},\g{z},\g{v},\g{y}) \land \g{v} \in {\flushvar}_k)
		\\
		Q_i(\g{x},\g{y})  := & \suc i {\g{x}} {\g{y}} \lor \flk i {\g{x}} {\g{y}}.
	\end{array}
\]
	E.g., with reference to Figures \ref{fig:log:avv} and \ref{fig:log:sets}, $\suc 2 5 6$, $\nex 3 5 9$, and $\flk 3 5 7$ hold.
\end{itemize}
		
 We can now show a meaningful sample of the various formulas that code the automaton's transition relation.
 
 \begin{itemize}
 \item
 The following formula states that if $\mathcal{A}$ is in position $\bm{x}$ and state $q_i$ and performs a push transition by reading the character in position  $\bm{y}$, it goes to state $q_k$ according to the transition relation $\delta$. 
 	\[
 	\varphi_{push\_fw} := \forall \g{x}, \g{y}  
 	\bigwedge_{i=0}^{N} 
 	\bigwedge_{k=0}^{N} 
 	\bigwedge_{c \in \Sigma}
 	\left(
    \begin{array}{c}
 	\g{x} \lessdot \g{y}  \land c(\g{y})
 	\land
 	Q_i({\g{x}}, {\g{y}})
 	\land \\
 	\delta_{\text{push}}(q_i, c) = q_k \\    
 	\Rightarrow 
    \g{y} \in \pushvar_k
    \end{array}
 	\right)
 	\]
 Notice that the original formula given in Section \ref{sec:reg:logic} for RL can be seen as a particular case of the above one.

\item
 Conversely, if $\mathcal{A}$ is in state $q_k$ after a push starting from position $\bm{x}$ and reading character $c$, in that position it must have been in a state $q_i$ such that $\delta(q_i, c) = q_k$:
 \begin{small}
 \[
 	\varphi_{push\_bw} := \forall \g{x},\g{y}  \bigwedge_{k=0}^{N} 
 	\bigwedge_{c \in \Sigma}
 	\left(
 	\begin{array}{c}
 	\g{x} \lessdot \g{y} \ \land \ 
 	c(\g{y}) \ \land \  \g{y} \in {\pushvar}_k
 	\land \\
 	(\g{x}+1=\g{y} \ \lor  \ \mu(\g{x} , \g{y}))  
 	\\ \Rightarrow \\
 	\bigvee_{i=0}^N  \left( Q_i ({\g{x}}, {\g{y}}) 
 	\land \delta_{\text{push}}(q_i,c) = q_k 
 	\right)
 	\end{array}
 	\right)
 	\]
    \end{small}
 \item
 The formulas coding the shift transitions are similar to the previous ones and therefore omitted.
 \item
 	To define $\varphi_{\delta_\text{pop}}$ we introduce the 
 	shortcut $\tree_{i,j} (\g{x},\g{z},\g{v},\g{y})$, which represents the fact that $\mathcal{A}$ is ready to
 	perform a pop transition from state $q_i$ having on top of the stack state $q_j$;
 	such pop transition corresponds to the reduction of the portion of string between positions $\g{x}$ and $\g{y}$ (excluded).
    \[
 		\tree_{i,j}( \g{x},\g{z},\g{v},\g{y}) :=
 		\tree(\g{x},\g{z},\g{v},\g{y}) \land
 		Q_i(\g{v}, {\g{y}}) 
 		\land 
 		Q_j({\g{x}}, {\g{z}}).
 	\]
 	%
 	Formula $\varphi_{\delta_\text{pop}}$ is thus defined as the conjunction of three formulas. 
 	As before, the forward (abbreviated with the subscript $fw$) formula gives the sufficient condition for two positions to be in the sets $\markvar_k$ and $\flushvar_k$, when performing a pop move,
 	and the backward formulas state symmetric necessary conditions.\\
    \begin{small}
    \[
 		\varphi_{pop\_fw} := 
 		\forall \g{x}, \g{z}, \g{v}, \g{y}
 		\bigwedge_{i=0}^N \bigwedge_{j=0}^N  
 		\bigwedge_{k=0}^{N} 
 		\left(
        \begin{array}{c}
 		\tree_{i,j}(\g{x},\g{z},\g{v},\g{y})
 		\land \\
 		\delta_{\text{pop}}(q_i,q_j) = q_k \\
 		\Rightarrow \\
 		\g{x} \in \markvar_k 
 		\land 
 		\g{v} \in \flushvar_k
        \end{array}
 		\right)
        \]
        \[
 		\varphi_{pop\_bwB} := 
 		\forall \g{x} \bigwedge_{k=0}^N \left( 
        \begin{array}{c}
        \g{x} \in {\markvar}_k
        \Rightarrow \\
 		\exists \g{y},\g{z},\g{v} 
 		\bigvee_{i=0}^N \bigvee_{j=0}^N  
 		\tree_{i,j}(\g{x},\g{z},\g{v},\g{y}) 
 		\land \\
 		\delta_{\text{pop}}(q_i,q_j) = q_k 
        \end{array}
 		\right) 
        \]
        \[
 		\varphi_{pop\_bwC} := 
 		\forall \g{v} \bigwedge_{k=0}^N \left( 
        \begin{array}{c}
        \g{v} \in {\flushvar}_k 
        \Rightarrow \\
 		\exists \g{x},\g{y},\g{z} 
 		\bigvee_{i=0}^N \bigvee_{j=0}^N  
 		\tree_{i,j}(\g{x},\g{z},\g{v},\g{y}) 
 		\land \\
 		\delta_{\text{pop}}(q_i,q_j) = q_k 
        \end{array}
 		\right)
        \]
    \end{small}
 \end{itemize}
 
\subsection{Local parsability for parallel parsers}\label{sec: OPParPars}
Let us now go back to the original motivation that inspired R. Floyd when he invented the OPG family, namely supporting efficient, deterministic parsing. In the introductory part of this section we noticed that the mechanism of precedence relations isolates  grammar's rhs from their context so that they can be reduced to the corresponding lhs independently from each other. This fact guarantees that, in whichever order such reductions are applied, at the end a complete  grammar derivation will be built; such a derivation corresponds to a visit of the syntax-tree, not necessarily leftmost or rightmost, and its \emph{construction has no risk of applying any back-track} as it happens instead in nondeterministic parsing. We call this property \emph{local parsability property}, which intuitively can be defined as the possibility of applying deterministically a bottom-up, shift-reduce parsing algorithm by inspecting only an a priori bounded portion of any string containing a rhs. Various formal definitions of this concept have been given in the literature, the first one probably being the one proposed by Floyd himself in \cite{DBLP:journals/cacm/Floyd64}; a fairly general definition of local parsability and a proof that OPG enjoy it can be found in \cite{BarenghiEtAl2015}.

Local parsability, however, has the drawback that it loses chances of deterministic parsing when the information on how to proceed with the parsing is arbitrarily far from the current position of the parser. We therefore have a trade-off between the size of the family of recognizable languages, which in the case of LR grammars is the whole DCFL class (see Section~\ref{LRPars}), and the constraint of proceeding rigorously left-to-right for the parser. So far this trade-off has been normally solved in favor of the generality in the absence of serious counterparts in favor of the other option. We argue however, that the massive advent of parallel processing, even in the case of small architectures such as those of tablets and smartphones, could dramatically change the present state of affairs. On the one side parallelizing parsers such as LL or LR ones requires reintroducing a kind of nondeterministic \textit{guess} on the state of the parser in a given position, which in most cases voids the benefits of exploiting parallel processors (see \cite{BarenghiEtAl2015} for an analysis of previous literature on various attempts to develop parallel parsers); on the contrary, OPL are from the beginning oriented toward parallel analysis whereas their previous use in compilation shows that they can be applied to a wide variety of practical languages, and further more as suggested by other examples given here and in \cite{LonatiEtAl2015}.

Next we show how we exploited the local parsability property of OPG to realize a complete and general parallel parser for these grammars.
A first consequence of the basic property is the following statement.
\begin{statement}\label{cor:locpars}
For every substring $a \delta b$ of  $\gamma a \delta b \eta$ $\in V^*$ derivable from $S$, there exists a unique string  $\alpha$, called the {\em irreducible string}, deriving $\delta$ such that 
$S \derives{*} \gamma a \alpha b \eta \derives{*} \gamma a \delta b \eta$, and the precedence relations between the consecutive terminals of $a \alpha b$ do not contain the pattern $\lessdot \left( \doteq \right)^* \gtrdot$.  
Therefore there exists a factorization $a \alpha b = \zeta  \theta$  into two possibly empty factors such that the left factor does not contain  $\lessdot$ and  the right factor does not contain $\gtrdot$.
\end{statement}

On the basis of the above statement, the original parsing algorithm is generalized in such a way that it may receive as input a portion of a string, not necessarily enclosed within the delimiters $\#$, and produces as output two stacks, one that stores the substring $\zeta$ and one that stores $\theta$ as defined in Statement \ref{cor:locpars}, and a partial derivation of $a \alpha b = \zeta  \theta \derives{*} a\,\delta\,b\, $. For instance, with reference to the grammar $GAE_{FNF}$, which is a FNF of $GAE_1$, if we supply to such a generalized parser the partial string $+e*e*e+e$ we obtain $\zeta = + T +$, $\theta = e$ and $+T+ \derives{*} +e*e*e+$ since $+ \gtrdot +$ and $+ \lessdot e$. We call $\mathcal{S}^L$ and $\mathcal{S}^R$ the two stacks produced by this partial parsing.

At this point it is fairly easy to let several such generalized parsers work in parallel:
\begin{itemize}
\item
Suppose to use $k$ parallel processors, also called \emph{workers}; then split the input into $k$ chunks; given that an OP parser needs a look-ahead-look-back of one character, the chunks must overlap by one character for each consecutive pair. For instance, the global input string $\#e+e+e*e*e+*e+e\#$, with $k=3$ could be split as shown below:
\[
\# \overbrace{e \; + \; e}^1 \; \; + \; \; \overbrace{e \; * \; e \; * \; e \; +}^2 \; \; e \; \; \overbrace{* \; e \; + \; e}^3 \#
\]
where the unmarked symbols $+$ and $e$ are shared by the adjacent segments. The splitting can be applied arbitrarily, although in practice it seems natural to use segments of approximately equal length and/or to apply some heuristic criterion (for instance, if possible one should avoid particular cases where only $\lessdot$ or $\gtrdot$ relations occur in a single chunk so that the parser could not produce any reduction).
\item
Each chunk is submitted to one of the workers which produces a partial result in the form of the pair ($\mathcal{S}^L , \mathcal{S}^R$) (notice that some of those partial stacks may be empty).
\item
The partial results are concatenated into a new string $\in V^*$ and the process is iterated until a short enough single chunk is processed and the original input string is accepted or rejected. In practice it may be convenient to build the new segments to be supplied to the workers by facing an $\mathcal{S}^R$ with the following $\mathcal{S}^L$ so that the likelihood of applying many new reductions in the next pass is increased. For instance the $\zeta = + T +$ part produced by the parsing of the second chunk could be paired with the $\mathcal{S}^R = \#E+$ part obtained from the parsing of the first chunk, producing the string $\#E+T+$ to be supplied to a worker for the new iteration. Some experience shows that quite often optimal results in terms of speed-up are obtained with 2, at most 3 passes of parallel parsing.
\end{itemize}

\cite{BarenghiEtAl2015} describes in detail \textit{PAPAGENO, a PArallel PArser GENeratOr} built on the basis of the above algorithmic schema. It has been applied to several real-life data definition, or programming, languages including JSON, XML, and Lua and different HW architectures. The paper also reports on the experimental results in terms of the obtained speed-up compared with standard sequential parser generators as Bison. PAPAGENO is freely available at 
\url{https://github.com/PAPAGENO-devels/papageno} under GNU license.

\section{Concluding remarks}\label{Concl}
The main goal of this paper has been to show that an old-fashioned and almost abandoned family of formal languages indeed offers considerable new benefits in apparently unrelated application fields of high interest in modern applications, i.e., automatic property verification and parallelization. In the first field OPL significantly extend the generative power of the successful class of VPL still maintaining all of their properties: to the best of our knowledge, OPL are the largest class of languages closed under all major language operations and provided with a complete classification in terms of MSO logic.

Various other results about this class of languages have been obtained or are under development, which have not been included in this paper for length limits. We mention here just the most relevant or promising ones with appropriate references for further reading.
\begin{itemize}
\item
The theory of OPL for languages of finite length strings has been extended in \cite{LonatiEtAl2015} to so called $\omega$-languages, i.e. languages of infinite length strings: the obtained results perfectly parallel those originally obtained by B\"uchi and others for RL and subsequently extended to other families, noticeably VPL \cite{jacm/AlurM09}; in particular, $\omega$-OPL lose determinizability in case of B\"uchi acceptance criterion as it happens for RL and VPL.
\item
Some investigation is going on to devise more tractable automatic verification algorithms than those allowed by the full characterization of these languages in terms of MSO logic. On this respect, the state of the art is admittedly still far from the success obtained with model checking exploiting various forms of temporal logics for FSA and several extensions thereof such as, e.g., timed automata \cite {AlurDill1994a}. Some interesting preliminary results have been obtained for VPL by \cite{lmcs/AlurABEIL08} and for a subclass of OPL in \cite{LMPP15}.
\item
The local parsability property can be exploited not only to build parallel parsers but also to make them \textit{incremental}, in such a way that when a large piece of text or software code is locally modified its analysis should not be redone from scratch but only the affected part of the syntax-tree is ``plugged'' in the original one with considerable saving; furthermore incremental and/or parallel parsing can be naturally paired with incremental and/or parallel semantic processing, e.g. realized through the classic schema of \emph{attribute evaluation} \cite {Knuth68,CBM13}. Some early results on incremental software verification by exploiting the locality property are reported in \cite{Bianculli2013ArXiv}. We also mention ongoing work on parallel XML-based query processing.
\item
A seminal paper by Sch\"utzemberger \cite{Schutzenberger61} introduced the concept of \emph{weighted languages} as  RL where each word is given a weight in a given algebra which may represent some ``attribute'' of the word such as importance or probability. 
Later, these weighted languages too have been characterized in terms of MSO logic \cite{DBLP:journals/tcs/DrosteG07} and such a characterization has also been extended to VPL \cite{DBLP:journals/ijfcs/DrosteP14} and $\omega$-VPL \cite{DBLP:journals/corr/DrosteD15}. Our two research groups are both confident that those results can also be extended to \emph{weighted OPL} and are starting a joint investigation on this promising approach.
\end{itemize}

\noindent
\small{{\em Acknowledgments.} We acknowledge the contribution to our research given by Alessandro Barenghi,
Stefano Crespi Reghizzi, Violetta Lonati, Angelo Morzenti, and Federica Panella.
}

\bibliography{opbib}

\end{document}